\pdfoutput=1

\newcommand*{\ATLASLATEXPATH}{}
\documentclass[UKenglish,texlive=2016,cernpreprint]{\ATLASLATEXPATH atlasdoc}

\usepackage[subfigure=true]{\ATLASLATEXPATH atlaspackage}
\usepackage{\ATLASLATEXPATH atlasbiblatex}
\usepackage{\ATLASLATEXPATH atlasphysics}

\usepackage{HIGG-2017-03-PAPER-defs}

\usepackage{multirow}
\usepackage{amssymb}

\AtlasTitle{Search for the Standard Model Higgs boson produced in association with top quarks and decaying into a $\boldsymbol{b\bar{b}}$ pair in $\boldsymbol{pp}$ collisions at $\sqrt{\textbf{s}}$~$\boldsymbol{=}$~13~\TeV\ with the ATLAS detector}

\author{The ATLAS Collaboration}
\AtlasRefCode{HIGG-2017-03}
\PreprintIdNumber{CERN-EP-2017-291}
\AtlasJournal{Phys.\ Rev.\ D.}
\AtlasJournalRef{Phys.\ Rev.\ D. 97 (2018) 072016}
\AtlasDOI{10.1103/PhysRevD.97.072016}

\AtlasAbstract{
A search for the Standard Model Higgs boson produced in association with a top-quark pair, $t\bar{t}H$, is presented.
The analysis uses 36.1~fb$^{-1}$~of $pp$ collision data at $\sqrt{s}$~=~13 \tev{} collected with the ATLAS detector at the Large Hadron Collider in 2015 and 2016.
The search targets the $H \to b\bar{b}$ decay mode.
The selected events contain either one or two electrons or muons from the top-quark decays, and are then categorized according to the number of jets and how likely these are to contain $b$-hadrons.
Multivariate techniques are used to discriminate between signal and background events, the latter being dominated by $t\bar{t}$ + jets production. 
For a Higgs boson mass of 125~\GeV, the ratio of the measured $t\bar{t}H$ signal cross-section to the Standard Model expectation is found to be $\mu = 0.84^{+0.64}_{-0.61}$. 
A value of $\mu$ greater than 2.0 is excluded at 95\% confidence level while the expected upper limit is $\mu < 1.2$ in the absence of a $t\bar{t}H$ signal.
}

\hypersetup{pdftitle={ATLAS document},pdfauthor={The ATLAS Collaboration}}

\begin{document}

\maketitle

\tableofcontents

\section{Introduction}
\label{sec:intro}
After the discovery of the Higgs boson~\cite{Englert:1964aa,Higgs:1964aa,Guralnik:1964aa} 
in 2012 by the ATLAS~\cite{HIGG-2012-27} and CMS~\cite{CMS-HIG-12-028} collaborations, attention has
turned to more detailed measurements of its properties and couplings as a means of testing the predictions
of the Standard Model (SM)~\cite{Glashow:1961aa,Weinberg:1967aa,Salam:1969aa}.
In particular, the coupling to the top quark, the heaviest particle in the SM, could be very sensitive to effects of physics beyond the SM (BSM)~\cite{Englert:2014uua}. 
Assuming that no BSM particle couples to the Higgs boson, the ATLAS and CMS experiments measured a value of the top-quark's Yukawa coupling equal to $0.87\pm0.15$ times the SM prediction by combining~\cite{HIGG-2015-07} their respective Higgs-boson measurements from the Run 1 
dataset collected at center-of-mass energies of 7~\TeV\ and 8~\TeV\ at the Large Hadron Collider (LHC). 
This measurement relies largely on the gluon--gluon fusion production mode and on the decay mode to photons, which both depend on loop contributions with a top quark.
If no assumption is made about the particle content of such loop contributions, then the top-quark coupling is only determined through tree-level processes, and a value of $1.4\pm0.2$ times the SM prediction is obtained.

\begin{figure*}[ht!]
\centering
\subfigure[]{\includegraphics[width=0.31\textwidth]{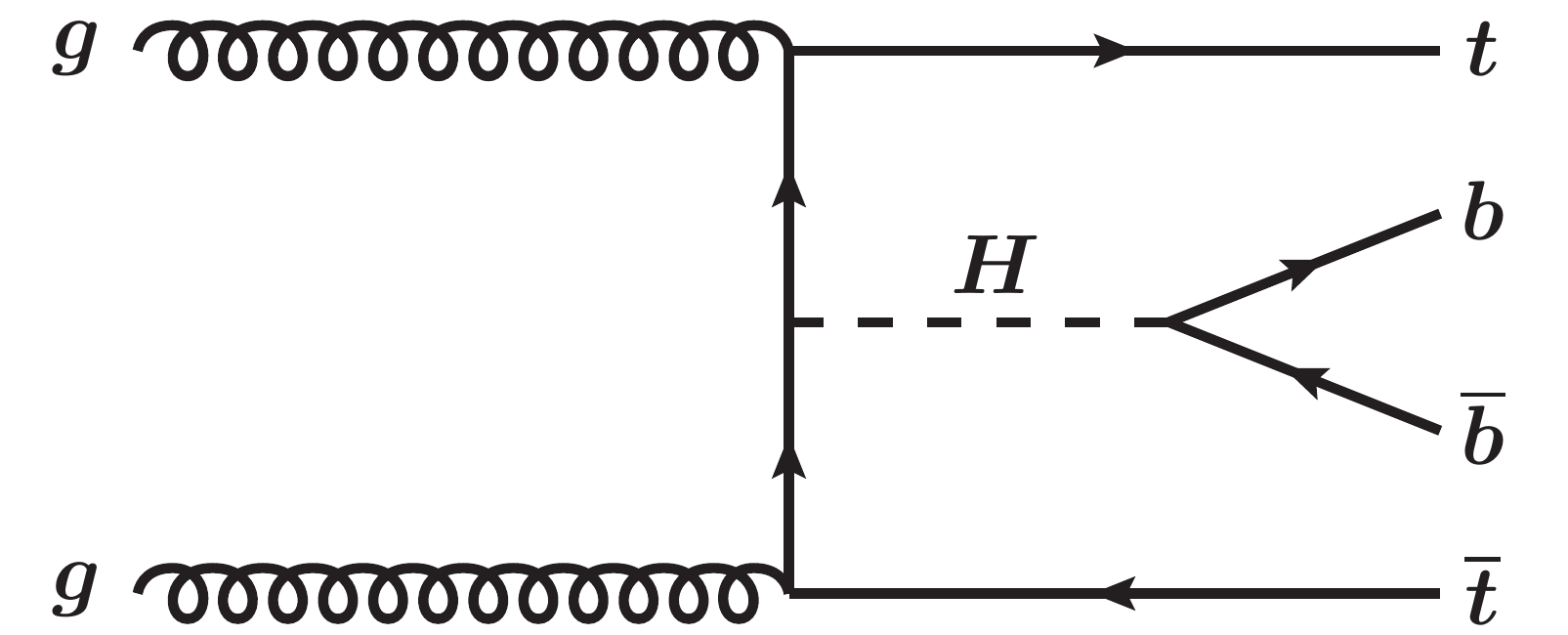}}\label{fig:feyman_a}\hspace{0.5cm}
\subfigure[]{\includegraphics[width=0.31\textwidth]{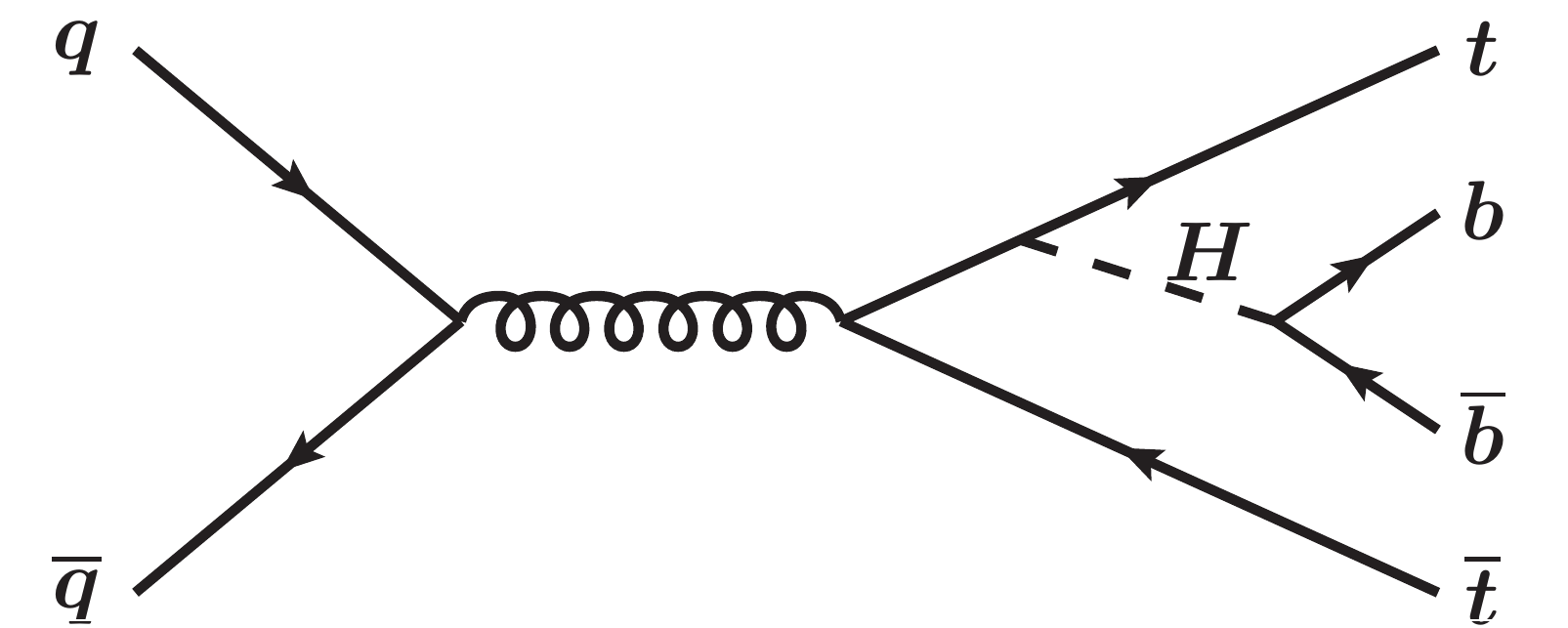}}\label{fig:feyman_b}\hspace{0.5cm}
\subfigure[]{\includegraphics[width=0.31\textwidth]{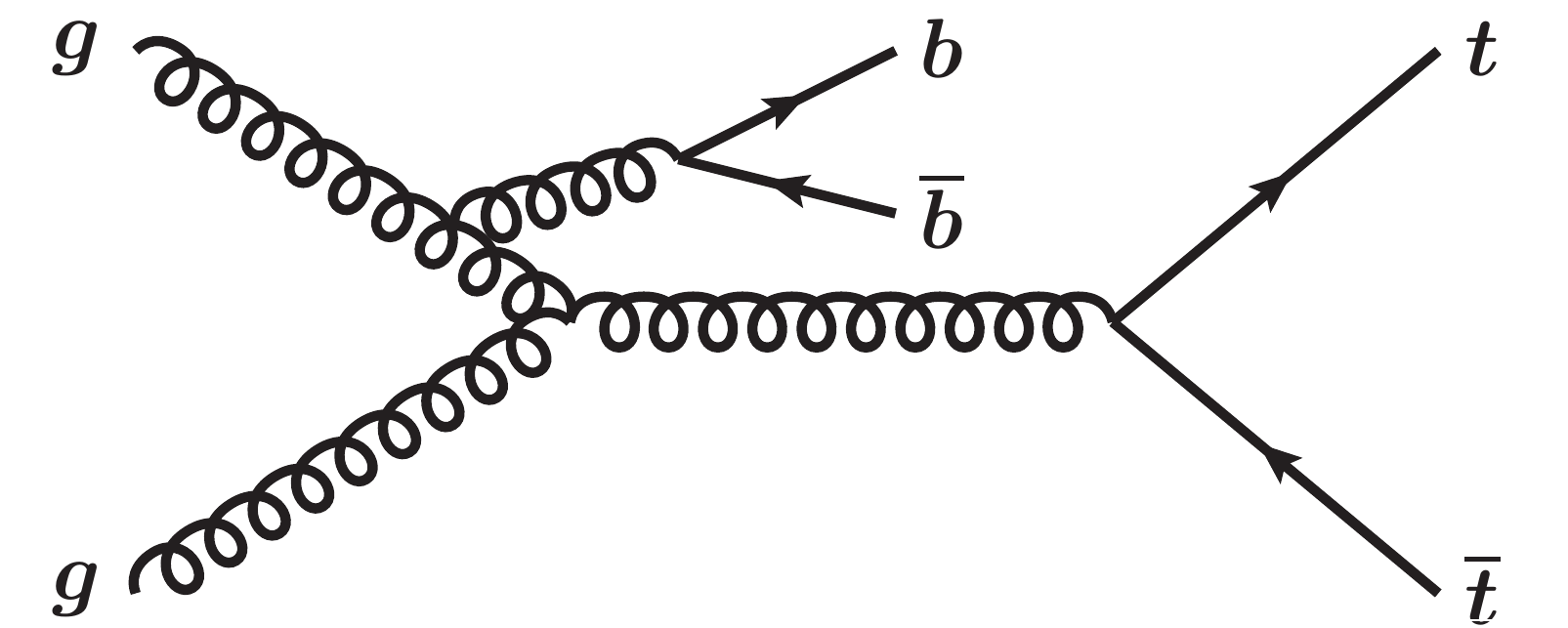}}\label{fig:feyman_c}
\caption{
Representative tree-level Feynman diagrams for (a) $t$-channel and (b) $s$-channel production of
the Higgs boson in association with a top-quark pair (\ttH) and the subsequent decay of the Higgs boson to \bbbar,
and (c) for the main background, $\ttbar+\bbbar$. 
\label{fig:feynman}}
\end{figure*}

Higgs-boson production in association with a pair of top quarks, \tth, is the most favorable production mode for 
a direct 
measurement of the 
top-quark's Yukawa coupling~\cite{ttH1,ttH2,ttH4,Beenakker:2001rj}. 
Although this production mode only contributes around 1\% of the total Higgs-boson production cross-section~\cite{deFlorian:2016spz},
the top quarks in the final state offer a distinctive signature and allow many
Higgs-boson decay modes to be accessed. Of these, the decay to two $b$-quarks is predicted to have a branching fraction of about 58\%~\cite{deFlorian:2016spz}, the largest Higgs-boson decay mode.
This decay mode is sensitive to the $b$-quark's Yukawa coupling, the second largest in the SM.
In order to select events at the trigger level and reduce the backgrounds,
the analysis targets events in which one or both top quarks decay semi-leptonically, producing an electron or a muon.%
\footnote{Throughout this document, `lepton' refers to 
electron or muon, unless otherwise specified. Electrons and muons 
from the decay of a $\tau$ itself originating from a $W$ boson are included.
} 
The main experimental challenges for this channel are the low combined
efficiency to reconstruct and identify all final-state particles, the combinatorial ambiguity from the many jets containing $b$-hadrons in the
final state which makes it difficult to reconstruct the Higgs boson, and the large backgrounds from the production of \ttbar{} + jets especially when the associated jets stem from $b$- or $c$-quarks.
Some representative Feynman diagrams for the \tth{} signal are shown in Figure~\ref{fig:feynman}, together with the dominant \ttbb\
background.

The ATLAS collaboration searched for \tth\ production with Higgs-boson decays to $b\bar{b}$ at $\sqrt{s}=8$~\TeV, using $\ttbar$ decays with
at least one lepton~\cite{HIGG-2013-27} or no leptons~\cite{HIGG-2015-05}. A combined signal strength $\mu = \sigma/\sigma_{\mathrm{SM}}$
of $1.4 \pm 1.0$ was measured.
The CMS collaboration searched for the same process at $\sqrt{s}=7$~\TeV\ and $\sqrt{s}=8$~\TeV\ using \ttbar{} decays with a single-lepton or dilepton in the final state, obtaining a signal strength of $0.7 \pm 1.9$~\cite{CMS-HIG-13-029}.
These results were combined with each other, and with results for Higgs boson decay to vector bosons, to $\tau$-leptons or to photons~\cite{HIGG-2013-26,HIGG-2013-25,CMS-HIG-13-029},
resulting in an observed (expected) significance of 4.4 (2.0) standard deviations for \tth{} production~\cite{HIGG-2015-07}.
The measured signal strength is $2.3^{+0.7}_{-0.6}$. 

In this article, a search for \tth\ production with 36.1~\ifb~of $pp$ collision data at $\sqrt{s}=13~\tev$ is presented.
The analysis targets Higgs-boson decays to $b$-quarks, but all the decay modes are considered and may contribute to the signal.
Events with either one or two leptons are taken into account, and exclusive 
analysis categories 
are defined 
according to the number of leptons, the number of jets, and the value of a $b$-tagging discriminant which provides a measure of how likely a jet is to contain a $b$-hadron.
In the single-lepton channel, a specific category,  referred to as `boosted’ in the following, is designed to select events containing a Higgs boson and with at least one of the two top quarks produced at high transverse momentum.
In the analysis categories with the largest signal contributions, multivariate discriminants are used to classify events as more or less signal-like. 
The signal-rich categories are analyzed together with the signal-depleted ones in a combined profile
likelihood fit that simultaneously determines the event yields for the signal and for the most important background components, while constraining the overall background model within the assigned systematic uncertainties.
The combination of the results presented in this article with the results from other analyses targeting \tth{} production with different final states is reported in Ref.~\cite{ttHML2017}.

The article is organized as follows. The ATLAS detector is described in Section~\ref{sec:detector}.
Section~\ref{sec:selection} summarizes the selection criteria applied to events and physics objects.
The signal and background modeling are presented in Section~\ref{sec:modelling}.
Section~\ref{sec:analysis} describes the event categorization while Section~\ref{sec:mvas} presents the multivariate analysis techniques. 
The systematic uncertainties are 
summarized in Section~\ref{sec:systs}.
Section~\ref{sec:results} presents the results and Section~\ref{sec:conclusion} gives the conclusions.

\section{ATLAS detector}
\label{sec:detector}
The ATLAS detector~\cite{PERF-2007-01} at the LHC 
covers nearly the entire solid angle\footnote{ATLAS 
uses a right-handed coordinate system with its origin at the nominal interaction point (IP) in the 
center of the detector and the $z$-axis coinciding with the axis of the beam pipe.  The $x$-axis points from
the IP to the center of the LHC ring, and the $y$-axis points upward. Cylindrical coordinates ($r$,$\phi$) are used 
in the transverse plane, $\phi$ being the azimuthal angle around the beam pipe. The pseudorapidity is defined in 
terms of the polar angle $\theta$ as $\eta = - \ln \tan(\theta/2)$. 
Unless stated otherwise, angular distance is measured in units 
of $\Delta R \equiv \sqrt{(\Delta\eta)^{2} + (\Delta\phi)^{2}}$.} around the collision point. 
It consists of an inner tracking detector surrounded by a thin superconducting solenoid magnet producing a
2 T axial magnetic field, electromagnetic and hadronic calorimeters, and an external muon spectrometer (MS)
incorporating three large toroid magnet assemblies.
The inner detector (ID) consists of a high-granularity silicon pixel detector
and a silicon microstrip tracker,
together providing precision tracking in the pseudorapidity range $|\eta|<2.5$, complemented by a straw-tube
transition radiation tracker providing tracking and electron identification information for $|\eta|<2.0$.
A new innermost silicon pixel layer, the insertable B-layer~\cite{IBL} (IBL),
was added to the detector between Run 1 and Run 2. The IBL improves the ability to identify displaced
vertices and thereby significantly improves the $b$-tagging performance~\cite{ATL-PHYS-PUB-2015-022}.
The electromagnetic sampling calorimeter uses lead or copper as the absorber material 
and liquid argon (LAr) as the active medium, and is divided into barrel ($|\eta|<1.475$), endcap ($1.375<|\eta|<3.2$) and forward ($3.1<|\eta|<4.9$) regions.
Hadron calorimetry is also based on the sampling technique and covers $|\eta|<4.9$, with either scintillator tiles or LAr as the active medium and with 
steel, copper or tungsten as the absorber material. 
The muon spectrometer measures the deflection 
of muons with $|\eta|<2.7$ using multiple layers of high-precision tracking chambers located in a toroidal field.
The field integral of the toroids ranges between 2.0 and 6.0 Tm across most of the detector.
The muon spectrometer is also 
instrumented with separate trigger chambers covering $|\eta|<2.4$.
A two-level trigger system~\cite{TRIG-2016-01}, using custom hardware followed by a software-based level, is used to reduce the trigger rate to an average of around one kHz for offline storage.

\section{Event selection}
\label{sec:selection}
Events are selected from $pp$ collisions at $\sqrt{s}=13$~TeV recorded by the ATLAS detector in 2015 and 2016.
Only events for which all relevant subsystems were operational are considered. Events are required
to have at least one vertex with two or more tracks with transverse momentum $\pt > 0.4$~\GeV. 
The vertex with the largest sum of the squares of the transverse momenta of associated tracks is taken as the primary vertex. 
The event reconstruction is affected by multiple $pp$ collisions in a single bunch crossing and by collisions in neighboring 
bunch crossings, referred to as `pileup'. 
The number of interactions per bunch crossing in this dataset ranges from about 8 to 45 interactions.
The dataset corresponds to an integrated luminosity of
$3.2 \pm 0.1$ fb$^{-1}$ recorded in 2015 and $32.9 \pm 0.7$ fb$^{-1}$ recorded in 2016, 
for a total of $36.1 \pm 0.8$ fb$^{-1}$~\cite{DAPR-2013-01}.

Events in both the single-lepton and dilepton channels were recorded using 
single-lepton triggers. 
Events are required to fire triggers 
with either low lepton \pt\ thresholds and a lepton isolation requirement, 
or with higher thresholds but with a looser identification criterion and without any isolation requirement. 
The lowest \pt\ threshold used for muons is 20 (26)~\GeV\ in 2015 (2016), 
while for electrons the threshold is 24 (26)~\GeV.

Electrons are reconstructed from energy deposits (clusters) in the electromagnetic calorimeter matched to tracks 
reconstructed in the ID~\cite{PERF-2013-03, ATLAS-CONF-2016-024} 
and are required to have $\pt>10$~\GeV\ and $|\eta|<2.47$. 
Candidates in the calorimeter 
barrel--endcap 
transition region 
($1.37 <|\eta| < 1.52$) are excluded. 
Electrons must satisfy the \textit{loose} identification criterion described in Ref.~\cite{ATLAS-CONF-2016-024}, based on
a likelihood discriminant combining observables related to the shower shape in the calorimeter and to the track matching the electromagnetic cluster.
Muons are reconstructed from either track segments or full tracks in the MS
which are matched to tracks in the ID~\cite{PERF-2015-10}.
Tracks are then re-fitted using information from both detector systems. 
Muons are required to have $\pt>10$~\GeV\ and $|\eta|<2.5$. 
To reduce the contribution of leptons from hadronic decays (non-prompt leptons),
both electrons and muons must satisfy isolation criteria based on information from both the tracker and the calorimeter.
The \textit{loose} lepton isolation working point~\cite{ATLAS-CONF-2016-024,PERF-2015-10} is used.
Finally, lepton tracks must match the primary vertex of the
event:
the longitudinal impact parameter $\mathrm{{IP}_{z}}$ is required to satisfy $|\mathrm{{IP}_{z}}|<0.5$~mm, while the transverse impact parameter significance, $|\mathrm{IP_{r\phi}}|/\sigma_{\mathrm{IP_{r\phi}}}$, must be less than 5 for electrons and 3 for muons.

Jets are reconstructed from three-dimensional topological energy clusters~\cite{PERF-2014-07} in the calorimeter 
using the \antikt\ jet algorithm~\cite{Cacciari:2008gp} implemented in the FastJet package~\cite{Cacciari:2011ma} with a radius parameter of 0.4. Each topological cluster 
is calibrated to the electromagnetic scale response prior to jet reconstruction. The reconstructed jets are then 
calibrated to the jet energy scale derived from simulation and in situ corrections based on 13 TeV 
data~\cite{PERF-2016-04}.
After energy calibration, jets are required to have $\pt > 25$~\gev{} and $|\eta| < 2.5$.  
Quality criteria are imposed to identify jets arising from non-collision sources or detector noise, and any event 
containing such a jet is removed~\cite{ATLAS-CONF-2015-029}. 
Finally, to reduce the effect of pileup, an additional requirement is made using an algorithm that matches jets with $\pt < 60$~\GeV\ and $|\eta| < 2.4$ to tracks with $\pt > 0.4$~\GeV\ to identify jets consistent with the primary vertex. This algorithm is known as jet vertex tagger~\cite{PERF-2014-03}, referred to as JVT in the remainder of this article.

Jets are tagged as containing $b$-hadrons through a multivariate $b$-tagging algorithm (MV2c10) that combines information from an impact-parameter-based algorithm, from the explicit reconstruction of an inclusive secondary vertex and from a multi-vertex fitter that attempts to reconstruct the $b$- to $c$-hadron decay chain~\cite{PERF-2012-04,ATL-PHYS-PUB-2016-012}.
This algorithm is optimized to efficiently select jets containing $b$-hadrons ($b$-jets) and separate them from jets containing $c$-hadrons ($c$-jets), jets containing hadronically decaying $\tau$-leptons ($\tau$-jets) and from other jets (light jets).
Four working points are defined by different MV2c10 discriminant output thresholds and are 
referred to in the following as 
\textit{loose}, \textit{medium}, \textit{tight} and \textit{very tight}.
The efficiency for $b$-jets with $\pT>20$~\GeV\ in simulated \ttbar\ events to pass the different working points are 85\%, 77\%, 70\% and 60\%, respectively, corresponding to rejection factors\footnote{The rejection factor is defined as the inverse of the efficiency to pass a given $b$-tagging working point.} of $c$-jets in the range 3--35 and of light jets in the range 30--1500.
A $b$-tagging discriminant value is assigned to each jet according to the tightest working point it satisfies, ranging from 1 for 
a jet that does not satisfy any of the $b$-tagging criteria defined by the considered working points 
up to 5 for jets satisfying the \textit{very tight} criteria. 
This $b$-tagging discriminant is used to categorize selected events as discussed in Section~\ref{sec:analysis} and as an input to multivariate analysis techniques described in Section~\ref{sec:mvas}.

Hadronically decaying $\tau$ leptons (\tauhad) are distinguished from jets using the track multiplicity
and a multivariate discriminant based on the track collimation, further jet substructure, and kinematic information~\cite{ATL-PHYS-PUB-2015-045}.
These \tauhad{} candidates 
are required to have $\pt>25$~\GeV, $|\eta| < 2.5$ and pass the \textit{Medium} $\tau$-identification working point.

To avoid counting a single detector response as more than one lepton or jet, an overlap removal procedure is adopted.
To prevent double-counting of electron energy deposits as jets,
the closest jet within 
$\Delta R_{y}=\sqrt{(\Delta y)^2+(\Delta \phi)^2}=0.2$ 
of a selected electron is removed.\footnote{The rapidity is defined as $y=\frac{1}{2} \ln\frac{E+p_{\text{z}}}{E-p_{\text{z}}}$ where $E$ is the energy and $p_{\text{z}}$ is the longitudinal component of the momentum along the beam pipe.}
If the nearest jet surviving that selection is within
$\Delta R_{y} = 0.4$ of the electron, the electron is discarded.
Muons are removed if they are separated from the nearest jet by $\Delta R_{y} < 0.4$,
which reduces the background from 
heavy-flavor decays inside jets.
However, if this jet has fewer than three associated tracks,
the muon is kept and the jet is removed instead;
this avoids an inefficiency for high-energy muons undergoing significant energy loss in the calorimeter. 
A \tauhad{} candidate is rejected if it is separated by $\Delta R_{y} < 0.2$ from any selected electron or muon. 

The missing transverse momentum in the event is defined as the negative vector sum of the \pt of 
all the selected electrons, muons and jets described above, 
with an extra term added to account for
energy in the event which is not associated with any of 
these. 
This extra term, referred to as the `soft term' in the following, 
is calculated from ID tracks matched to the primary vertex to make it resilient to pileup 
contamination~\cite{ATL-PHYS-PUB-2015-027,PERF-2014-04}. The missing transverse momentum is not used for event selection but 
it is included in the 
inputs to the multivariate discriminants that are built in the most sensitive analysis categories.

For the boosted category, the selected jets are 
used as inputs for further jet reclustering~\cite{Nachman:2014kla} 
through an \antikt\ algorithm with a radius parameter of $R = 1.0$, resulting in a collection of large-$R$ jets. 
Large-$R$ jets with a reconstructed invariant mass lower than 50~\GeV\
are removed.
The resulting large-$R$ jets are used to identify top quarks and Higgs bosons in signal events when these have high transverse momenta (boosted) and decay into collimated hadronic final states.
Boosted Higgs-boson candidates are required to have $\pt>200$~\GeV\ 
and contain at least two constituent jets, among which at least two are $b$-tagged at the \textit{loose} working point.
If more than one boosted Higgs-boson candidate is identified, 
the one with the highest sum of constituent-jet $b$-tagging discriminants is selected.
Additional large-$R$ jets are considered as potential boosted top-quark candidates. 
Boosted top-quark candidates are required to have $\pt>250$~\GeV,
exactly one constituent jet satisfying the \textit{loose} $b$-tagging working point plus at least one additional constituent jet which is not $b$-tagged.
If more than one boosted top-quark candidate is identified, the one with the highest mass is selected.

Events are required to have at least one reconstructed lepton with $\pt > 27$~\GeV\ matching a lepton with the same flavor reconstructed by the trigger algorithm within $\Delta R < 0.1$.
Events in the dilepton channel must have exactly two leptons with opposite electric charge.
The subleading lepton \pt\ 
must be above 15~\GeV\ in the $ee$ channel or above 10~\GeV\ in the $e\mu$ and $\mu\mu$ channels. In the $ee$ and $\mu\mu$ channels, the dilepton 
invariant mass must be above 15 \GeV\ and outside of the \Zboson-boson mass window 83--99 \GeV.
To maintain orthogonality with other \tth\ search channels~\cite{ttHML2017}, 
dilepton events are vetoed if they contain one or more \tauhad{} candidates.
Events enter the single-lepton channel if they contain exactly one lepton with $\pt > 27$~\GeV\ and no other selected leptons with $\pt > 10$~\GeV. 
In the single-lepton channel, events are removed if they contain two or more \tauhad{} candidates.

To improve the purity in events passing the above selection, selected leptons are further required to satisfy additional identification and isolation criteria, otherwise the corresponding events are removed.
For electrons, the \textit{tight} identification criterion based on a likelihood discriminant~\cite{ATLAS-CONF-2016-024} is used,
while for muons the \textit{medium} identification criterion~\cite{PERF-2015-10} is used.
Both the electrons and muons are required to satisfy the Gradient isolation criteria~\cite{ATLAS-CONF-2016-024,PERF-2015-10}, which become more stringent as the \pt{} of the leptons considered drops.

Finally, events in the dilepton channel must have at least three jets, of which at least two must be $b$-tagged at the \textit{medium} working point. 
Single-lepton events containing at least one boosted Higgs-boson candidate, at least one boosted top-quark candidate 
and at least one additional jet $b$-tagged at the \textit{loose} working point enter the boosted category. 
Events that do not enter the boosted category and have at least five jets, with at least two of them $b$-tagged at the \textit{very tight} working point or three of them $b$-tagged at the \textit{medium} working point, are classified as `resolved' single-lepton events.
The fraction of simulated \ttH(\htobb) events passing the dilepton event selection is 2.5\%. These fractions are 8.7\% for the resolved single-lepton channel and 0.1\% for the boosted category.

\section{Signal and background modeling}
\label{sec:modelling}
This section describes the simulation and data-driven techniques used to model the \ttH\ signal and
the background processes, to train the multivariate discriminants and to define the templates for the signal extraction fit.
In this analysis, most Monte Carlo (MC) samples were produced using the full ATLAS detector simulation~\cite{Aad:2010ah} 
based on {\GEANT}4~\cite{Agostinelli:2002hh}.
A faster simulation, where the full \GEANT4 simulation of the calorimeter response is replaced by a detailed parameterization of the shower shapes~\cite{ATL-PHYS-PUB-2010-013}, was adopted for some of the samples used to estimate modeling systematic uncertainties.
To simulate the effects of pileup, additional interactions were generated using \textsc{Pythia}~8.186~\cite{Pythia8} and overlaid onto the simulated hard-scatter event.
Simulated events are reweighted to match the pileup conditions observed in the data.
All simulated events are processed 
through 
the same reconstruction algorithms and analysis chain as 
the data.
In the simulation, the top-quark mass is assumed to be $m_t=172.5$~\gev. 
Decays of $b$- and $c$-hadrons were 
performed by 
\textsc{EvtGen} v1.2.0~\cite{Lange:2001uf}, except in samples simulated
by the \textsc{Sherpa} event generator.

\subsection{Signal modeling}

The \ttH\ signal process was modeled using \textsc{MadGraph5}\_aMC@NLO~\cite{Alwall:2014hca} 
(referred to in the following as \amcnlo) version 2.3.2 for the matrix element (ME) calculation at next-to-leading-order (NLO) accuracy in quantum chromodynamics (QCD), 
interfaced to the \textsc{Pythia}~8.210
parton shower (PS) and hadronization model using the A14 set of tuned parameters~\cite{ATL-PHYS-PUB-2014-021}. 
The NNPDF3.0NLO parton distribution function (PDF) set~\cite{NNPDF} was used, and the factorization 
and renormalization scales were set to $\mu_{\textrm{F}} = \mu_{\textrm{R}} = \HT / 2$, 
with $\HT$ defined as 
the scalar sum of the transverse masses $\sqrt{\pT^2 + m^2}$ of all final-state particles. 
The top quarks were decayed using \textsc{MadSpin}~\cite{Artoisenet:2012st}, preserving all spin correlations.
The Higgs-boson mass was set to 125 \GeV\ and all decay modes were considered. The \ttH\ cross-section of $507^{+35}_{-50}$~fb
was computed~\cite{Raitio:1978pt,Beenakker:2002nc,Dawson:2003zu,Yu:2014cka,Frixione:2015zaa,deFlorian:2016spz} at NLO accuracy in QCD  and includes NLO electroweak corrections.
The branching fractions were calculated using {\textsc{HDECAY}}~\cite{Djouadi:1997yw,deFlorian:2016spz}.

\subsection{\texorpdfstring{$\ttbar$}{ttbar} + jets background}
\label{sec:modeling_ttbar}

The nominal sample used to model the \ttbar\ background
was generated using the \powhegbox\ v2 NLO event
generator~\cite{Nason:2004rx,Frixione:2007vw,Alioli:2010xd,Powhegttbar}, 
referred to as \powheg\ in the remainder of this article,
with the NNPDF3.0NLO PDF set. 
The \hdamp\ parameter, which controls 
the transverse momentum of the first gluon emission beyond the Born configuration, was set to 1.5 times 
the top-quark mass \cite{ATL-PHYS-PUB-2016-020}. The parton shower and the hadronization were modeled by  
\textsc{Pythia}~8.210
with the A14 set of tuned parameters.
The renormalization and factorization scales were set to the transverse mass of the top quark, 
defined as 
$m_{\textrm{T},t} = \sqrt{m_t^2 + p_{\textrm{T},t}^2}$, 
where $p_{\textrm{T},t}$ 
is the transverse momentum of the top quark in the \ttbar\ center-of-mass reference frame.
The sample is normalized using the predicted cross-section of $832^{+46}_{-51}$~pb, calculated with the Top++2.0 program~\cite{Czakon:2011xx} at next-to-next-to-leading order (NNLO) in perturbative QCD including resummation of next-to-next-to-leading logarithmic (NNLL) soft gluon terms~\cite{Cacciari:2011hy,Cacciari:2011hy,Czakon:2012zr,Czakon:2012pz,Czakon:2013goa}.
Alternative \ttbar\ samples 
used 
to derive systematic uncertainties are described in Section \ref{sec:systs}.

The \ttbar + jets background is categorized according to the flavor of additional jets in the event, using the same
procedure as described in Ref.~\cite{HIGG-2013-27}. Generator-level particle jets are 
reconstructed from stable particles (mean lifetime $\tau > 3 \times 10^{-11}$ seconds)
using the anti-$k_t$ algorithm with a radius 
parameter $R=0.4$, and are required to have $\pt>15$~\GeV\ and $|\eta|<2.5$.
This categorization employs a jet flavor-labeling procedure that is more refined than the one described in Section~\ref{sec:selection}.
The flavor of a jet is determined by counting the number of $b$- or $c$-hadrons 
within $\Delta R < 0.4$ of the jet axis. 
Jets matched to exactly one $b$-hadron, with \pt\ above 5~\GeV, are labeled single-$b$-jets, while those matched to two
or more $b$-hadrons are labeled $B$-jets (with no \pt\ requirement on the second hadron); single-$c$- and $C$-jets are 
defined analogously, 
only considering jets not already defined as single-$b$- or $B$-jets. 
Events that have at least one single-$b$- or $B$-jet, not counting heavy-flavor jets from top-quark or $W$-boson decays,
are labeled as \ttbin; those with no single-$b$- or $B$-jet but
at least one single-$c$- or $C$-jet are labeled as \ttcin. Finally, events not containing any heavy-flavor jets aside from those from top-quark or $W$-boson decays are labeled as \ttlight.
This classification is used to define the background categories in the likelihood fit.
A finer classification is then used to assign correction factors and estimate uncertainties:
events with exactly
two single-$b$-jets are labeled as \ttbb, those with only one single-$b$-jet are labeled as \ttb, and those with only one $B$-jet
are labeled as \ttB, the rest of the \ttbin{} events 
being 
labeled as \ttbbb.
Events with additional $b$-jets entirely originating from multi-parton interactions (MPI) or $b$-jets from final-state radiation (FSR), 
i.e.\ originating from gluon radiation from the top-quark decay products, are considered separately in the \ttbmpifsr{} subcategory.
Background events from \ttbar\ containing extra $c$-jets are divided analogously.

To model the dominant \ttbin\ background with the highest available precision, 
the relative contributions of the different subcategories, \ttbbb, \ttbb, \ttB\ and \ttb, in 
the \powheg+\pythiaeight\ sample described above
are scaled 
to match those predicted by an NLO 
$t\bar{t}b\bar{b}$ 
sample including 
parton showering and hadronization~\cite{Cascioli:2013era}, generated with \textsc{Sherpa+OpenLoops}~\cite{Gleisberg:2008ta, Cascioli:2011va}.  
The sample was produced with \sherpa\ version 2.1.1 and the CT10 four-flavor (4F) scheme PDF set~\cite{Guzzi:2011sv,Gao:2013xoa}.
The renormalization scale for this sample was set
to the CMMPS value, $\mu_{\mathrm{CMMPS}} = \prod_{i=t,\bar{t},b,\bar{b}} E_{\mathrm{T},i}^{1/4}$~\cite{Cascioli:2013era}, while the factorization
scale was set to $H_\mathrm{T}/2 = \frac{1}{2} \sum_{i=t,\bar{t},b,\bar{b}} E_{\mathrm{T},i}$. The resummation scale $\mu_{\mathrm{Q}}$, which sets an upper bound for
the hardness of the parton-shower emissions, was also set to $H_\mathrm{T}/2$. 
This sample, referred to as `\ShOL{}' in the remainder of this article, 
employs a description of the kinematics of the two additional $b$-jets with NLO precision in QCD, taking into account the $b$-quark mass, and is therefore the most precise MC prediction for the 
\ttbin\
process available at present.
Topologies that are not included in this NLO calculation but are labeled as \ttbin{}, i.e. events in the \ttbmpifsr{} subcategory, are not scaled.

Figure~\ref{fig:ttbbFrac} shows the predicted fractions for each of the \ttbin\ subcategories, 
with the \powheg+\pythiaeight\ inclusive \ttbar\ sample compared to 
the \ttbb\ \ShOL\ sample. The \ttbmpifsr{} subcategory is not present in the \ttbb\ \ShOL\ sample and accounts for 10\% of the events in the \powheg+\pythiaeight\ \ttbin{} sample.

\begin{figure}[!ht]
\begin{center}
\includegraphics[width=0.5\textwidth]{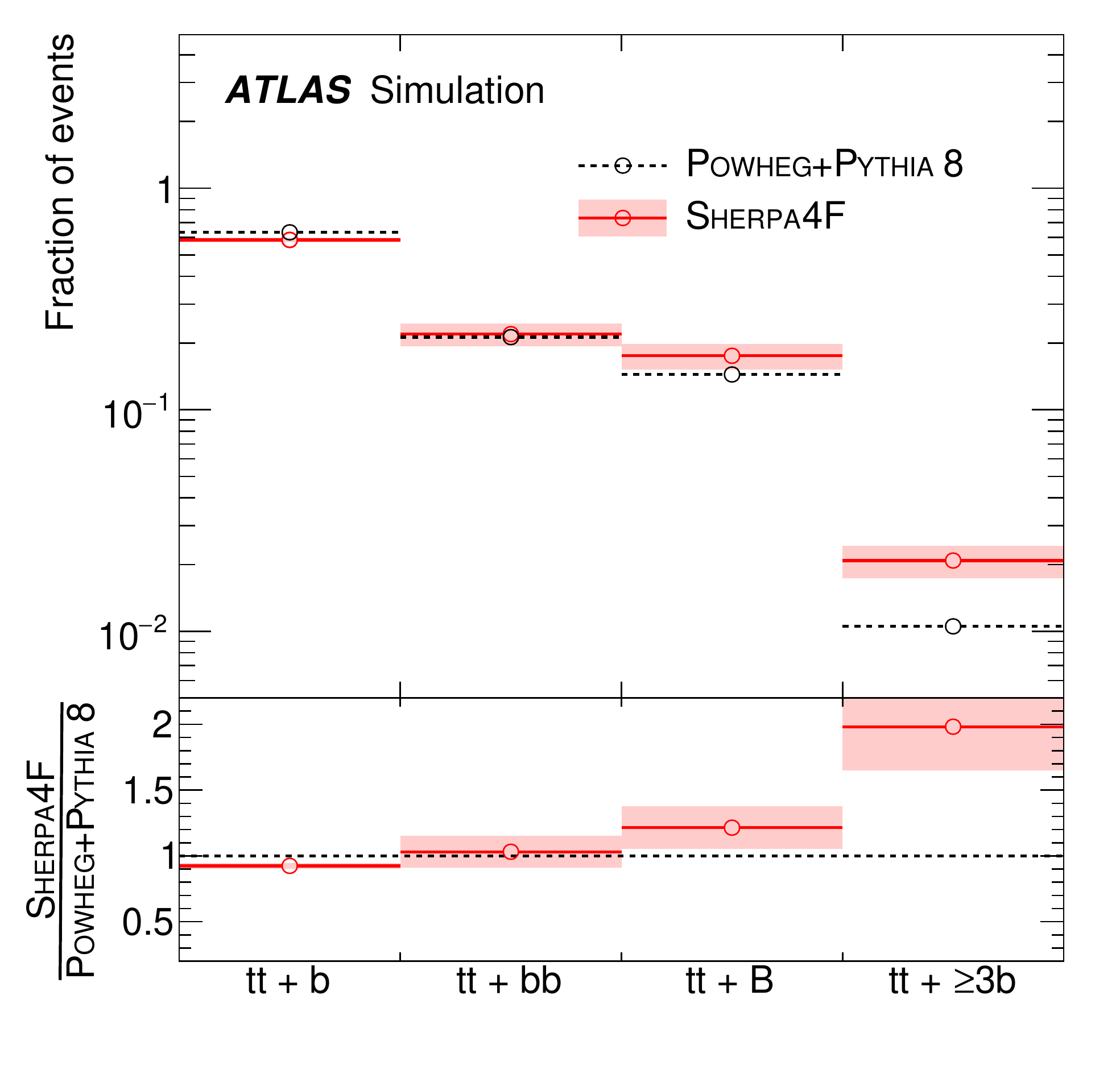}
\caption{The relative predicted fractions of the \ttb, \ttbb, \ttB{} and \ttbbb{} subcategories before any event selection. 
The prediction from  the inclusive \powheg+\pythiaeight\ sample is compared to
the four-flavor $t\bar{t}b\bar{b}$ calculation from \ShOL, 
with its uncertainties (from a combination of the sources discussed in Section \ref{sec:systs}) shown as the shaded area. 
The fractions are normalized to the sum of the four contributions shown here, without considering the \ttbmpifsr{} subcategory as part of the total.
}
\label{fig:ttbbFrac} 
\end{center}
\end{figure}

\subsection{Other backgrounds}

Samples of $\ttbar \Wboson$ and $\ttbar \Zboson$ ($\ttbar V$) events were generated with an NLO matrix element 
using \amcnlo\ interfaced to \textsc{Pythia}~8.210 with the NNPDF3.0NLO PDF and the A14 parameter set.

Samples of $Wt$ and $s$-channel single-top-quark backgrounds were generated with 
\powhegbox\ v1 at NLO accuracy using the CT10 PDF set.
Overlap between the \ttbar\ and $Wt$ final states was handled using the `diagram removal' scheme~\cite{Frixione:2008yi}.
The 
$t$-channel single-top-quark events 
were generated using the \powhegbox\ v1 event generator at NLO accuracy with 
the four-flavor PDF set CT10 4F. For this process, the top quarks were decayed using \textsc{MadSpin}. 
All single-top-quark samples were interfaced to \textsc{Pythia}~6.428~\cite{Sjostrand:2006za} with the Perugia 2012 set of tuned parameters~\cite{Skands:2010ak}.
The single-top-quark $Wt$, $t$- and $s$-channel samples are normalized using 
the approximate NNLO theoretical cross-sections~\cite{Kidonakis:2010ux,Kidonakis:2010tc,Kidonakis:2011wy}.

Samples of $W/Z$ production in association with jets were generated using \textsc{Sherpa} 2.2.1. 
The matrix elements were calculated for up to two partons at NLO and four partons 
at leading order (LO) using \textsc{Comix}~\cite{Gleisberg:2008fv} and \textsc{OpenLoops},
and merged with the \textsc{Sherpa} parton shower~\cite{Schumann:2007mg} 
using the ME+PS@NLO prescription~\cite{Hoeche:2012yf}. 
The NNPDF3.0NNLO PDF set was used in conjunction with dedicated parton-shower tuning. 
The $W/Z$ + jet events are normalized using the NNLO cross-sections~\cite{ATLAS-CONF-2015-039}. 
For $Z$ + jet events, the normalization of the heavy-flavor component is corrected by a factor 1.3, extracted from dedicated control regions in data, 
defined by requiring two opposite-charge same-flavor leptons ($e^+e^-$ or $\mu^+\mu^-$) with an invariant mass, $m_{\ell\ell}$, 
inside the \Zboson-boson mass window 83--99~\GeV.
The diboson + jet samples were generated using \textsc{Sherpa} 2.1.1 as described in Ref.~\cite{ATL-PHYS-PUB-2016-002}.

Higgs-boson production in association with a single top quark is 
rare 
in the SM, but is included in the analysis and treated as background. 
Samples of single top quarks produced in association with a $W$ boson and with a Higgs boson, $tWH$, 
were produced with \amcnlo\ interfaced to \herwigpp~\cite{Herwigpp} with the CTEQ6L1 PDF set. 
Samples of single top quarks plus Higgs boson plus jets, $tHqb$, 
were produced at LO with \amcnlo\ 
interfaced to \pythiaeight, using the CT10 4F scheme PDF set. 
The other Higgs-boson production modes were found to be negligible and are not considered.
Four-top production ($t\bar{t}t\bar{t}$) as well as $t\bar{t}WW$ events
were generated with \amcnlo{} with LO accuracy and interfaced with \pythiaeight. 
Events from $tZ$ production were also generated with \amcnlo\ with LO accuracy, but interfaced with \pythia.
The process $tZW$ was also generated with \amcnlo\ interfaced with \pythiaeight, 
but with NLO accuracy.

In the single-lepton channel, the background from events with a jet or a photon misidentified as a lepton (hereafter referred to as fake lepton) or non-prompt lepton is estimated directly from data using a matrix method~\cite{ATLAS-CONF-2014-058}. 
A data sample enhanced in fake and non-prompt leptons is selected by removing the lepton isolation requirements and, 
for electrons, loosening the identification criteria. 
Next, the efficiency for these `loose' leptons to
satisfy the nominal selection (`tight') criteria is measured in data, 
separately for real prompt leptons and for fake or non-prompt leptons. 
For real prompt leptons the efficiency is measured in $Z$-boson events, 
while for fake and non-prompt leptons it is estimated from events with low missing transverse momentum 
and low values of the reconstructed leptonic $W$-boson transverse mass.\footnote{
The reconstructed leptonic $W$-boson transverse mass is defined as $\sqrt{2\pt^{\textrm{lepton}}E_{\textrm{T}}^{\textrm{miss}}(1-\cos\Delta\phi)}$, 
where $\pt^{\textrm{lepton}}$ is the transverse momentum of the selected lepton, 
$E_{\textrm{T}}^{\textrm{miss}}$ is the magnitude of the missing transverse momentum 
and $\Delta\phi$ is the azimuthal angle between the lepton and the missing transverse momentum.}
With this information, the number of fake or non-prompt leptons satisfying the tight criteria can be calculated by inverting the matrix defined by the two equations:
\[
N^{\textrm{l}} = N_{\textrm{r}}^{\textrm{l}} + N_{\textrm{f}}^{\textrm{l}},
\qquad
N^{\textrm{t}} = \varepsilon_{\textrm{r}} N_{\textrm{r}}^{\textrm{l}} + \varepsilon_{\textrm{f}} N_{\textrm{f}}^{\textrm{l}},
\]
where $N^{\textrm{l}}$ ($N^{\textrm{t}}$) is the number of events observed in data passing the loose (tight) lepton selection, 
$N_{\textrm{r}}^{\textrm{l}}$ ($N_{\textrm{f}}^{\textrm{l}}$) is the number of events with a real prompt (fake or non-prompt) lepton in the loose lepton sample, 
and $\varepsilon_{\textrm{r}}$ ($\varepsilon_{\textrm{f}}$) is the efficiency for these events to pass the tight lepton selection.
By generalizing the resulting formula to extract $\varepsilon_{\textrm{f}} N_{\textrm{f}}^{\textrm{l}}$, 
a weight is assigned to each event selected in the loose lepton data sample, 
providing a prediction for both the yields and the kinematic distribution shapes for the fake and non-prompt lepton background.
In the three most sensitive single-lepton signal regions, \srsixjone, \srsixjtwo\ and \srfivejone\ (see Section~\ref{sec:analysis}), 
the contribution from events with a 
fake or    
non-prompt lepton is found to be very small, consistent with zero, and is neglected.
In the dilepton channel, 
this background
is estimated from simulation and is normalized to data in a control region with two same-sign leptons.

All background samples described in this section, apart from the $\ttbar V$ samples, are referred to as `non-\ttbar' and grouped together in the figures and tables. 
The contribution to the total background prediction from non-\ttbar{} varies between 4\% and 15\% depending on the considered signal or control region, 
as can be seen in Appendix~\ref{app:yields}.

\section{Event categorization}
\label{sec:analysis}
After the selection, the data sample is dominated by background from \ttbar events. 
In order to take advantage of the higher jet and $b$-jet multiplicities of the \ttH\ signal process, 
events are classified into non-overlapping 
analysis categories 
based on the total number of jets, 
as well as the number of $b$-tagged jets at the four working points. 
Events in the boosted single-lepton category are not further categorized due to the small number of selected events in this category.
Events in the dilepton (resolved single-lepton) channel are first classified according to whether the number of jets is exactly three (five) 
or at least four (six). 
These events are then further subdivided into
analysis categories, 
depending on the number of jets tagged at the four $b$-tagging working points, 
or, equivalently, on the values of the $b$-tagging discriminant for the jets. 
The $b$-tagging requirements are optimized in order to obtain categories enriched in one of the relevant sample components:
\tth\ plus \ttbb, \ttb, 
\ttcin\ and \ttlight. 
The analysis categories where \tth\ and \ttbb\ are enhanced relative to the other backgrounds are referred to as `signal regions';
in these, multivariate techniques are used to further separate the \ttH{} signal from the background events. 
The remaining 
analysis categories 
are referred to as `control regions'; no attempt is made to separate the signal from the background in these 
analysis categories, 
but they provide stringent constraints on backgrounds and systematic uncertainties in a combined fit with the signal regions.

\begin{figure}[ht!]
\centering
\subfigure[]{\includegraphics[width=0.47\textwidth]{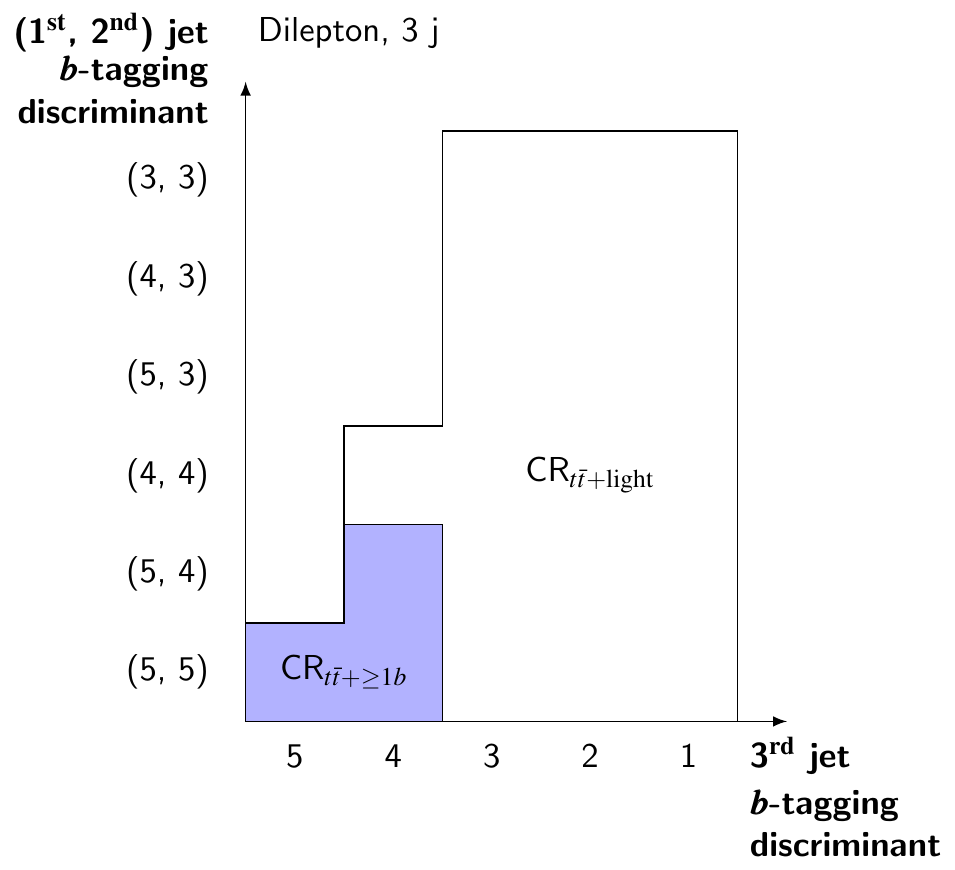}\label{fig:regions_dl_a}}\\
\subfigure[]{\includegraphics[width=\textwidth]{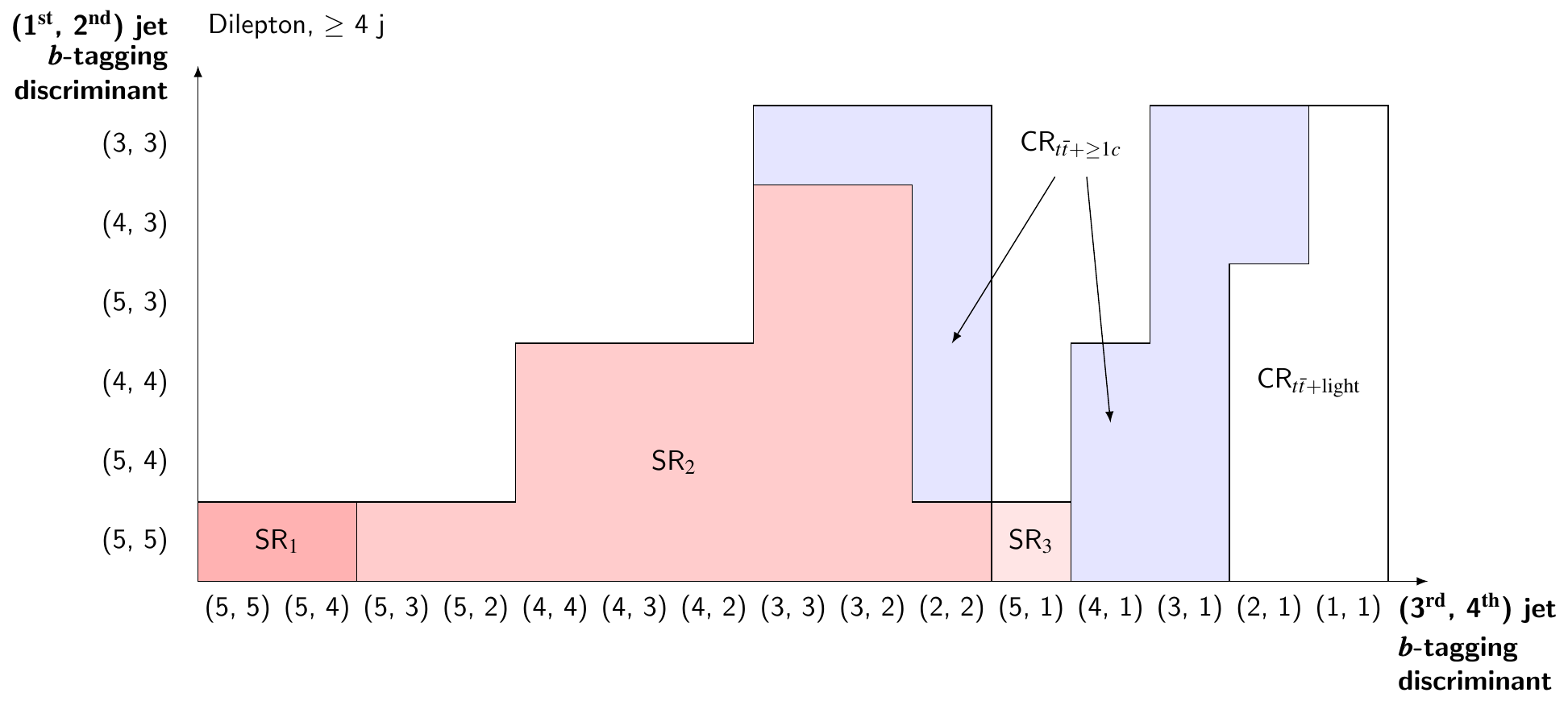}\label{fig:regions_dl_b}}
\caption{Definition of the (a) three-jet and (b) four-jet signal and control regions in the dilepton channel, 
as a function of the $b$-tagging discriminant defined in Section \ref{sec:selection}. 
The vertical axis shows the values of the $b$-tagging discriminant for the first two jets, 
while the horizontal axis shows these values for (a) the third jet or (b) the third and fourth jets. 
The jets are ordered according to their value of the $b$-tagging discriminant in descending order. 
}
\label{fig:regions_dl}
\end{figure}

\begin{figure}[ht!]
\subfigure[]{\includegraphics[width=\textwidth]{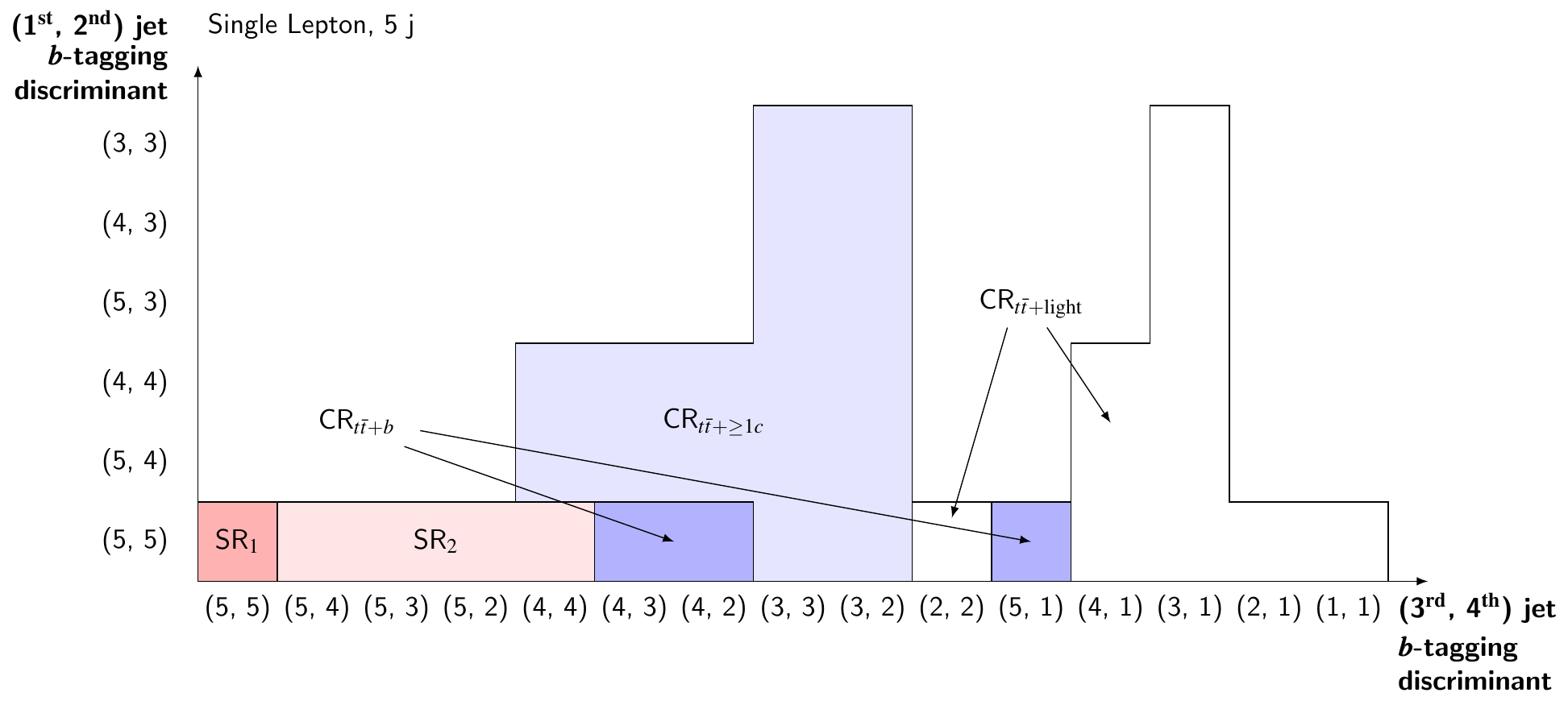}\label{fig:regions_sl_a}}\\
\subfigure[]{\includegraphics[width=\textwidth]{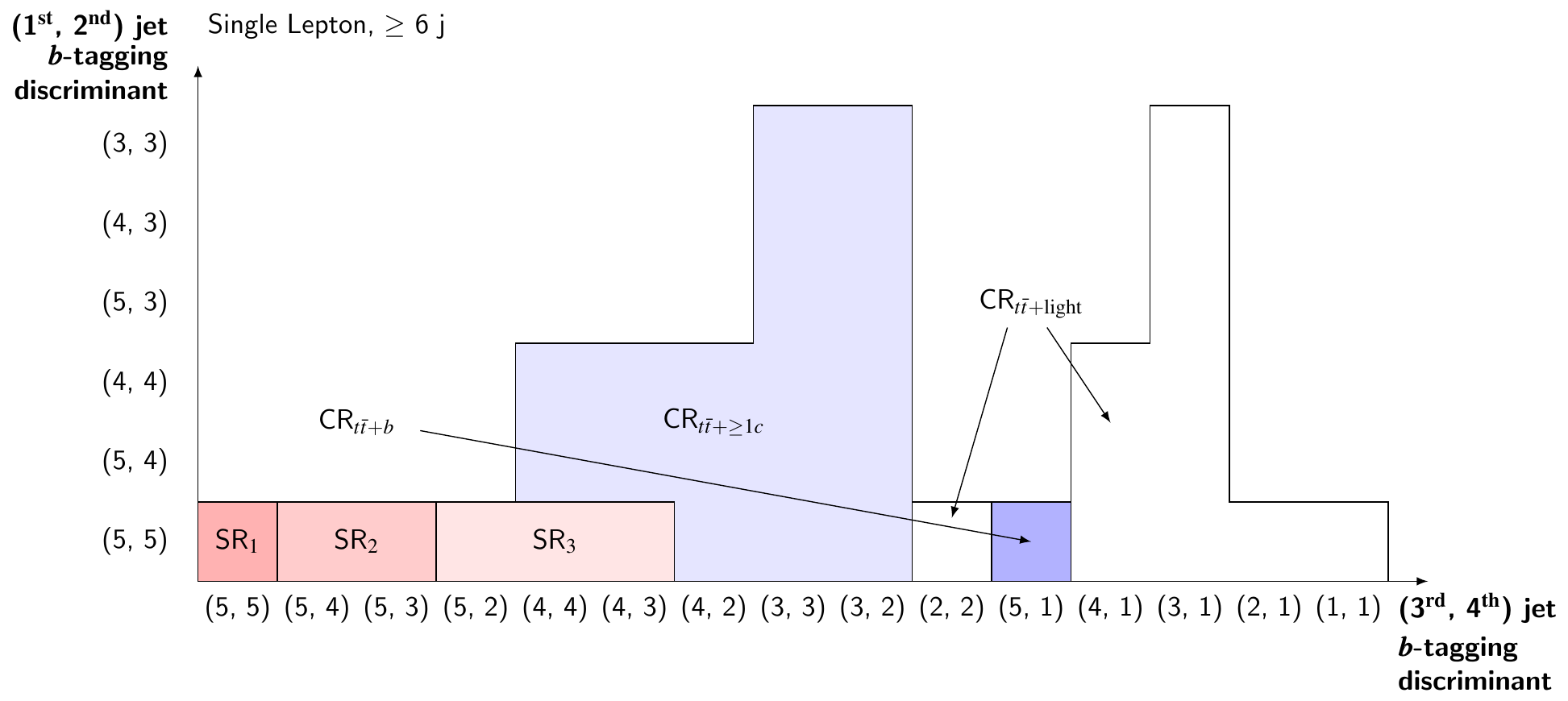}\label{fig:regions_sl_b}}
\caption{Definition of the (a) five-jet and (b) six-jet signal and control regions in the single-lepton resolved channel, 
as a function of the $b$-tagging discriminant defined in Section \ref{sec:selection}. 
The vertical axis shows the values of the $b$-tagging discriminant for the first two jets, 
while the horizontal axis shows these values for the third and fourth jets. 
The jets are ordered according to their value of the $b$-tagging discriminant in descending order.
}
\label{fig:regions_sl}
\end{figure}

In the dilepton channel, three signal regions are defined, 
with different levels of purity for the \tth\ and \ttbb{} components. 
The signal region with the highest \tth{} signal purity, referred to as \srfourjone, 
is defined by requiring at least four jets of which three are $b$-tagged at the \textit{very tight} working point and another one is $b$-tagged at the \textit{tight} working point.
The other two signal regions, \srfourjtwo\ and \srfourjthree, are defined with looser $b$-tagging requirements. 
The remaining dilepton events with at least four jets are divided into two control regions, 
one enriched in \ttlight, \crfourjttlight, 
and one in \ttcin, \crfourjttc. 
Dilepton events with three jets are split into two control regions, \crthreejttlight\ and \crthreejttb, 
enriched in \ttlight{} and \ttbin, respectively.
The detailed definition of the signal and control regions for the dilepton channel is presented in Figure \ref{fig:regions_dl}.

In the single-lepton channel, 
five signal regions are formed from events passing the resolved selection, 
three 
requiring 
at least six jets, and the other two 
requiring exactly five jets. 
They 
are referred to as \srsixjone, \srsixjtwo, \srsixjthree, \srfivejone\ and \srfivejtwo.
The two purest signal regions, \srsixjone\ and \srfivejone, require four $b$-tagged jets at the \textit{very tight} working point,
while looser requirements are applied in the other signal regions.
Events passing the boosted single-lepton selection form a sixth signal region, \srboosted. 
The remaining events with at least six jets are then categorized into three control regions enriched in \ttlight, \ttcin\ and \ttb, referred to as \crsixjttlight, \crsixjttc, \crsixjttoneb, respectively. Analogously, remaining events with exactly five jets are categorized into other three control regions, referred to as \crfivejttlight, \crfivejttc\ and \crfivejttoneb. 
The detailed definition of the signal and control regions for the resolved single-lepton channel is presented in Figure~\ref{fig:regions_sl}.

Figures~\ref{fig:pieCharts} and~\ref{fig:SoBplots} show, respectively, 
the fraction of the different background components as well as the \tth{} signal purity 
for each of the signal and control regions in the dilepton and single-lepton channels. 
The \htobb{} decay represents 89\% of the \tth{} signal events in the signal regions of the dilepton channel, 96\% in the signal regions of the resolved single-lepton channel and 86\% in the boosted signal region.

\begin{figure*}[ht]
\centering
\subfigure[]{\includegraphics[width=0.48\textwidth]{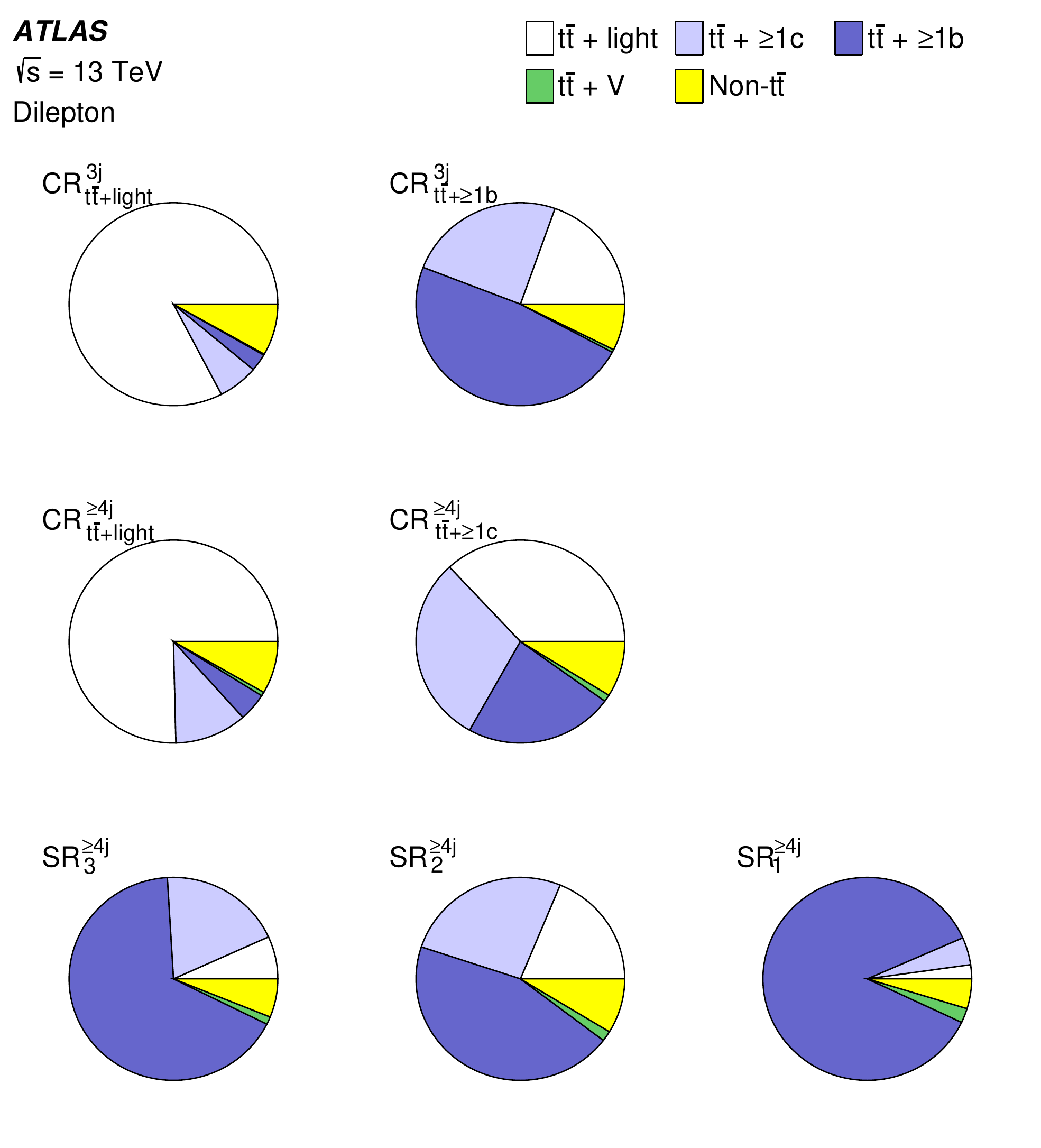}}\label{fig:pieChart_a} \hspace{0.02\textwidth}
\subfigure[]{\includegraphics[width=0.48\textwidth]{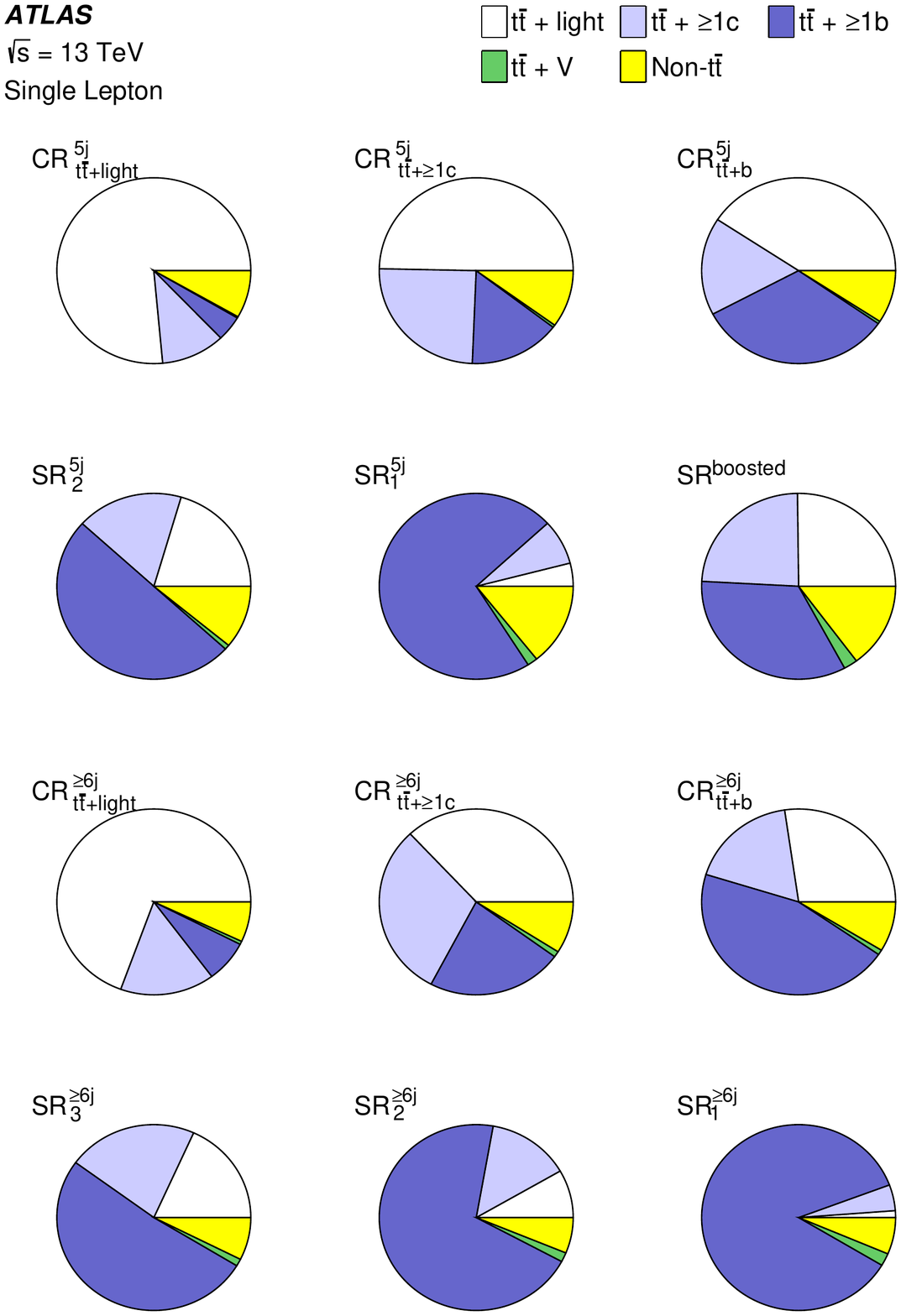}}\label{fig:pieChart_b}
\caption{Fractional contributions of the various backgrounds to the total background prediction in each analysis category 
(a) in the dilepton channel and (b) in the single-lepton channel. 
The predictions for the various background contributions are obtained through the simulation and the data-driven estimates described in Section~\ref{sec:modelling}. 
The $\ttbar$ background is divided as described in Section~\ref{sec:modelling}.
The predicted event yields in each of the analysis categories, broken down into the different signal and background contributions, are reported in Appendix~\ref{app:yields}.
}
\label{fig:pieCharts}
\end{figure*}

\begin{figure*}[ht]
\centering
\subfigure[]{\includegraphics[width=0.48\textwidth]{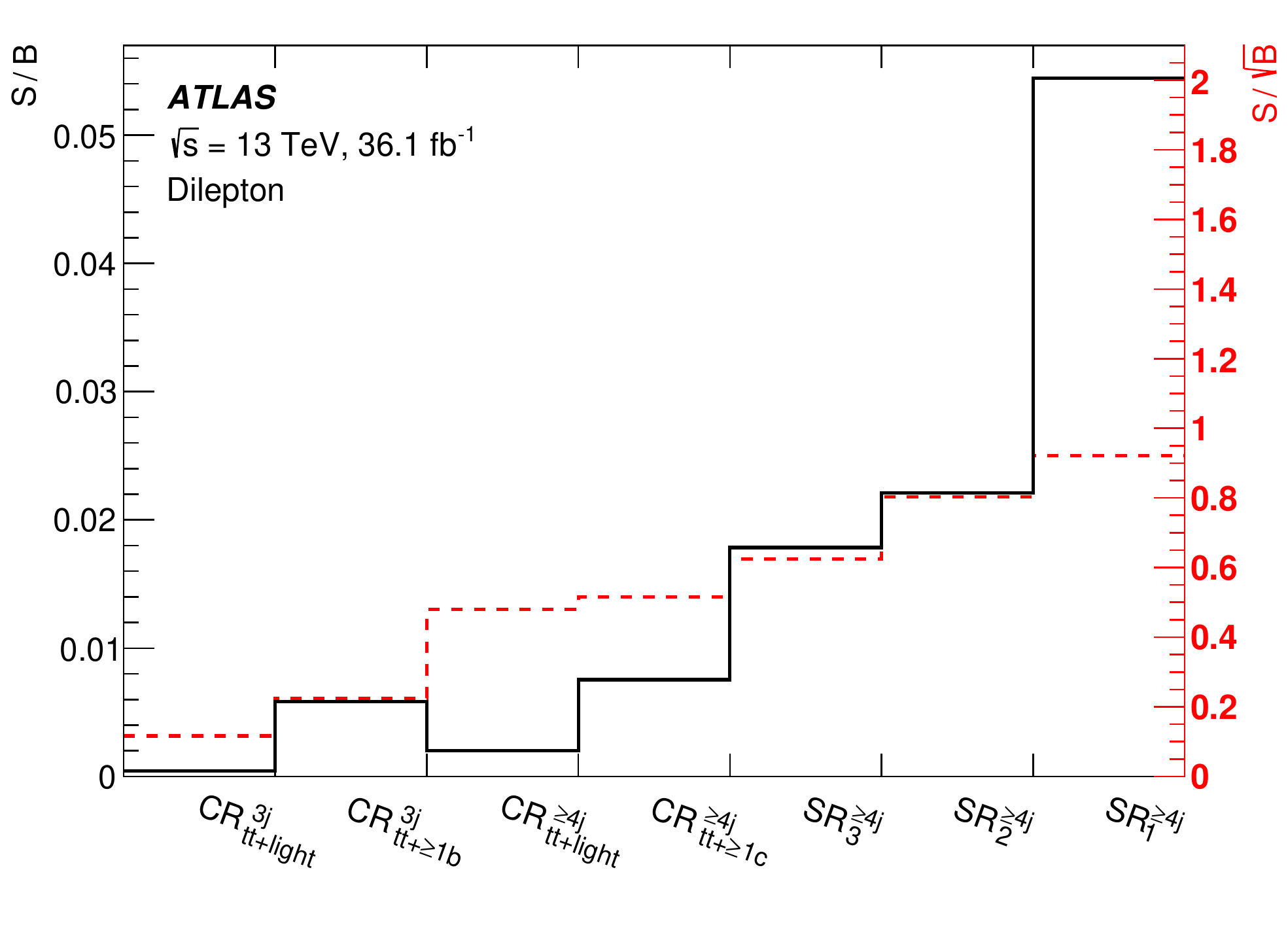}}\label{fig:SoBplots_a} \hspace{0.02\textwidth}
\subfigure[]{\includegraphics[width=0.48\textwidth]{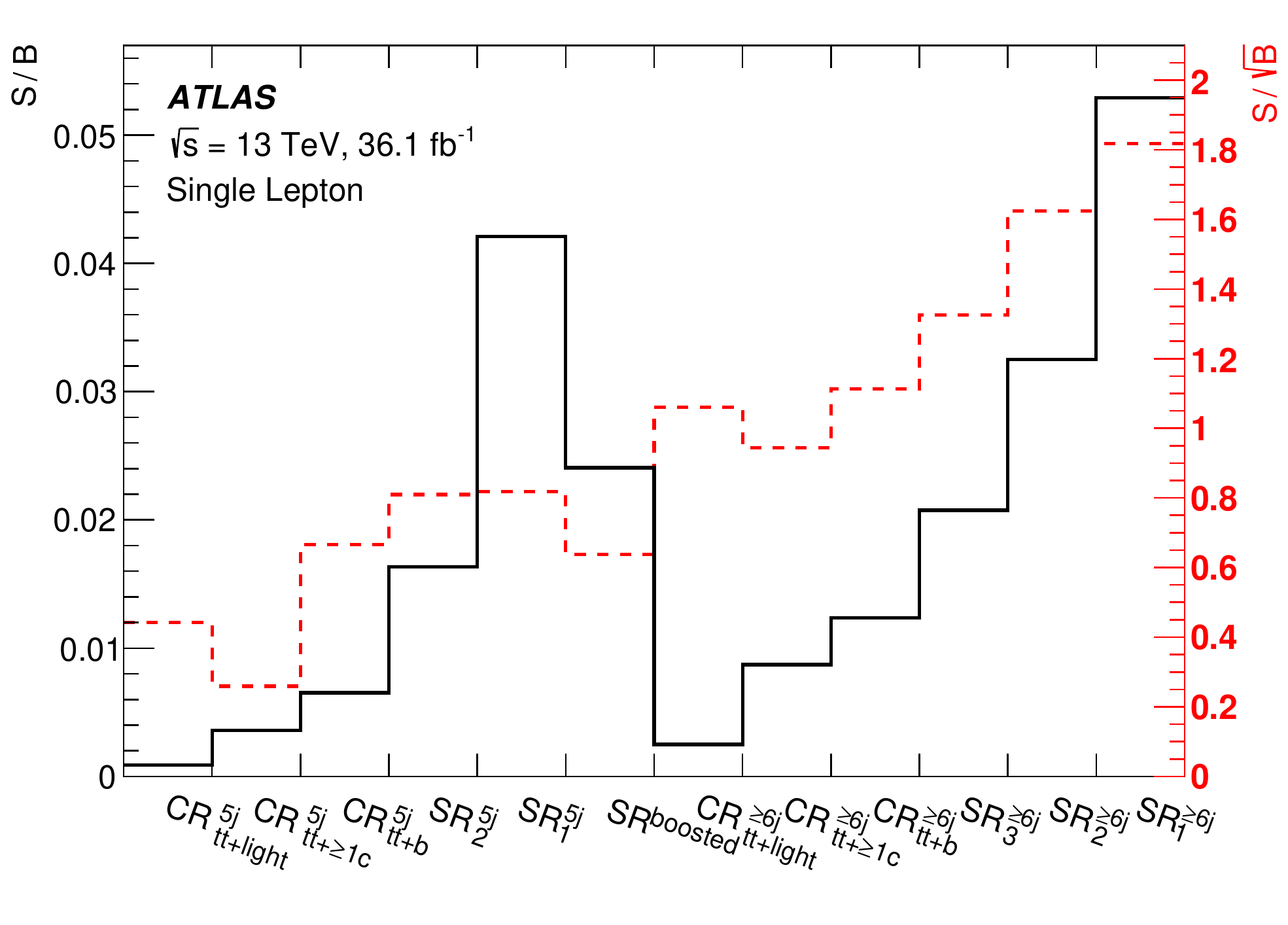}}\label{fig:SoBplots_b}
\caption{
The ratios $S/B$ 
(black solid line, 
referring to the vertical axis on the left) 
and $S/\sqrt{B}$ 
(red dashed line, 
referring to the vertical axis on the right) 
for each of the analysis categories 
(a) in the dilepton channel and (b) in the single-lepton channel, 
where $S$ ($B$) is the number of selected signal (background) events predicted by the simulation or through the data-driven estimates as described in Section~\ref{sec:modelling}. 
}
\label{fig:SoBplots}
\end{figure*}

\section{Multivariate analysis techniques}
\label{sec:mvas}
In each of the signal regions, a boosted decision tree (BDT) is exploited to discriminate between the \tth\ signal and the backgrounds. 
This BDT is referred to as the `classification BDT' in the following.
The distributions of the classification BDTs in the signal regions are used as the final discriminants 
for the profile likelihood fit described in Section~\ref{sec:results}.
In the control regions, the overall event yield is used as input to the fit,
except in those enriched in \ttcin\ in the single-lepton channel, \crfivejttc\ and \crsixjttc; 
in these two control regions, the distribution of the scalar sum of the \pt\ of the jets, \hthad, is used to further control the \ttcin\ background.

The final state of the \tth(\htobb) process is composed of many jets stemming from the Higgs-boson and top-quark decay products, as well as from additional radiation.
Many combinations of these jets are possible when reconstructing the Higgs-boson and top-quark candidates to explore their properties and the signal event topology. 
To enhance the signal separation, three intermediate multivariate techniques are implemented prior to the classification BDT: (a) the `reconstruction BDT' used to select the best combination of jet--parton assignments in each event and to build the Higgs-boson and top-quark candidates, (b) a likelihood discriminant (LHD) method that combines the signal and background probabilities of all possible combinations in each event, (c) a matrix element method (MEM) that exploits the full matrix element calculation to separate the signal from the background.
The outputs of the three intermediate multivariate methods are used as input variables to the classification BDT in one or more of the signal regions.
The properties of the Higgs-boson and top-quark candidates from the reconstruction BDT are used to define additional input variables to the classification BDT.
Although the intermediate techniques exploit similar information, they make use of this information from different perspectives and based on different assumptions, so that their combination further improves the separation power of the classification BDT.
Details of the implementation of these multivariate techniques are described in Sections~\ref{sub:classification}--\ref{sub:mem}.

\subsection{Classification BDT}
\label{sub:classification}
The classification BDT is trained to separate the signal from the \ttbar{} background on a sample that is statistically independent 
of
the sample used for the evaluation.
The toolkit for multivariate analysis (TMVA)~\cite{Hoecker:TMVA} 
is used to train both this and the reconstruction BDT. 
The classification BDT is built by combining several input variables that exploit the different kinematics of signal and background events, as well as the $b$-tagging information.
General kinematic variables, such as invariant masses and angular separations of pairs of reconstructed jets and leptons, 
are combined with outputs of the intermediate multivariate discriminants and the $b$-tagging discriminants of the selected jets.
In the case of 
the boosted single-lepton signal region, 
kinematic variables are built from the properties of the large-$R$ jets and their jet constituents.
The input variables to the classification BDT in each of the signal regions are listed in Appendix \ref{sec:app_classBDTinputs}.
The input variables are selected to maximize the performance of the classification BDT; however, only variables with good modeling of data by simulation are considered.
The output of the reconstruction BDT, the LHD and the MEM represent the most powerful variables in the classification BDT.

\subsection{Reconstruction BDT}
\label{sub:recobdt}
The reconstruction BDT is employed in all dilepton and resolved single-lepton signal regions.
It is trained to match reconstructed jets to the partons emitted from top-quark and Higgs-boson decays.
For this purpose, \Wboson-boson, top-quark and Higgs-boson candidates are built from combinations of jets and leptons.
The $b$-tagging information is used to discard combinations containing jet--parton assignments inconsistent with the correct parton candidate flavor. 

In the single-lepton channel, leptonically decaying \Wboson-boson candidates are assembled from the lepton four-momentum ($p_{\ell}$) and the neutrino four-momentum ($p_{\nu}$);
the latter is built from the missing transverse momentum, its $z$ component being inferred by solving the equation $m_{W}^{2}=(p_{\ell}+p_{\nu})^{2}$, where $m_W$ represents the \Wboson-boson mass. 
Both solutions of this quadratic equation are used in separate combinations. 
If no real solutions exist, the discriminant of the quadratic equation is set to zero, giving a unique solution.
The hadronically decaying \Wboson-boson and the Higgs-boson candidates are each formed from a pair of jets.
The top-quark candidates are formed from one \Wboson-boson candidate and one jet.
The top-quark candidate containing the hadronically (leptonically) decaying \Wboson\ boson is referred to as the hadronically (leptonically) decaying top-quark candidate.
In the single-lepton signal regions with exactly five selected jets, more than 70\% of the events do not contain both jets from the hadronically decaying \Wboson{} boson.
Therefore, the hadronically decaying top-quark candidate is assembled from two jets, 
one of which is $b$-tagged.
In the dilepton channel, no attempt to build leptonically decaying \Wboson-boson candidates is made and the top-quark candidates are formed by one lepton and one jet. 

Simulated \tth\ events are used to iterate over all allowed combinations. The reconstruction BDT is trained to distinguish between correct and incorrect jet assignments, 
using invariant masses and angular separations in addition to other kinematic variables as inputs. 
In each event a specific combination of jet--parton assignments, corresponding to the best BDT output, 
is chosen in order to compute kinematic and topological information of the top-quark and Higgs-boson candidates to be input to the classification BDT.
However, although the best possible reconstruction performance can be obtained by including information related to the Higgs boson, such as the candidate Higgs-boson invariant mass, in the reconstruction BDT, this biases the background distributions of these Higgs-boson-related observables in the chosen jet--parton assignment towards the signal expectation,
reducing their ability to separate signal from background.
For this reason, two versions of the reconstruction BDT are used, 
one with and one without 
the Higgs-boson information 
and the resulting jet--parton assignments from one, the other or both are considered when computing input variables for the classification BDT, as detailed in Appendix \ref{sec:app_classBDTinputs}.

The Higgs boson is correctly reconstructed in 48\% (32\%) of the selected \tth{} events in the single-lepton channel \srsixjone\ 
using the reconstruction BDT with (without) information about the Higgs-boson kinematics included.
For the dilepton channel, the corresponding reconstruction efficiencies are 49\% (32\%) in \srfourjone. 
The reconstruction techniques are not needed in the signal region \srboosted, as the Higgs-boson and the top-quark candidates are chosen as the selected large-$R$ jets described in Section \ref{sec:selection}. The large-$R$ jet selected as a Higgs-boson candidate contains two $b$-tagged jets stemming from the decay of a Higgs boson in 47\% of the selected \tth{} events.

\subsection{Likelihood discriminant}
\label{sub:lhd}
In the resolved single-lepton signal regions, the output from a likelihood discriminant is 
included as an additional input variable for the classification BDT.
The LHD is computed analogously to Ref.~\cite{TOPQ-2014-14} as a product of one-dimensional probability density functions, pdfs, for the signal and the background hypotheses.
The pdfs are built for various invariant masses and angular distributions from reconstructed jets and leptons and from the missing transverse momentum, in a similar way to those used in the reconstruction BDT.

Two background hypotheses are considered, corresponding to the production of \ttbar\ $+\geq2$ $b$-jets and \ttbar\ + exactly one $b$-jet, respectively.
The likelihoods for both hypotheses are averaged, weighted by their relative fractions in simulated \ttbar\ + jets events.
In a significant fraction of both the \ttH\ and \ttbar\ simulated events with at least six selected jets, only one jet stemming from the hadronically decaying \Wboson\ boson is selected.
An additional hypothesis, for both the signal and the background, is considered to account for this topology.
In events with exactly five selected jets, variables including the hadronically decaying top-quark candidate are built similarly to those for the reconstruction BDT.

The probabilities $p^{\textrm{sig}}$ and $p^{\textrm{bkg}}$, for signal and background hypotheses, respectively, 
are obtained as the product of the pdfs for the different kinematic distributions, averaged among all possible jet--parton matching combinations.
Combinations are weighted using the $b$-tagging information to suppress the impact from parton--jet assignments that are inconsistent with the correct parton candidates flavor.
For each event, the discriminant is defined as the ratio of the probability $p^{\textrm{sig}}$ to the sum of $p^{\textrm{sig}}$ and $p^{\textrm{bkg}}$, and added as an input variable to the classification BDT.
As opposed to the reconstruction BDT method, the LHD method takes advantage of all possible combinations in the event, but it does not fully account for correlations between variables in one combination, as it uses a product of one-dimensional pdfs.

\subsection{Matrix element method}
\label{sub:mem}
A discriminant (MEM$_{D1}$) based on the MEM is computed following a method similar to the one described in Ref.~\cite{HIGG-2013-27} and is included as another input to the classification BDT.
The MEM consumes a significant amount of computation time and thus is implemented only in the most sensitive single-lepton signal region, \srsixjone.
The degree to which each event is consistent with
the signal and background hypotheses is expressed via signal and background likelihoods, referred to as $L_S$ and $L_B$, respectively.
These are computed using matrix element calculations at the parton level rather than using simulated MC samples as for the LHD method.
The matrix element evaluation is performed with \amcnlo{} at the LO accuracy.
The \ttH(\htobb) process is used as a signal hypothesis, while \ttbb{} is used as a background hypothesis.
To reduce the computation time, only diagrams representing gluon-induced processes are considered.
The parton distribution functions are modeled with the CT10 PDF set, interfaced via the LHAPDF package \cite{Whalley:2005nh}.
Transfer functions, that map the detector quantities to the parton level quantities, are derived from a \ttbar{} sample generated with {\textsc{Powheg}}+\pythia{} and validated with the nominal {\textsc{Powheg}}+\pythiaeight{} \ttbar{} sample.
The directions in $\eta$ and $\phi$ of all visible final-state objects are assumed to be well measured, and their transfer functions are thus represented by $\delta$-functions. The neutrino momentum is constrained by imposing transverse momentum conservation in each event, while its $p_\mathrm{z}$ is integrated over.
The integration is performed using VEGAS~\cite{1978JCoPh..27..192L}, 
following the implementation described in Ref.~\cite{SCHOUTEN201554}. 
As in the reconstruction BDT, $b$-tagging information is used to reduce the number of jet--parton assignments considered in the calculation.
The discriminating variable, MEM$_{D1}$, is defined as the difference between the logarithms of the signal and background likelihoods: MEM$_{D1} = \log_{10}(L_S) - \log_{10}(L_B)$.

\section{Systematic uncertainties}
\label{sec:systs}
Many sources of systematic uncertainty affect the search, including those related to the luminosity, 
the reconstruction and identification of 
leptons and jets, 
and the theory modeling of signal and background processes. 
Different uncertainties may affect only the overall normalization of the samples, 
or also the shapes of the distributions used to categorize the events and to build the final discriminants. 
All the sources of experimental uncertainty considered, with the exception of the uncertainty in the luminosity, affect both the normalizations and the shapes of distributions in all the simulated samples.
Uncertainties related to modeling of the signal and the backgrounds affect both the normalizations 
and the shapes of the distributions for the processes involved, 
with the exception of cross-section and normalization uncertainties that affect only the normalization of the considered sample.
Nonetheless, the normalization uncertainties modify the relative fractions of the different samples leading to a 
shape uncertainty in the distribution of the final discriminant 
for the total prediction in the different analysis categories.

A single independent nuisance parameter is assigned to each source of systematic uncertainty, as described in Section~\ref{sec:results}. 
Some of the systematic uncertainties, in particular most of the experimental uncertainties, 
are decomposed into several independent sources, as specified in the following. 
Each individual source then has a 
correlated 
effect across all the channels, analysis categories, signal and background samples.
For modeling uncertainties, especially \ttbar{} modeling, additional nuisance parameters are included to split some uncertainties into several sources independently affecting different subcomponents of a particular process.

\subsection{Experimental uncertainties}
The uncertainty of the combined 2015+2016 integrated luminosity is 2.1\%. It is derived, following 
a methodology similar to that detailed in Ref.~\cite{DAPR-2013-01}, from a calibration of the luminosity 
scale using $x$--$y$ beam-separation scans performed in August 2015 and May 2016.
A variation in the pileup reweighting of MC events is included to cover the uncertainty in the ratio of the predicted and measured inelastic cross-sections in the fiducial volume defined by $M_X$ > 13~\GeV\ where $M_X$ is the mass of the hadronic system~\cite{STDM-2015-05}.

The jet energy scale and its uncertainty are derived by combining information from test-beam data, LHC collision 
data and simulation~\cite{PERF-2016-04}. The uncertainties from these measurements are factorized into
eight independent sources. 
Additional uncertainties are considered, related to jet flavor, pileup corrections, 
$\eta$ dependence, and high-\pt\ jets, 
yielding a total of 20 independent sources. 
Although the uncertainties are not
large, totaling 1\%--6\% per jet (depending on the jet \pt), the effects are amplified by the large number of jets in the final state. 
Uncertainties in 
the jet energy resolution and in the efficiency to pass the JVT requirement that is meant to remove jets from pileup 
are also considered. 
The jet energy resolution is divided into two independent components.

The efficiency to correctly tag $b$-jets is measured in data using dileptonic \ttbar\ events. 
The mis-tag rate for $c$-jets is also measured 
in \ttbar\ events, identifying hadronic decays of $W$ bosons including $c$-jets~\cite{ATLAS-CONF-2018-001}, 
while for light jets it is measured 
in multi-jet events 
using jets containing secondary vertices and tracks with impact parameters consistent 
with a negative lifetime~\cite{PERF-2012-04}. 
The $b$-tagging efficiencies and mis-tag rates are first extracted for each of the four working points used in the analysis as a function of jet kinematics,
and then combined into a calibration of the $b$-tagging discriminant distribution, 
with corresponding uncertainties that correctly describe correlations across multiple working points. 
The uncertainty associated with the $b$-tagging efficiency, whose size ranges between 2\% and 10\% depending on the working point and on the jet \pt, is factorized into 30 independent sources.
The size of the uncertainties associated with the mis-tag rates is 5\%--20\% for $c$-jets depending on the working point and on the jet \pt, and 10\%--50\% for light jets depending on the working point and on the jet \pt and $\eta$. These uncertainties are factorized into 15 (80) independent sources for $c$-jets (light jets). 
Jets from \tauhad{} candidates are treated as $c$-jets for the mis-tag rate corrections and systematic uncertainties. 
An additional source of systematic uncertainty is considered on the extrapolation between $c$-jets and these $\tau$-jets.

Uncertainties associated with leptons arise from the trigger, reconstruction, identification, and isolation efficiencies,
as well as the lepton momentum scale and resolution. 
These are measured in data using leptons in 
$Z\to \ell^+\ell^-$, $J/\psi \to \ell^+\ell^-$ and $W\to e\nu$ events~\cite{ATLAS-CONF-2016-024,PERF-2015-10}. 
Uncertainties of these measurements account for a total of 24 independent sources, 
but have only a small impact on the result. 

All uncertainties in energy scales or resolutions are propagated to the missing transverse momentum. 
Additional uncertainties in the scale and resolution of the soft term are considered, 
for a total of three additional sources of systematic uncertainty.

\subsection{Modeling uncertainties}
The predicted \ttH\ signal cross-section uncertainty is $^{+5.8\%}_{-9.2\%}\textrm{(scale)}\pm3.6\%\textrm{(PDF)}$, 
the first component representing the QCD scale uncertainty and the second the PDF+$\alpha_\mathrm{S}$ uncertainty~\cite{deFlorian:2016spz,Raitio:1978pt,Beenakker:2002nc,Dawson:2003zu,Yu:2014cka,Frixione:2015zaa}. 
These two components are treated as 
uncorrelated 
in the fit.
The effect of QCD scale and PDF variations on the shape of the distributions considered in this analysis is found to be negligible.
Uncertainties in the Higgs-boson branching fractions are also considered; these amount to 2.2\% for the $b\bar{b}$ decay mode~\cite{deFlorian:2016spz}.
An additional uncertainty associated with the choice of parton shower and hadronization model is derived by comparing the nominal prediction
from \amcnlo+\pythiaeight\ to the one from \amcnlo\ interfaced to \herwigpp. 

The systematic uncertainties affecting the modeling of the \ttbar+jets background are summarized in Table~\ref{tab:ModUnc}. 
An uncertainty of $\pm6\%$ is assumed for the inclusive \ttbar\ NNLO+NNLL production cross-section~\cite{Czakon:2011xx}, 
including effects from varying the factorization and renormalization scales, the PDF, $\alpha_{\textrm{S}}$, and the top-quark mass. 
The \ttbin, \ttcin\ and \ttlight{} processes 
are 
affected by different types of uncertainties: \ttlight{} has additional diagrams and profits from relatively precise measurements in data; \ttbin{} and \ttcin\ can have similar or different diagrams depending on the flavor scheme used for the PDF, and
the mass differences between $c$- and $b$-quarks contribute to additional differences between these two processes.
For these reasons,
all uncertainties in \ttbar\ + jets background modeling, 
except the uncertainty in the inclusive cross-section, are assigned independent nuisance parameters for the \ttbin, \ttcin\ and \ttlight{} processes.
The normalizations of \ttbin\ and \ttcin\ are
allowed to float freely in the fit.
Systematic uncertainties in the shapes are extracted from the comparison between the nominal sample and various alternative samples.
For all these uncertainties, alternative samples are reweighted 
in such a way that they have the same fractions of \ttcin\ and \ttbin\ as the nominal sample. 
In the case of the \ttbin\ background, 
separate uncertainties are applied to the relative normalization of the \ttbin\ subcomponents as described later. 
Therefore, for all the alternative samples used to derive uncertainties that are not specifically associated with these fractions,
the relative contributions of the \ttbin\ subcategories are scaled to match the predictions of \ShOL, 
in the same way as for the nominal sample.
This scaling is not applied to the \ttbmpifsr{} subcategory, as explained in Section \ref{sec:modelling}.

\begin{table}
\caption{Summary of the sources of systematic uncertainty for \ttbar\ + jets modeling. 
The systematic uncertainties listed in the second 
section
of the table 
are evaluated in such a way as to have no impact on the relative fractions 
of \ttbin, \ttcin\ and \ttlight{} events, 
as well as on the relative fractions of the \ttb, \ttbb, \ttB{} and \ttbbb{} subcategories, 
which are all kept at their nominal values.
The systematic uncertainties listed in the third section of the table 
affect only the fractions of the various \ttbin\ subcategories. 
The last column of the table indicates the \ttbar\ category to which a systematic uncertainty is assigned. 
In the case where all three categories (\ttlight, \ttcin\ and \ttbin) are involved (marked with `all'), 
the last column also specifies whether the uncertainty is considered as correlated or uncorrelated across them.
}
\begin{center}
\begin{footnotesize}
\begin{tabular}{ lll }
\toprule
\hline
Systematic source                  & Description                                                                                  & \ttbar\ categories \\
\hline                            
\ttbar cross-section               & Up or down by 6\%                                                                            & All, correlated   \\
$k(\ttcin)$                        & Free-floating \ttcin\ normalization                                                          & \ttcin\           \\
$k(\ttbin)$                        & Free-floating \ttbin\ normalization                                                          & \ttbin\           \\
\hline                            
\Sh\ vs. nominal                   & Related to the choice of NLO event generator                                                 & All, uncorrelated \\
PS \& hadronization                & \powheg+\herwigseven\ vs. \powheg+\pythiaeight\                                              & All, uncorrelated \\
ISR / FSR                          & Variations of $\mu_{\mathrm{R}}$, $\mu_{\mathrm{F}}$, \hdamp\ and A14 Var3c parameters       & All, uncorrelated \\
\ttcin\ ME vs. inclusive           & \amcnlo+\herwigpp: ME prediction (3F) vs. incl. (5F)                                         & \ttcin \\
\ttbin\ \ShOL\ vs. nominal         & Comparison of \ttbb\ NLO (4F) vs. \powheg+\pythiaeight\ (5F)                                     & \ttbin \\
\hline                            
\ttbin\ renorm. scale              & Up or down by a factor of two                                                                & \ttbin \\
\ttbin\ resumm. scale              & Vary $\mu_{\mathrm{Q}}$ from $H_{\mathrm{T}}/2$ to $\mu_{\mathrm{CMMPS}}$                    & \ttbin \\
\ttbin\ global scales              & Set $\mu_{\mathrm{Q}}$, $\mu_{\mathrm{R}}$, and $\mu_{\mathrm{F}}$ to $\mu_{\mathrm{CMMPS}}$ & \ttbin \\
\ttbin\ shower recoil scheme       & Alternative model scheme                                                                     & \ttbin \\
\ttbin\ PDF (MSTW)                 & MSTW vs. CT10                                                                                & \ttbin \\
\ttbin\ PDF (NNPDF)                & NNPDF vs. CT10                                                                               & \ttbin \\
\ttbin\ UE                         & Alternative set of tuned parameters for the underlying event                                 & \ttbin \\
\ttbin\ MPI                        & Up or down by 50\%                                                                           & \ttbin \\
\ttbbb\ normalization              & Up or down by 50\%                                                                           & \ttbin \\
\hline
\bottomrule
\end{tabular}
\end{footnotesize}
\end{center}
\label{tab:ModUnc}
\end{table}

Uncertainties associated with 
the choice of \ttbar\ inclusive NLO event generator 
as well as the choice of parton shower and hadronization model 
are derived by comparing the prediction from {\powheg}+{\pythiaeight} 
with the \sherpa{}
predictions (hence varying simultaneously the NLO event generator and the parton shower and hadronization model) 
and with the predictions from \powheg\ interfaced with \herwigseven~\cite{Bellm:2015jjp} (varying just the parton shower and hadronization model). 
The former alternative sample was generated using \sherpa\ version 2.2.1 
with the ME+PS@NLO setup, interfaced with \textsc{OpenLoops}, 
providing NLO accuracy for up to one additional parton
and LO accuracy for up to four additional partons.
The NNPDF3.0NNLO PDF set was used 
and both the renormalization and factorization scales were set to $\sqrt{0.5 \times (m^2_{\textrm{T},t}+m^2_{\textrm{T},\bar{t}})}$. 
This sample is referred to as `\Sh' in the remainder of this 
article, 
which should not be confused with the \ShOL\ sample defined in Section~\ref{sec:modelling}.
The comparison with the latter alternative sample is considered as an independent source of uncertainty, 
related to the parton shower and hadronization model choice.
This 
sample was generated with the same settings for \powheg\ as the nominal \ttbar\ sample in terms of \hdamp, 
PDF and renormalization and factorization scales, but it was interfaced with \herwigseven\ version 7.0.1, with the H7-UE-MMHT 
set of tuned parameters for the underlying event. 
Additionally, the uncertainty in the modeling of
initial- and final-state radiation (ISR / FSR) is assessed with two alternative {\powheg}+{\pythiaeight} samples~\cite{ATL-PHYS-PUB-2017-007}. 
One sample with the amount of radiation increased 
has the renormalization and factorization scales 
decreased by a factor of two, the \hdamp\ parameter doubled, 
and uses the Var3c upward variation of the A14 parameter set. 
A second sample with the amount of radiation decreased has the scales increased by a factor of two and 
uses the Var3c downward variation of the A14 set. 
The uncertainties described in this paragraph correspond to three independent sources for each of the \ttlight, \ttcin\ and \ttbin\ components.

For the background from \ttcin, there is little guidance from theory or experiment to determine whether the nominal approach of
using charm jets produced primarily in the parton shower is more or less accurate than a prediction with \ttcc\ calculated at NLO
in the matrix element. 
For this reason, an NLO prediction with \ttcc\ in the matrix element, 
including massive $c$-quarks and therefore using the 3F scheme for the PDFs, 
is produced with \amcnlo\ interfaced to \herwigpp, 
as described in Ref.~\cite{ATL-PHYS-PUB-2016-011}. 
The difference between this sample and an inclusive \ttbar\ sample produced with the same event generator and a 5F scheme PDF set, 
in which the \ttcin\ process originates through the parton shower only, 
is taken as an additional uncertainty in the \ttcin\ prediction. 
This uncertainty is 
related to the choice between the \ttcc\ ME calculation and the prediction from the inclusive \ttbar\ production with $c$-jets via parton shower
and is applied as one additional independent source to the \ttcin{} background.

For the \ttbin\ process, the difference between the predictions from {\powheg}+{\pythiaeight} 
and \ShOL\ is considered as one additional source of uncertainty.
This uncertainty accounts for the difference between the description 
of the \ttbin\ process
by the NLO \ttbar\ inclusive MC sample with a 5F scheme and a description at NLO of \ttbb\ in the ME with a 4F scheme.
This uncertainty is not applied to the  \ttbmpifsr{} subcategory since it is not included in the 4F calculation.

The uncertainties described above do not affect the relative fractions of the \ttb, \ttbb, \ttB{} and \ttbbb{} subcomponents as these fractions are fixed to the prediction of \ShOL.
The uncertainties in these fractions 
in \ShOL\ are assessed separately and are divided into seven independent sources.
Three of these sources are evaluated by varying the renormalization scale up and down by a factor of two, changing the functional form 
of the resummation scale to $\mu_{\mathrm{CMMPS}}$, and adopting a global scale choice, 
$\mu_{\mathrm{Q}} = \mu_{\mathrm{R}} = \mu_{\mathrm{F}} = \mu_{\mathrm{CMMPS}}$. Additionally, two alternative PDF sets, {\textsc MSTW2008NLO}~\cite{MSTW}
and {\textsc NNPDF2.3NLO}, are considered, as well as an alternative shower recoil scheme and an alternative set of tuned parameters for the underlying event. 
These sources of uncertainty contribute to the uncertainty band shown in Figure~\ref{fig:ttbbFrac} for the \ShOL\ prediction. 
Given the large difference between the 4F prediction and the various 5F predictions for the \ttbbb\ process, 
which is not covered by the uncertainties described above, 
this sub-process is given an extra 50\% normalization uncertainty. 

The relative fraction of the \ttbmpifsr{} subcategory is not fixed in the alternative samples 
used to derive the systematic uncertainties related to the choice of NLO event generator, 
parton shower and hadronization model and to ISR/FSR.
These sources already incorporate variations related to the fraction and shape of the \ttbmpifsr{} subcategory.
In addition, a 50\% normalization uncertainty is assumed for the contribution from MPI, based on studies of different underlying event sets of tuned parameters.  

In total, thirteen independent sources of modeling uncertainties are assigned to the \ttbin{} component, four to the \ttcin{} component and three to the \ttlight{} component in addition to the one source that corresponds to the inclusive \ttbar\ production cross-section uncertainty.

An uncertainty of 40\% is assumed for the $W$ + jets cross-section, 
with
an additional 30\% normalization uncertainty used for $W$ + heavy-flavor jets, 
taken as uncorrelated between 
events with two and more than two heavy-flavor jets. 
These uncertainties are based on variations of 
the factorization and renormalization scales and of the matching parameters in the \textsc{Sherpa} simulation.
An uncertainty of 35\% is then applied to the $Z$~+~jets normalization, 
uncorrelated 
across jet bins, to account for both
the variations of the scales and matching parameters in \textsc{Sherpa} simulation and
the uncertainty in the extraction from data of the correction factor for the heavy-flavor component.

An uncertainty of $^{+5\%}_{-4\%}$ is considered for each of the three single-top production mode cross-sections 
~\cite{Kidonakis:2011wy,Kidonakis:2010ux,Kidonakis:2010tc}. 
For the $Wt$ and $t$-channel production modes, 
uncertainties associated with the choice of parton shower and hadronization model and with initial- and final-state radiation
are evaluated 
according to a set of alternative samples analogous to those used for the \ttbar\ process: 
the nominal prediction is compared with samples generated with \powheg\ interfaced with \herwigpp\ 
and with 
alternative \powhegbox\ v1 + \pythia\ samples 
with factorization and renormalization scale variations and appropriate variations of the Perugia 2012 set of tuned parameters. 
The uncertainty in the amount of interference between $Wt$ and \ttbar\ 
production at NLO~\cite{Frixione:2008yi} is assessed by comparing the default `diagram removal' scheme to the alternative
`diagram subtraction' scheme.

A 50\% normalization uncertainty in the diboson background is assumed, which includes uncertainties
in the inclusive cross-section and additional jet production~\cite{ATL-PHYS-PUB-2016-002}. 
The uncertainty of the $\ttbar V$ NLO cross-section prediction is 15\%~\cite{Campbell:2012dh}, 
split into PDF and scale uncertainties as for \ttH.
An additional $\ttbar V$ modeling uncertainty, related to the choice of event generator, parton shower and hadronization model, 
is assessed 
by comparing the nominal sample with alternative ones generated with \sherpa.
Uncertainties in $\ttbar V$ production are all treated as uncorrelated between $\ttbar Z$ and $\ttbar W$.
A total 50\% normalization uncertainty is considered for the $t\bar{t}t\bar{t}$ background.
The small backgrounds from $tZ$, $t\bar{t}WW$, $tHjb$ and $WtH$  
are each assigned two cross-section uncertainties, split into PDF and scale uncertainties, while $tWZ$ is assigned one cross-section uncertainty that accounts for both the scale and PDF effects.

Finally, a 50\% uncertainty is assigned to the overall estimated yield of non-prompt lepton events in the single-lepton channel, 
taken as uncorrelated between electron-plus-jet and muon-plus-jet events, 
between boosted and resolved analysis categories, 
and between the resolved analysis categories with exactly five jets and those with six or more jets. 
In the dilepton channel, the non-prompt lepton background is assigned a 25\% uncertainty, 
correlated across lepton flavors and all analysis categories.

\section{Results}
\label{sec:results}
The distributions of the discriminants from each of the 
analysis categories 
are combined in a profile likelihood fit
to test for the presence of a signal,
while simultaneously determining the normalization and constraining the differential distributions of the most important background components.
As described in Section~\ref{sec:mvas}, 
in the signal regions, the output of the classification BDT is used as the discriminant 
while only the total event yield is used in the control regions, 
with the exception of \crfivejttc\ and \crsixjttc, where the \hthad\ distribution is used.
No distinction is made in the fit between signal and control regions, other than a different choice of discriminant variables. 
The binning of the classification BDT is optimized to maximize the analysis sensitivity while keeping the total MC statistical uncertainty 
in each bin to a level adjusted to avoid biases due to fluctuations in the predicted number of events.

The likelihood function, ${\cal L}(\mu,\theta)$, 
is constructed as a product of Poisson probability terms over all bins in each distribution.
The Poisson probability depends on the predicted number of events in each bin, which in turn is a function of the
signal-strength parameter $\mu = \sigma/\sigma_{\mathrm{SM}}$ and $\theta$, where $\theta$ is the set of nuisance parameters that encode 
the effects of systematic uncertainties,
and of the two free floating normalization factors $k(\ttbin)$ and $k(\ttcin)$ for the \ttbin\ and \ttcin{} backgrounds, respectively.
The nuisance parameters are implemented in the likelihood function as 
Gaussian, log-normal or Poisson priors, 
with the exception of the normalization factors $k(\ttbin)$ and $k(\ttcin)$, 
for which no prior knowledge from theory or subsidiary measurements is assumed 
and hence which are only 
constrained by the profile likelihood fit to the data.
The statistical uncertainty of the prediction, 
that incorporates the statistical uncertainty of the MC events and 
of the data-driven fake and non-prompt lepton estimate, is included in the likelihood in the form of additional nuisance parameters, one for each of the included bins.
The test statistic 
$t_\mu$ 
is defined as the profile likelihood ratio: 
$t_\mu = -2\ln({\cal L}(\mu,\hat{\hat{\theta}}_\mu)/{\cal L}(\hat{\mu},\hat{\theta}))$,
where $\hat{\mu}$ and $\hat{\theta}$ are the values of the parameters which maximize the likelihood function, 
and $\hat{\hat{\theta}}_\mu$ are the values of the 
nuisance parameters which maximize the likelihood function for a given value of $\mu$. 
This test statistic is used 
to measure the probability that the observed data is compatible with the background-only hypothesis, 
and to perform statistical inferences about $\mu$, 
such as upper limits using the CL$_{\textrm{s}}$ method~\cite{Cowan:2010js,Read:2002hq,Junk:1999kv}. 
The uncertainty of the best-fit value of the signal strength, $\hat{\mu}$, 
is obtained varying 
$t_\mu$ 
by one unit. 

Figure~\ref{fig:summary_all} shows the 
observed event yield 
compared to the 
prediction in each control and signal region, 
both before the fit to data (`pre-fit') and after the fit to data (`post-fit'), 
performed in all the analysis categories in the two channels and with the signal-plus-background hypothesis. 
For the pre-fit prediction, the normalization factors for the \ttbin\ and \ttcin\ processes are set to 1, 
which corresponds to considering the prediction from \powheg\ + \pythiaeight\ for the fraction of each of these components relative to the total \ttbar{} prediction. 
Figure~\ref{fig:CR_slttc} shows the \hthad\ distributions 
in the \ttcin-enriched control regions of the single-lepton channel,
while Figures~\ref{fig:SR_dl},~\ref{fig:SR_sl5j} and~\ref{fig:SR_sl6j} show the distributions of the classification BDTs in the dilepton
and single-lepton signal regions, both before and after the fit. 
All these distributions are reasonably well modeled pre-fit within the assigned uncertainties.
The level of agreement is improved post-fit due to the nuisance parameters being adjusted by the fit. 
In particular, the best-fit values of $k(\ttbin)$ and $k(\ttcin)$ are  $1.24\pm0.10$ and $1.63\pm0.23$, respectively. 
The uncertainties in these measured normalization factors do not include the theory uncertainty of the corresponding \ttbin\ and \ttcin\ cross-sections. 
The post-fit uncertainty is also significantly reduced, as a result of the nuisance-parameter constraints and the correlations generated by the fit. 

In addition to the distributions that are given as input to the fit, 
all the distributions of the input variables to the classification BDTs in the signal regions are checked post-fit, 
and no significant deviations of the predictions from data are found. 
Figure~\ref{fig:mass} shows the data compared to the post-fit prediction for three of these distributions, 
namely 
the Higgs-boson candidate mass distributions
in the most sensitive signal regions in the dilepton channel and the single-lepton resolved channels 
as well as in the single-lepton boosted signal region. 

\begin{figure}[!ht]
\centering
\subfigure[]{\includegraphics[width=0.49\textwidth]{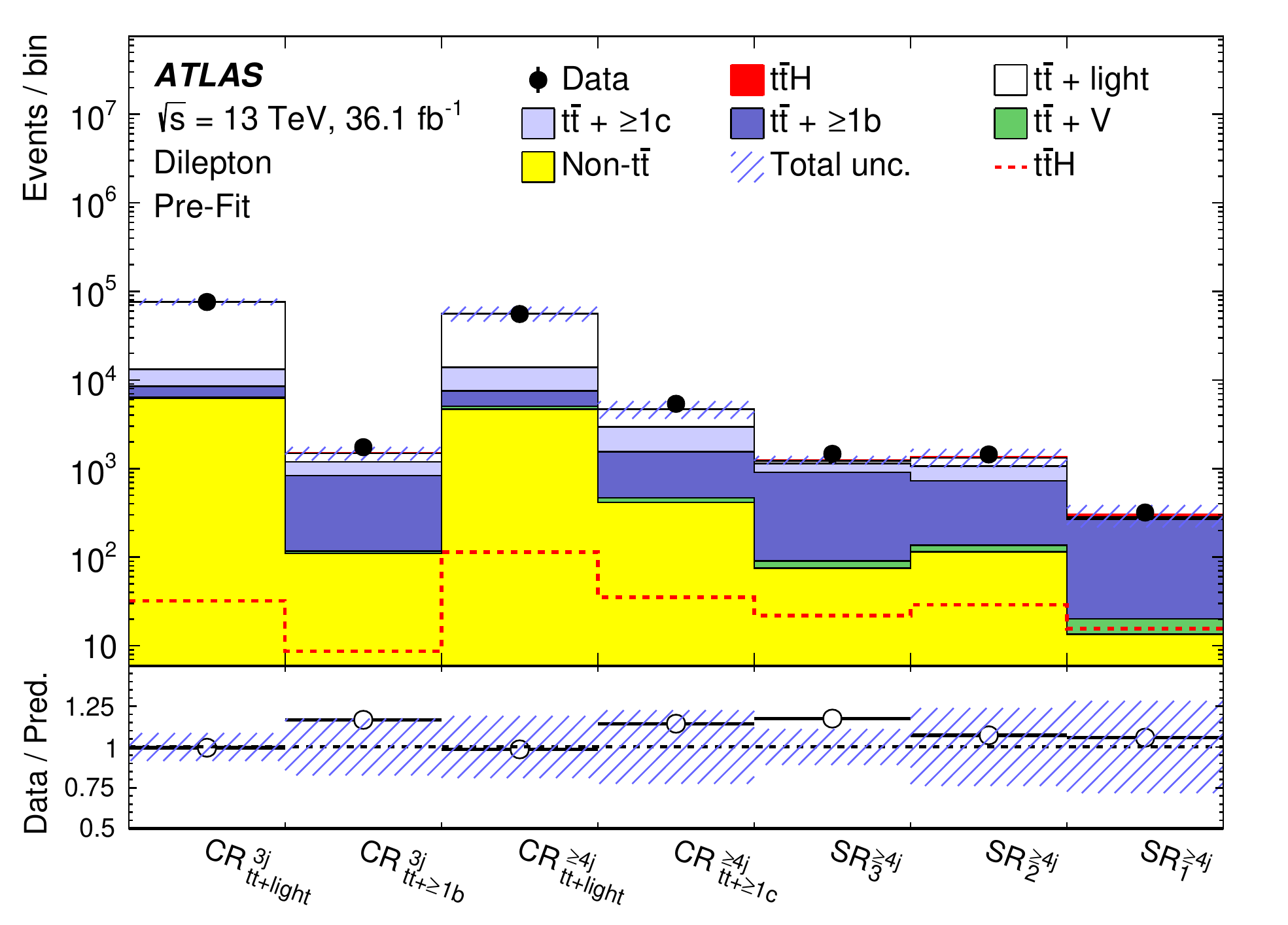}}\label{fig:summary_ljets_a}
\subfigure[]{\includegraphics[width=0.49\textwidth]{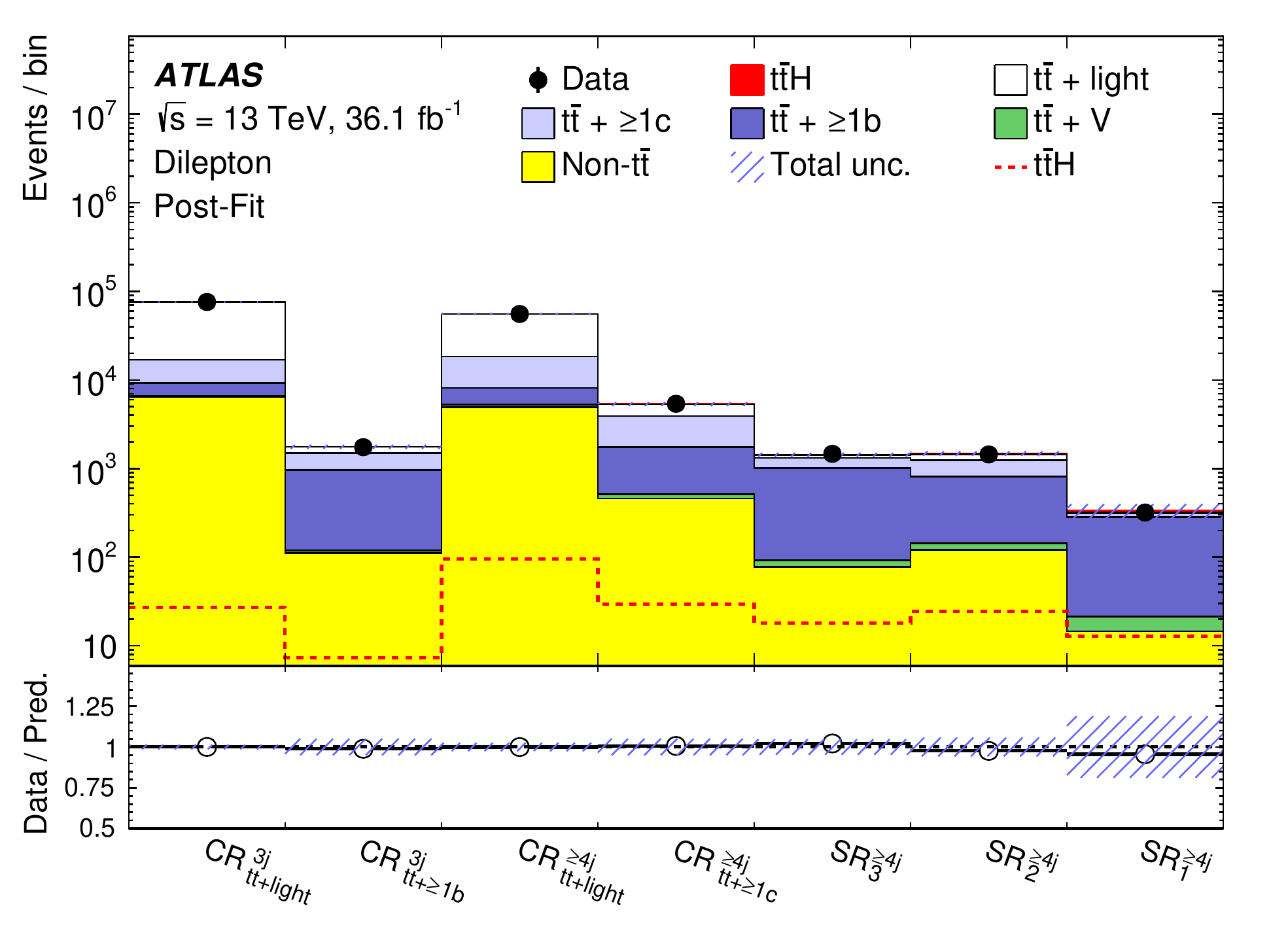}}\label{fig:summary_ljets_b}\\
\subfigure[]{\includegraphics[width=0.49\textwidth]{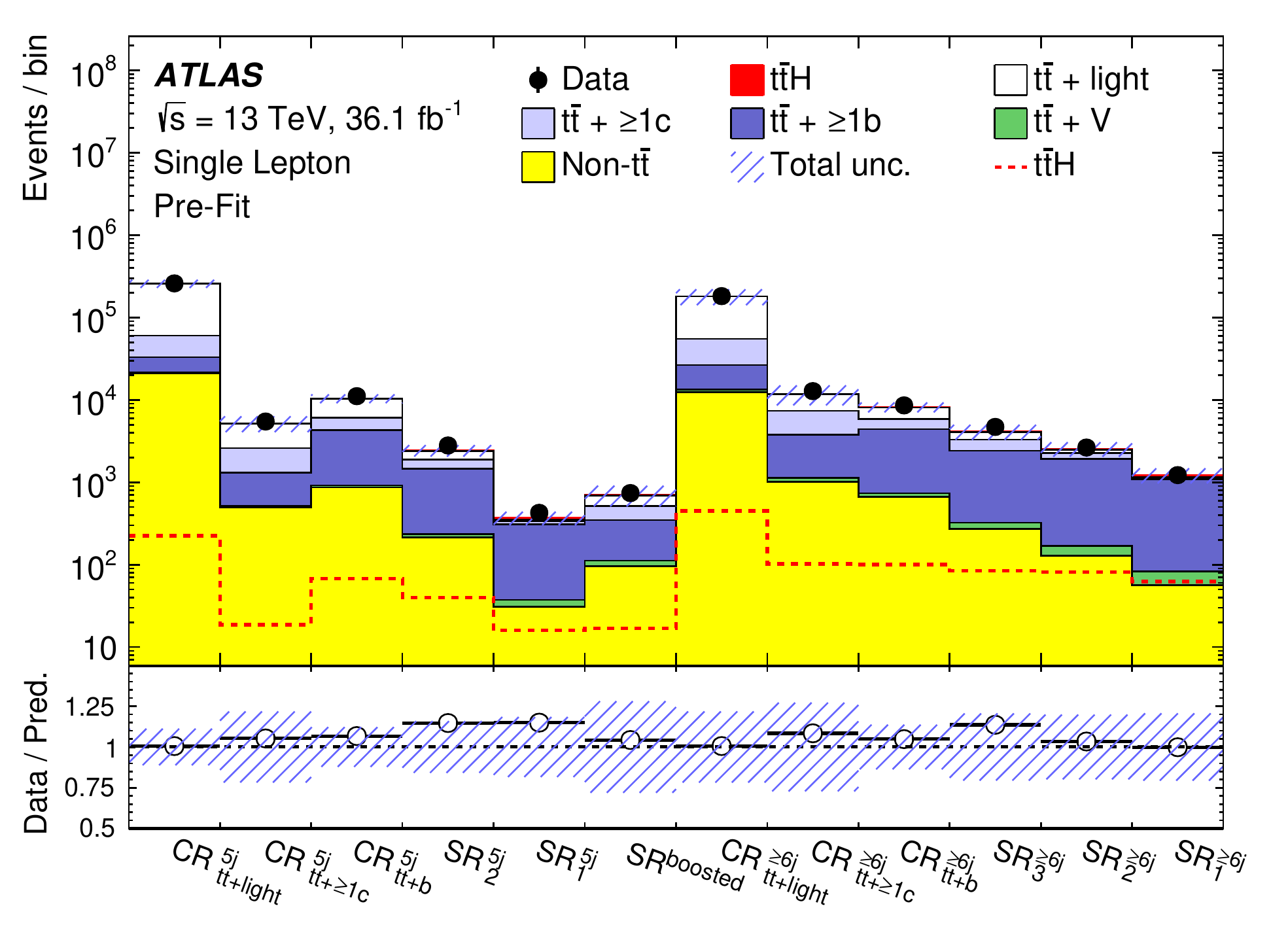}}\label{fig:summary_ljets_c}
\subfigure[]{\includegraphics[width=0.49\textwidth]{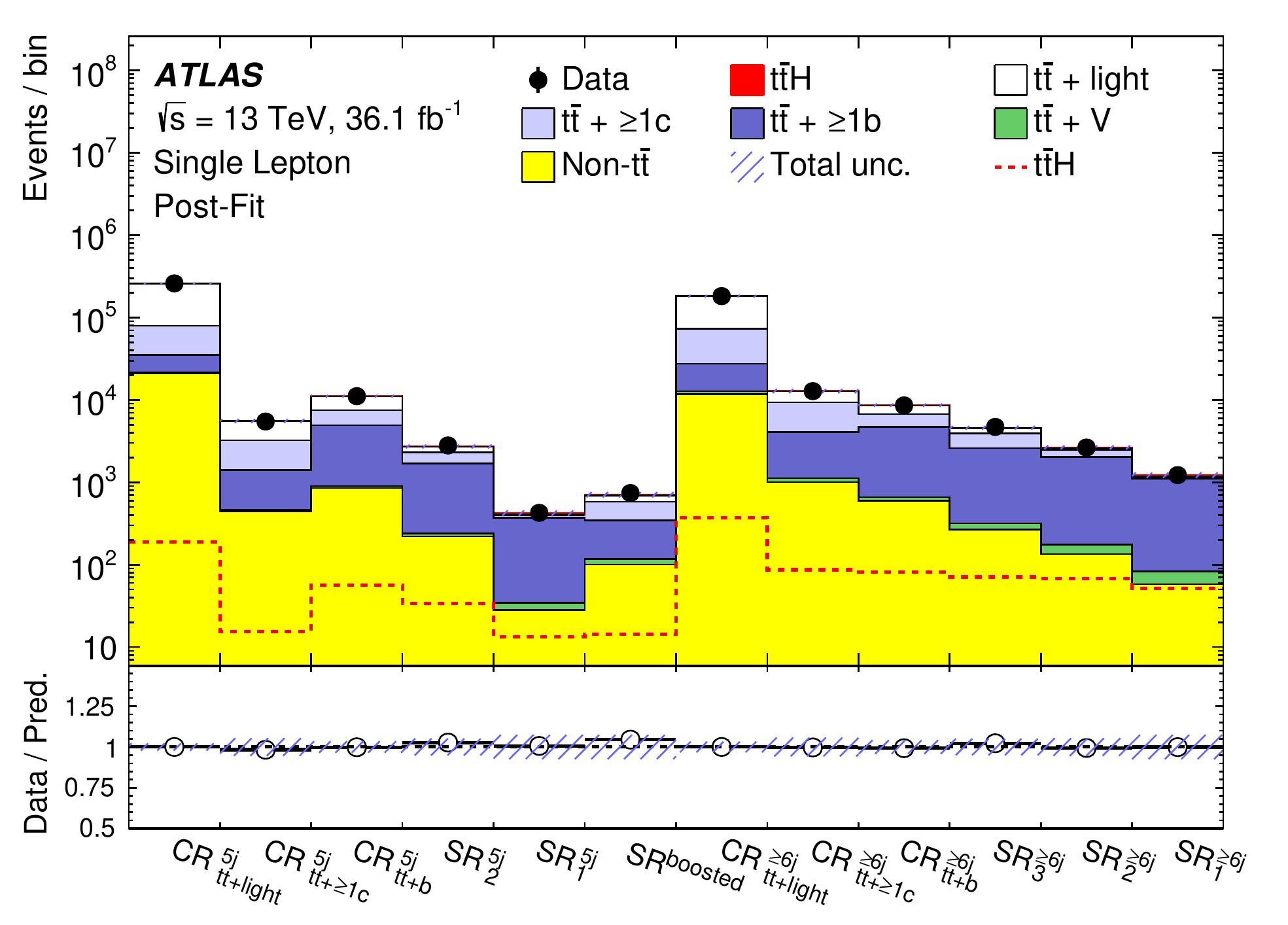}}\label{fig:summary_ljets_d}
\caption{Comparison of predicted and observed event yields in each of the control and signal regions, 
in the dilepton channel (a) before and (b) after the fit to the data, 
and in the single-lepton channel (c) before and (d) after the fit to the data. 
The $\ttH$ signal
is shown both as a filled red area stacked on the backgrounds
and separately for visibility as a dashed red line, 
normalized to the SM cross-section before the fit and to the 
fitted $\mu$ after the fit. 
The hatched area corresponds to the fitted uncertainty in the total prediction.
The pre-fit plots do not include an uncertainty for the \ttbin\ or \ttcin\ normalization.}
\label{fig:summary_all}
\end{figure}

\begin{figure}[ht!]
\begin{center}
\subfigure[]{\includegraphics[width=0.42\textwidth]{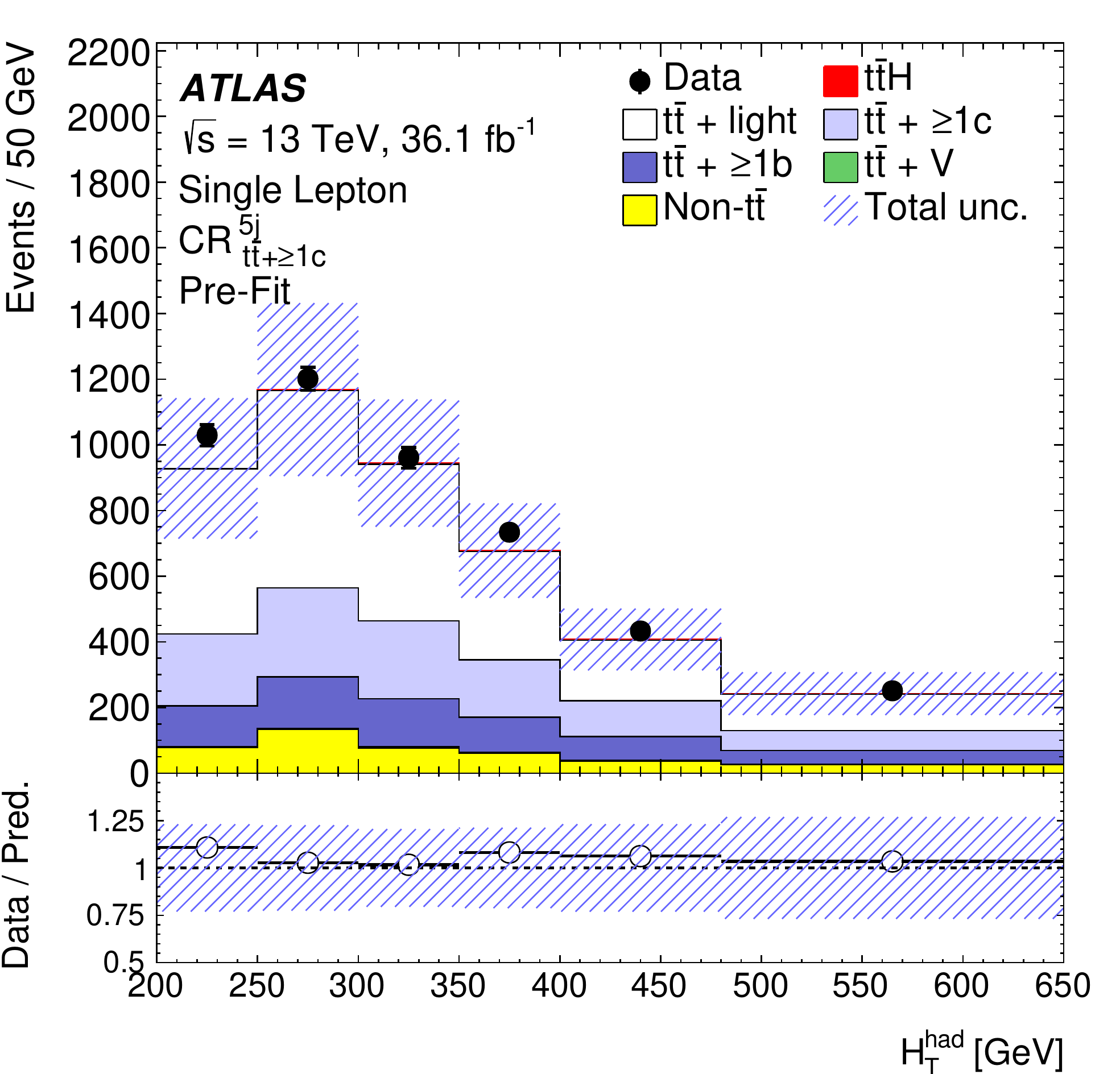}        }\label{fig:CR_sl5j_a}\hspace{1cm}
\subfigure[]{\includegraphics[width=0.42\textwidth]{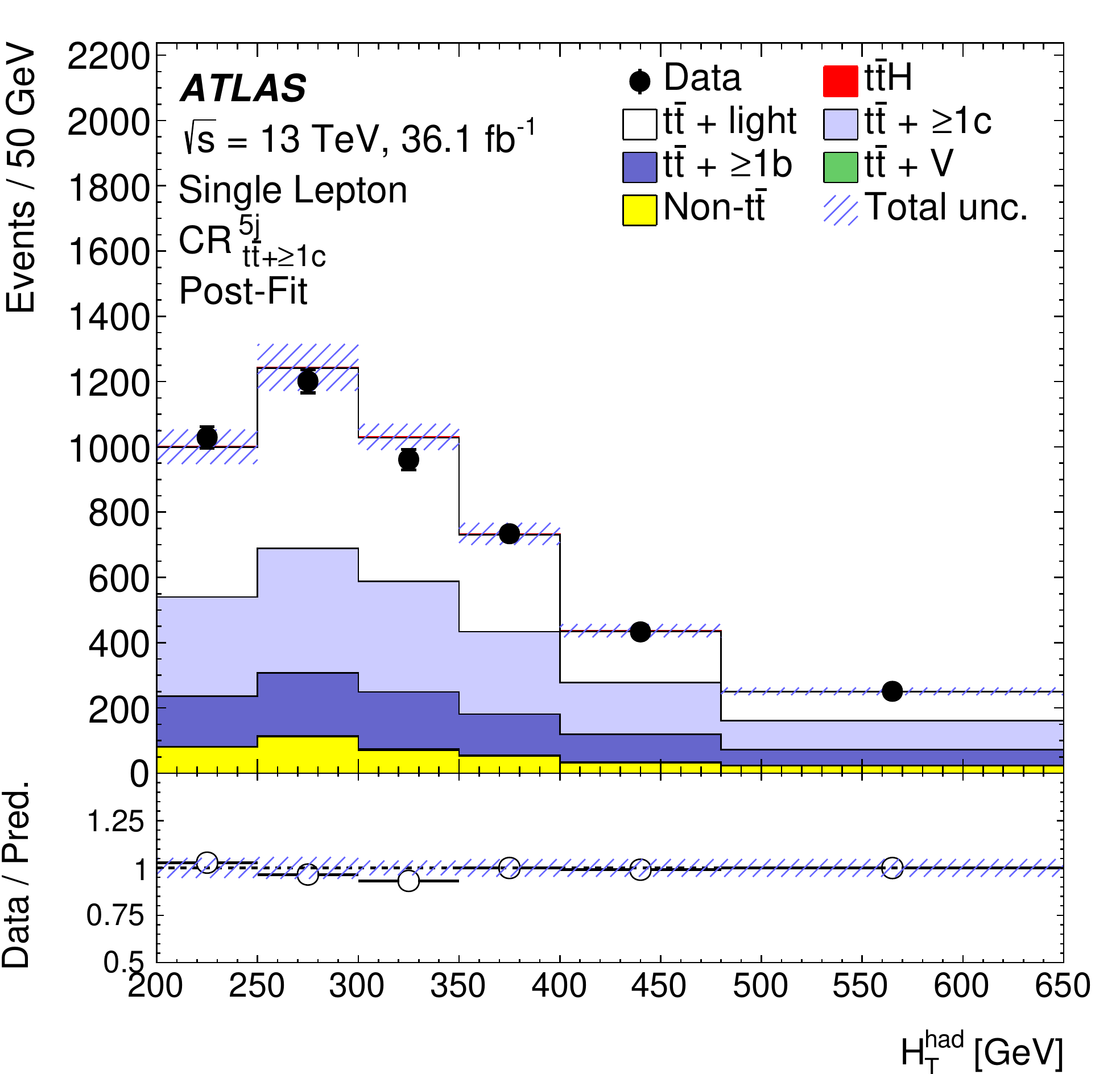}}\label{fig:CR_sl5j_b} \\
\subfigure[]{\includegraphics[width=0.42\textwidth]{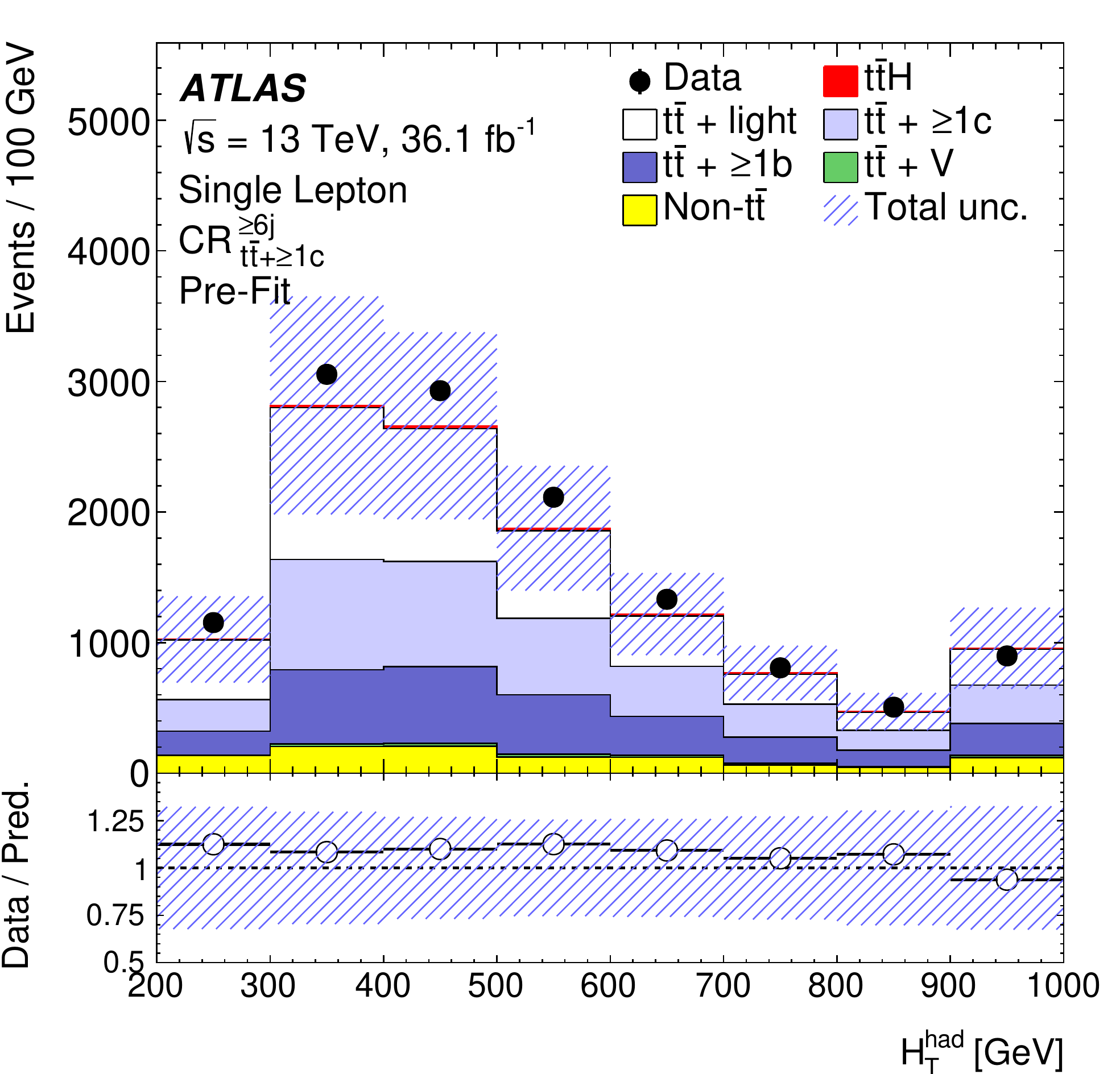}        }\label{fig:CR_sl6j_c}\hspace{1cm}
\subfigure[]{\includegraphics[width=0.42\textwidth]{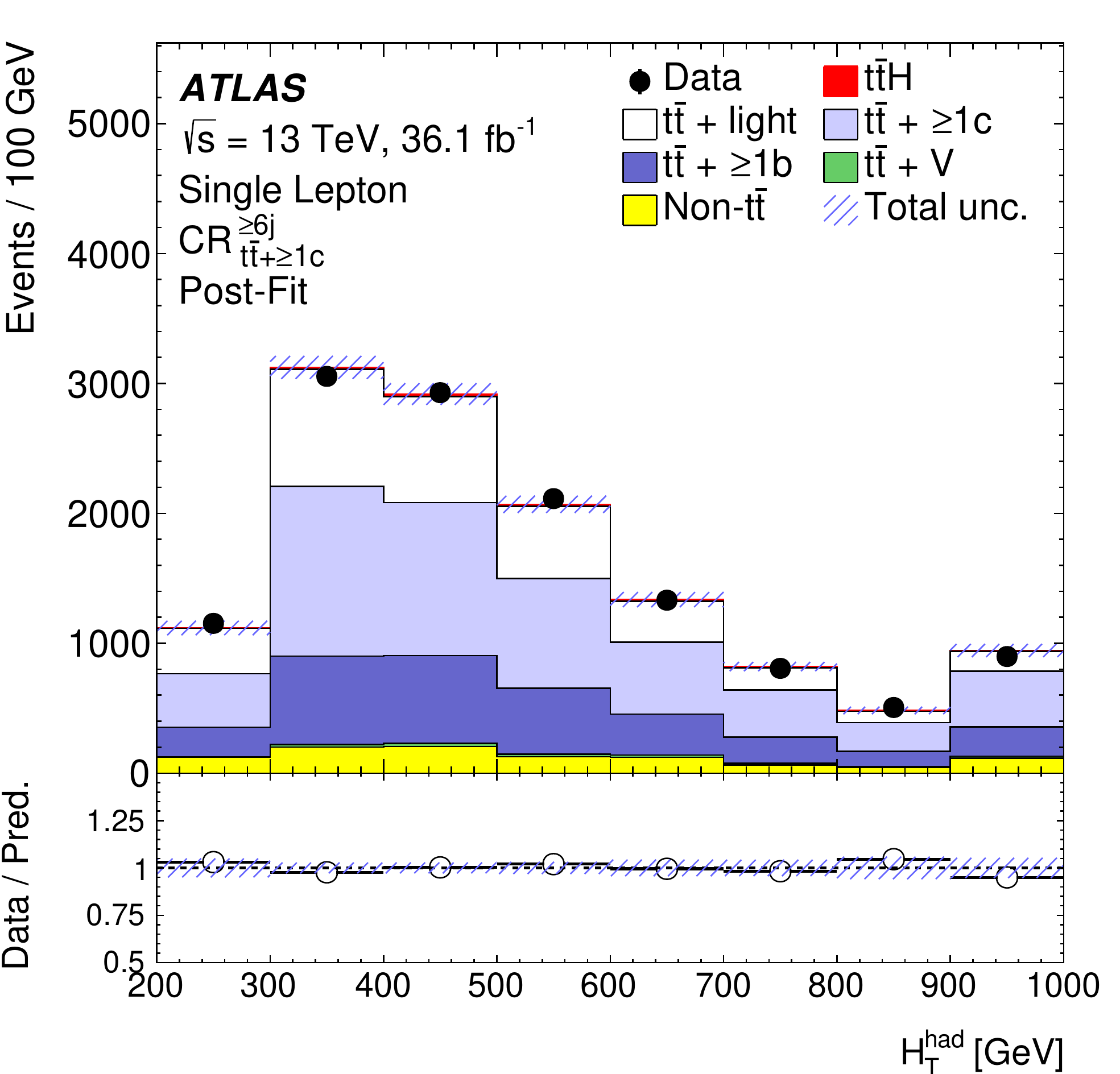}}\label{fig:CR_sl6j_d} \\
\caption{Comparison between data and prediction for the \hthad\ distributions in the single-lepton 
\ttcin-enriched control regions
(a, c) before, and (b, d) after the combined dilepton and single-lepton fit to the data. 
Despite its small contribution in these control regions, 
the \ttH\ signal prediction is shown stacked at the top of the background prediction, 
normalized to the SM cross-section before the fit and to the 
fitted $\mu$ after the fit. 
The pre-fit plots do not include an uncertainty for the \ttbin\ or \ttcin\ normalization.}
\label{fig:CR_slttc}
\end{center}
\end{figure}

\begin{figure}[ht!]
\begin{center}
\subfigure[]{\includegraphics[width=0.42\textwidth]{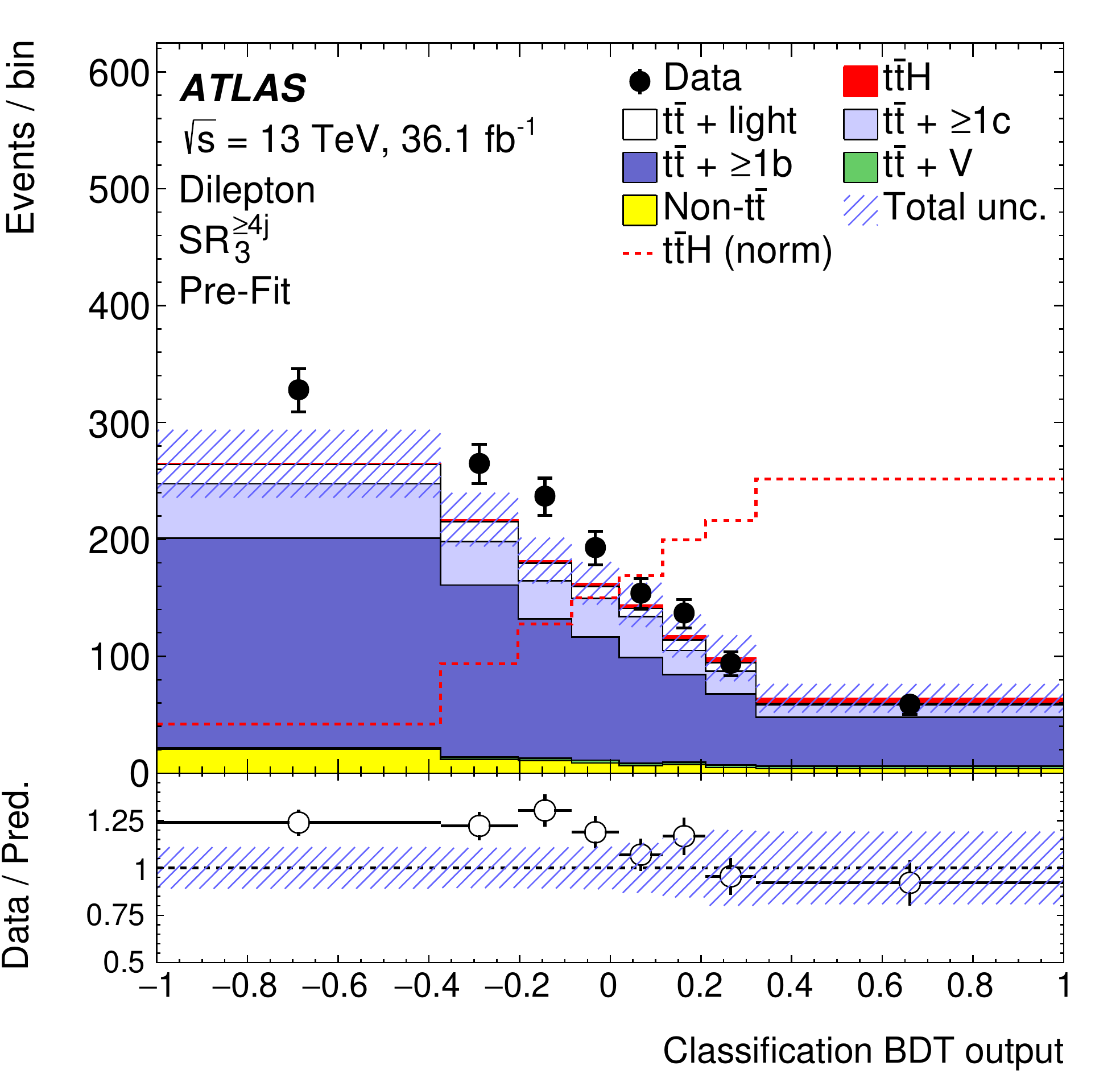}        }\label{fig:SR_dl_a}\hspace{1cm}
\subfigure[]{\includegraphics[width=0.42\textwidth]{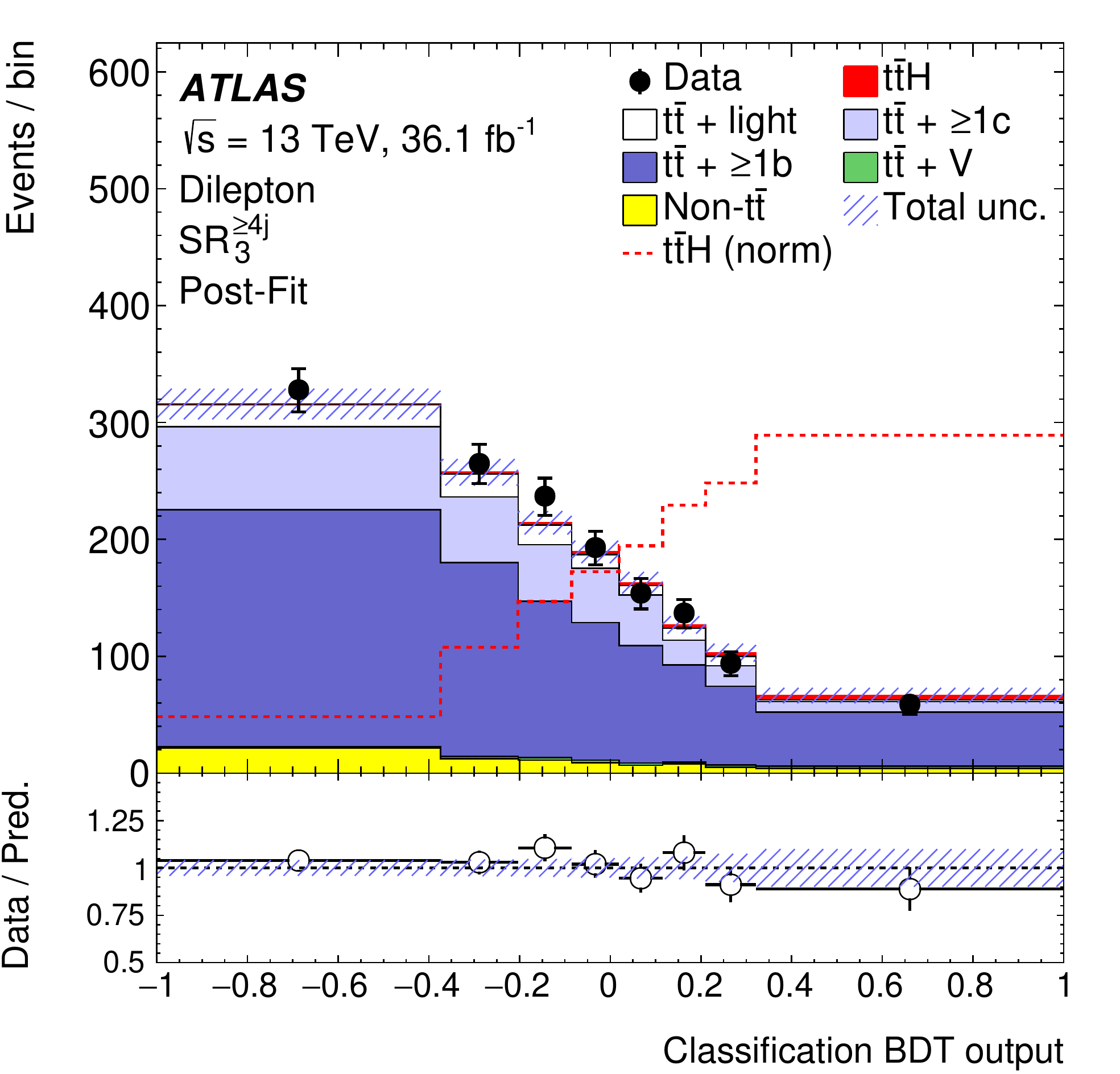}}\label{fig:SR_dl_b} \\[-0.3cm]
\subfigure[]{\includegraphics[width=0.42\textwidth]{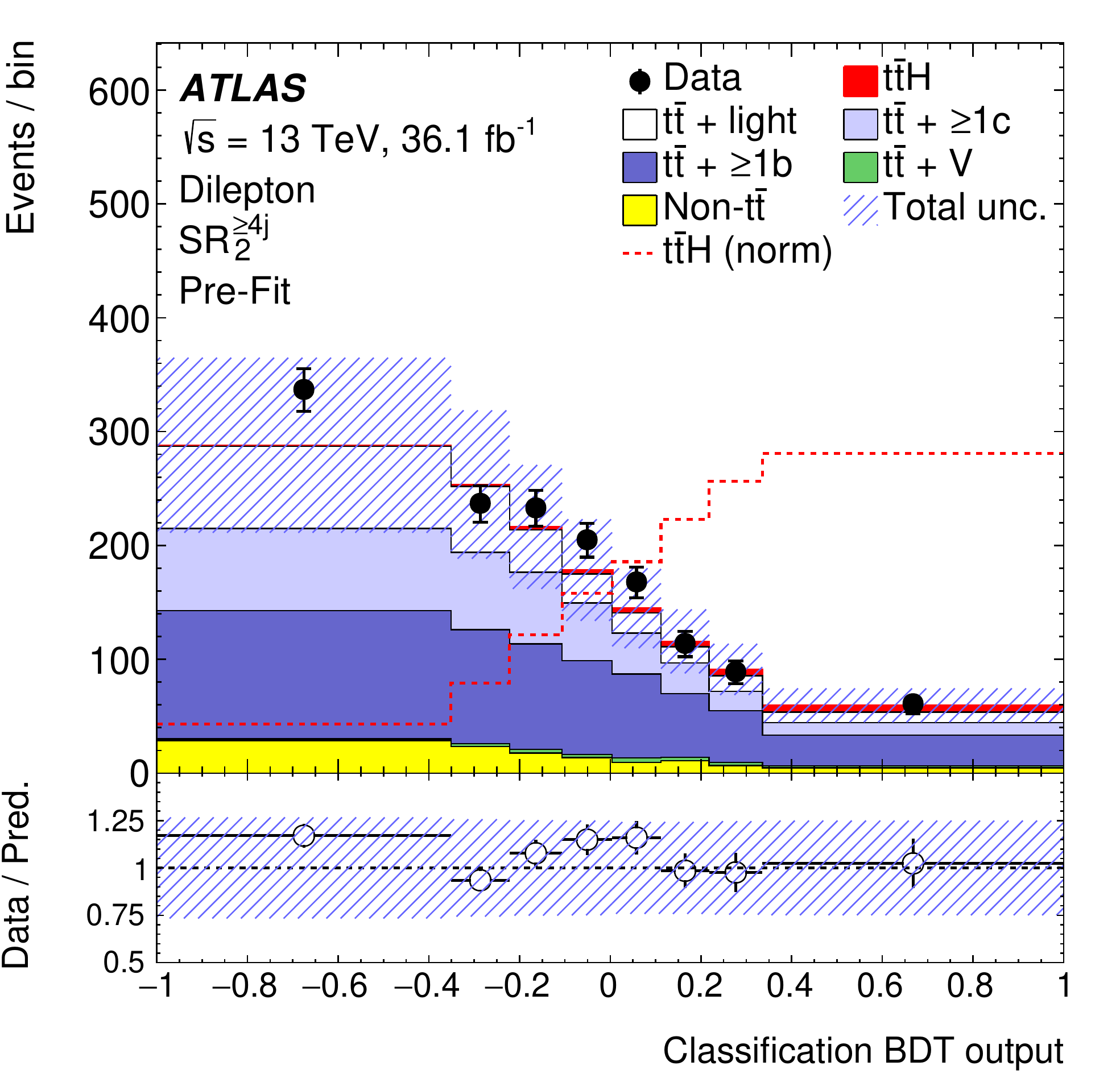}        }\label{fig:SR_dl_c}\hspace{1cm}
\subfigure[]{\includegraphics[width=0.42\textwidth]{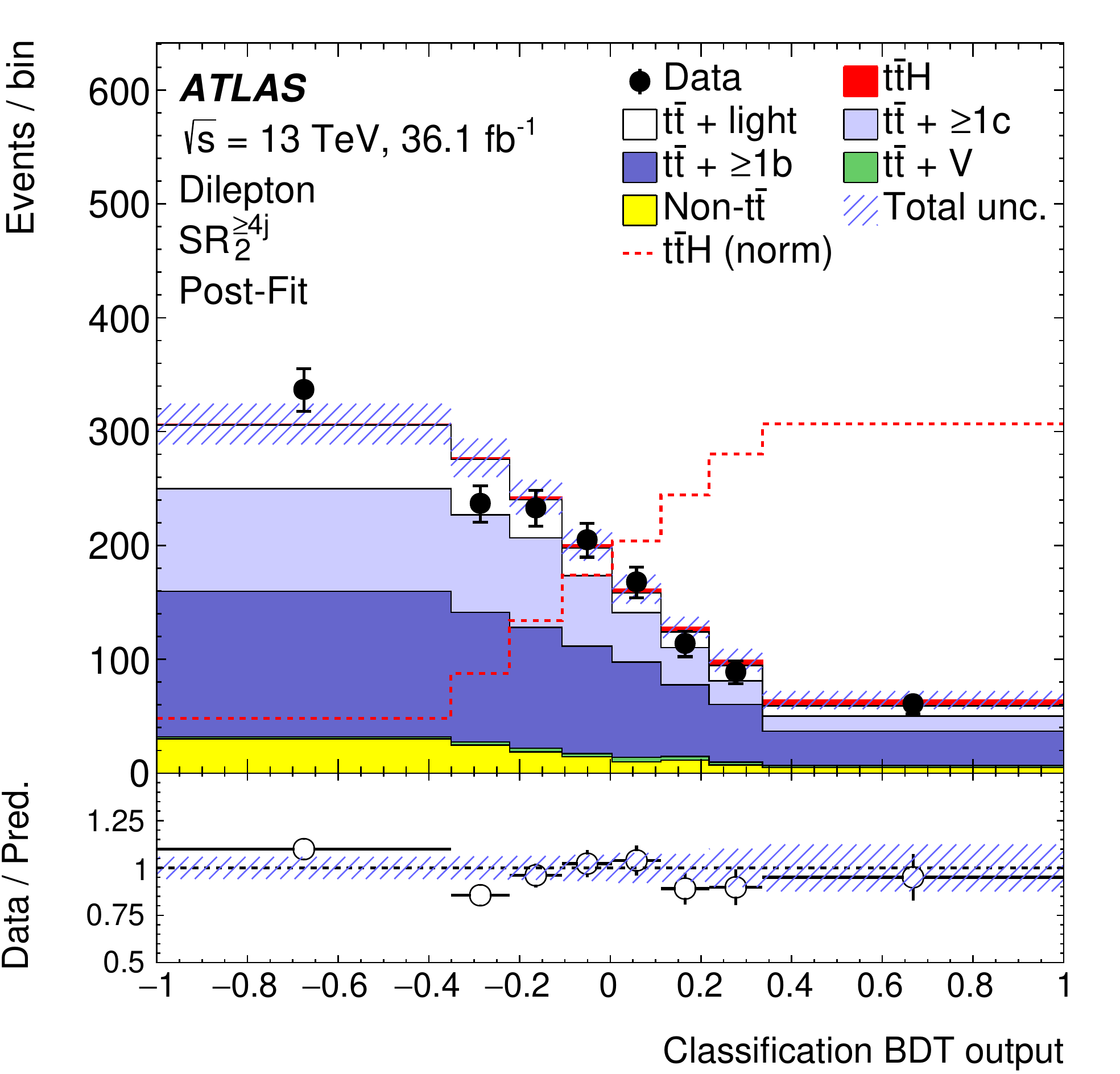}}\label{fig:SR_dl_d} \\[-0.3cm]
\subfigure[]{\includegraphics[width=0.42\textwidth]{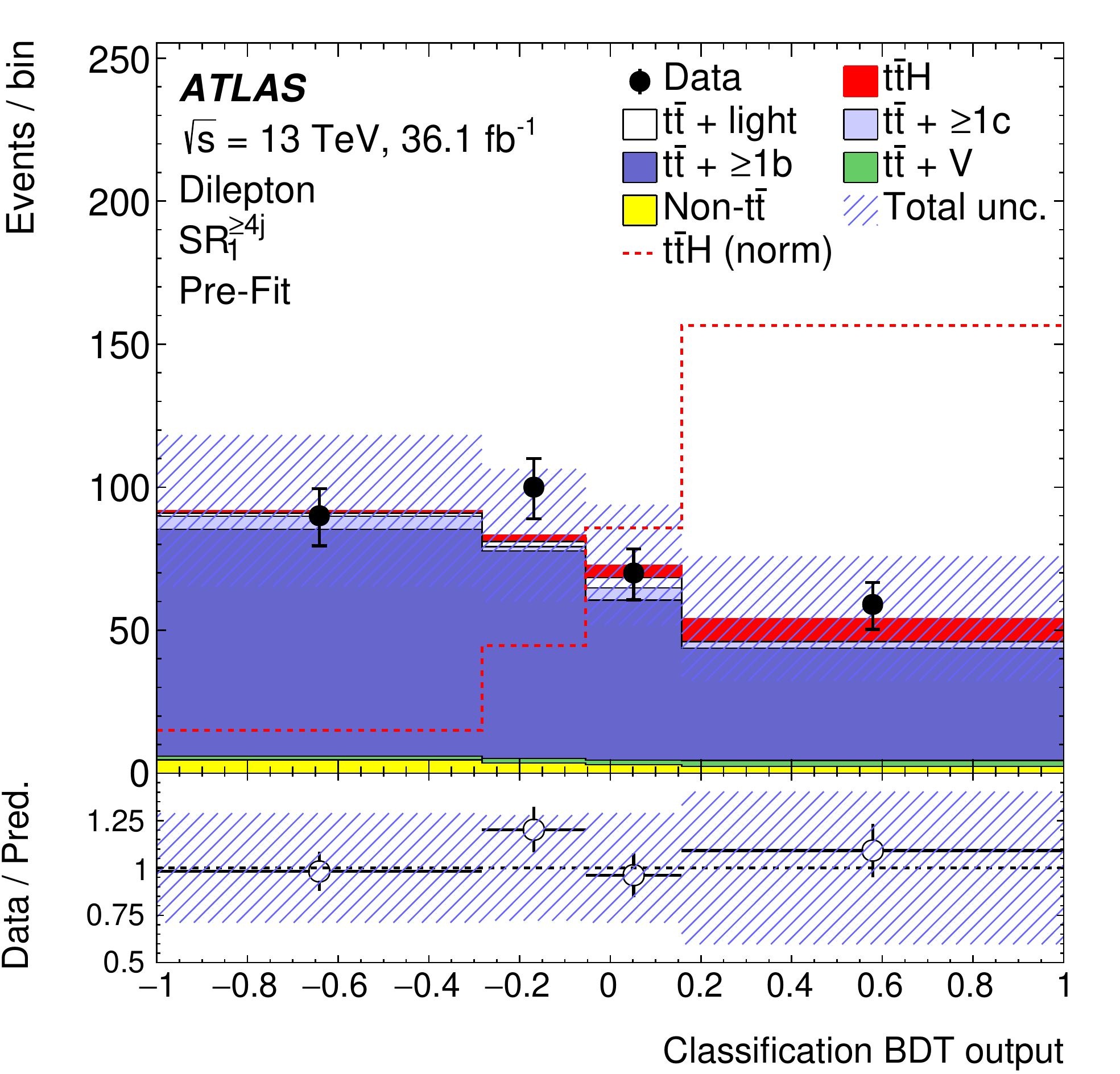}        }\label{fig:SR_dl_e}\hspace{1cm}
\subfigure[]{\includegraphics[width=0.42\textwidth]{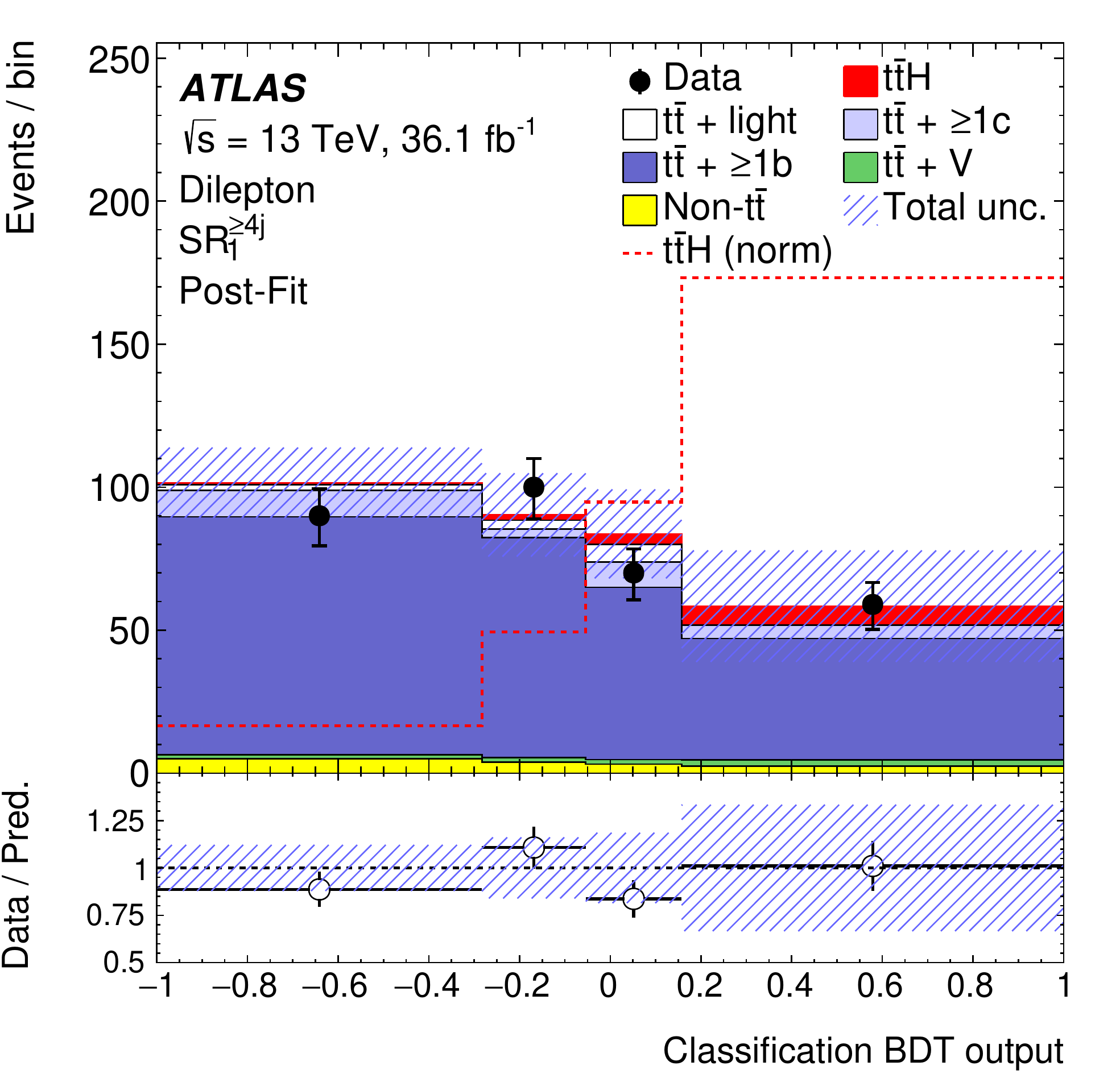}}\label{fig:SR_dl_f}
\vspace{-0.5cm}
\caption{Comparison between data and prediction for the BDT discriminant in the dilepton signal regions 
(a, c, e) before, and (b, d, f) after the combined dilepton and single-lepton fit to the data. 
The \ttH\ signal yield (solid red) is normalized to the SM cross-section before the fit and to the 
fitted $\mu$ after the fit. The dashed line shows the \ttH\ signal distribution normalized to the total 
background prediction. The pre-fit plots do not include an uncertainty for the \ttbin\ or \ttcin\ normalization.}
\label{fig:SR_dl}
\end{center}
\end{figure}

\begin{figure}[ht!]
\begin{center}
\subfigure[]{\includegraphics[width=0.42\textwidth]{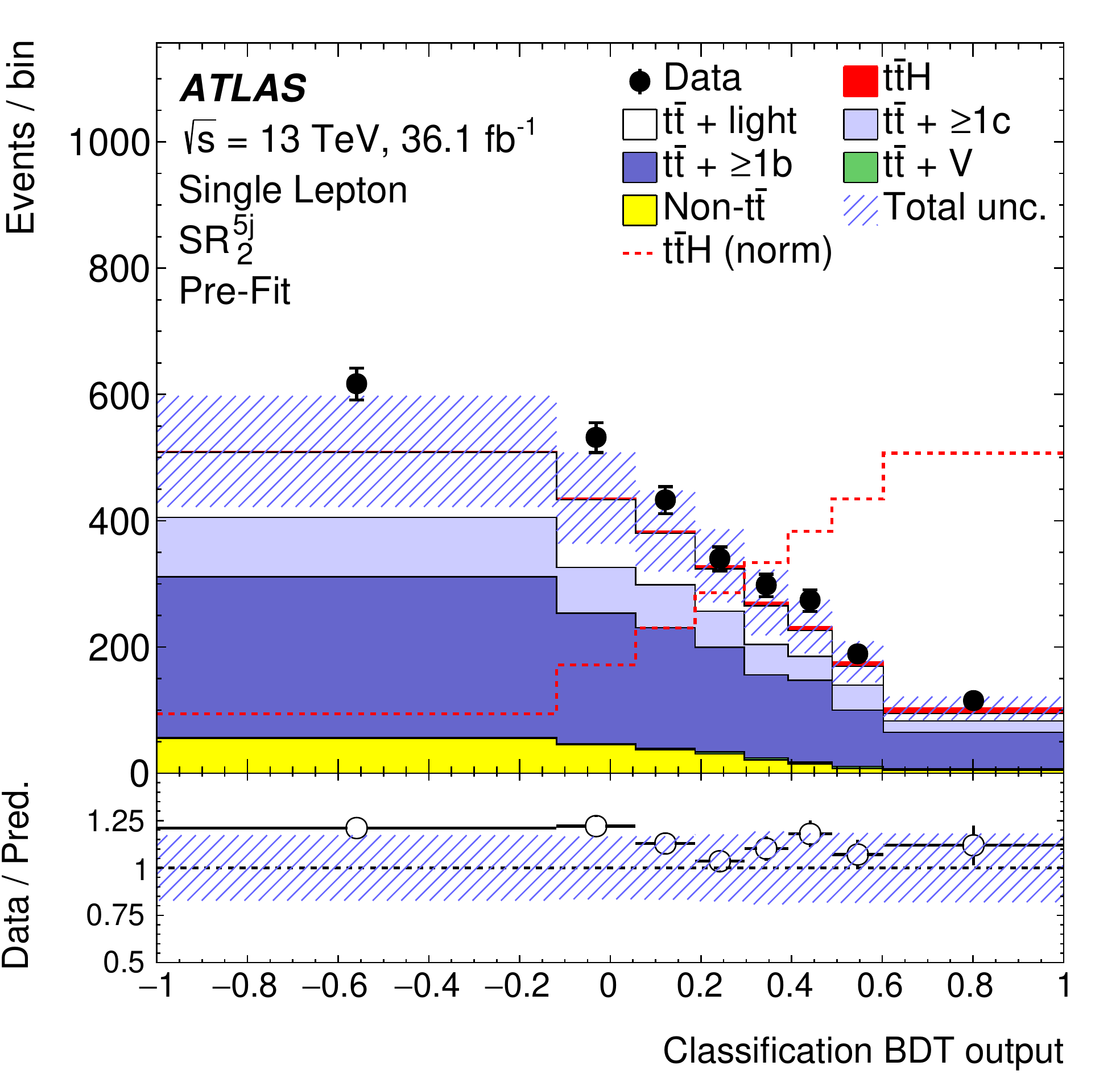}         }\label{fig:SR_sl5j_a}\hspace{1cm}
\subfigure[]{\includegraphics[width=0.42\textwidth]{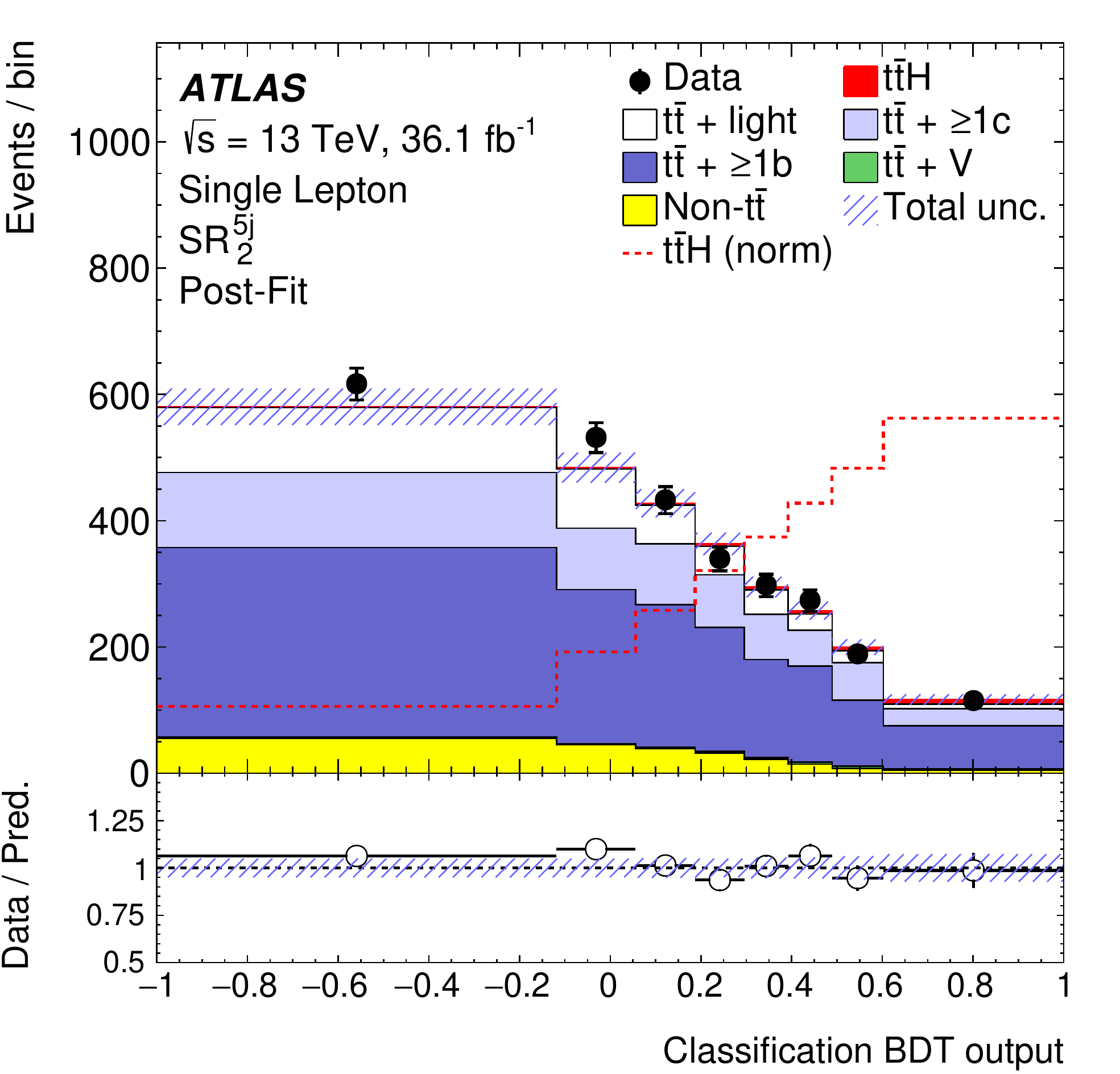} }\label{fig:SR_sl5j_b}\\[-0.3cm]
\subfigure[]{\includegraphics[width=0.42\textwidth]{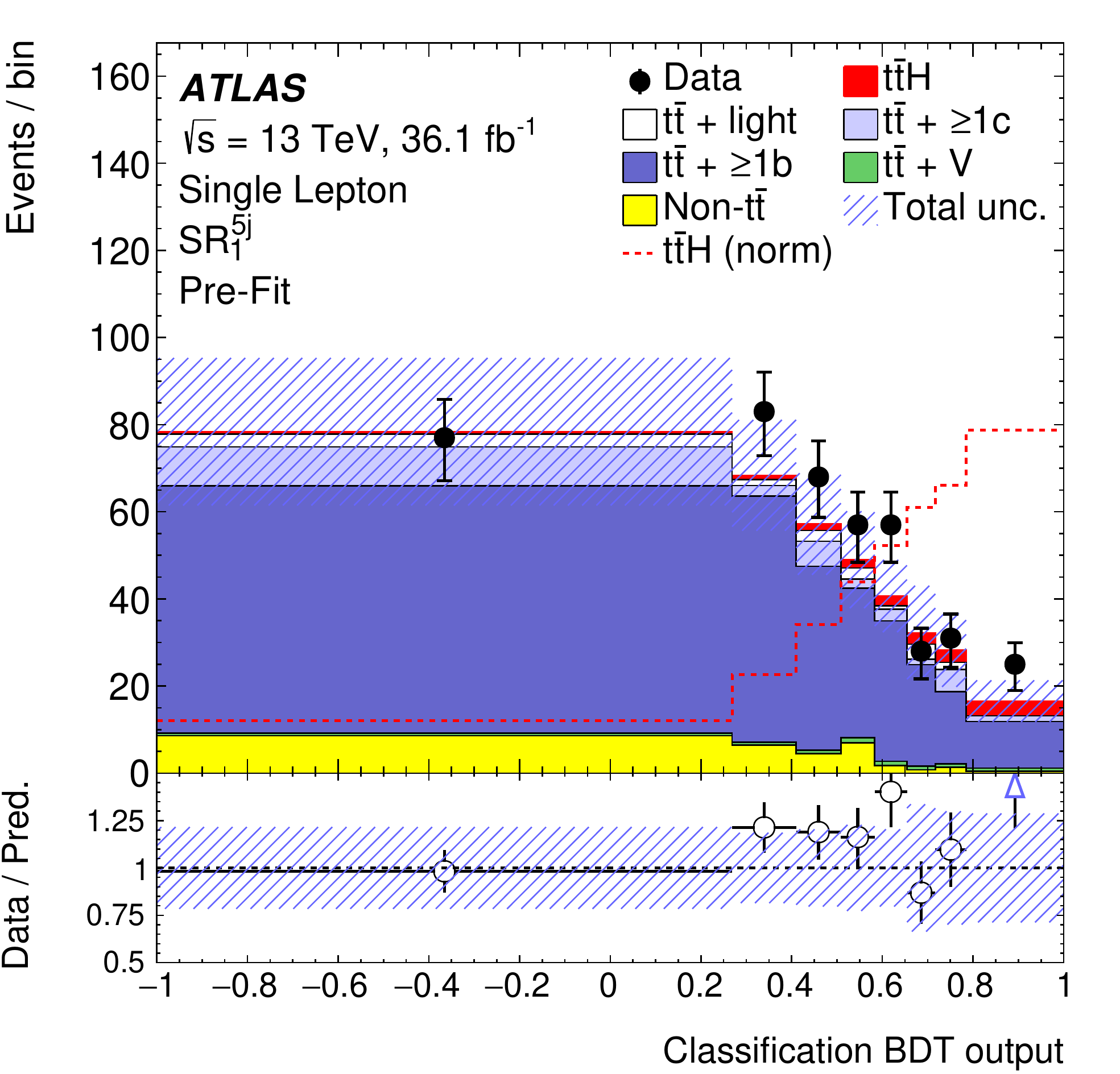}          }\label{fig:SR_sl5j_c}\hspace{1cm}
\subfigure[]{\includegraphics[width=0.42\textwidth]{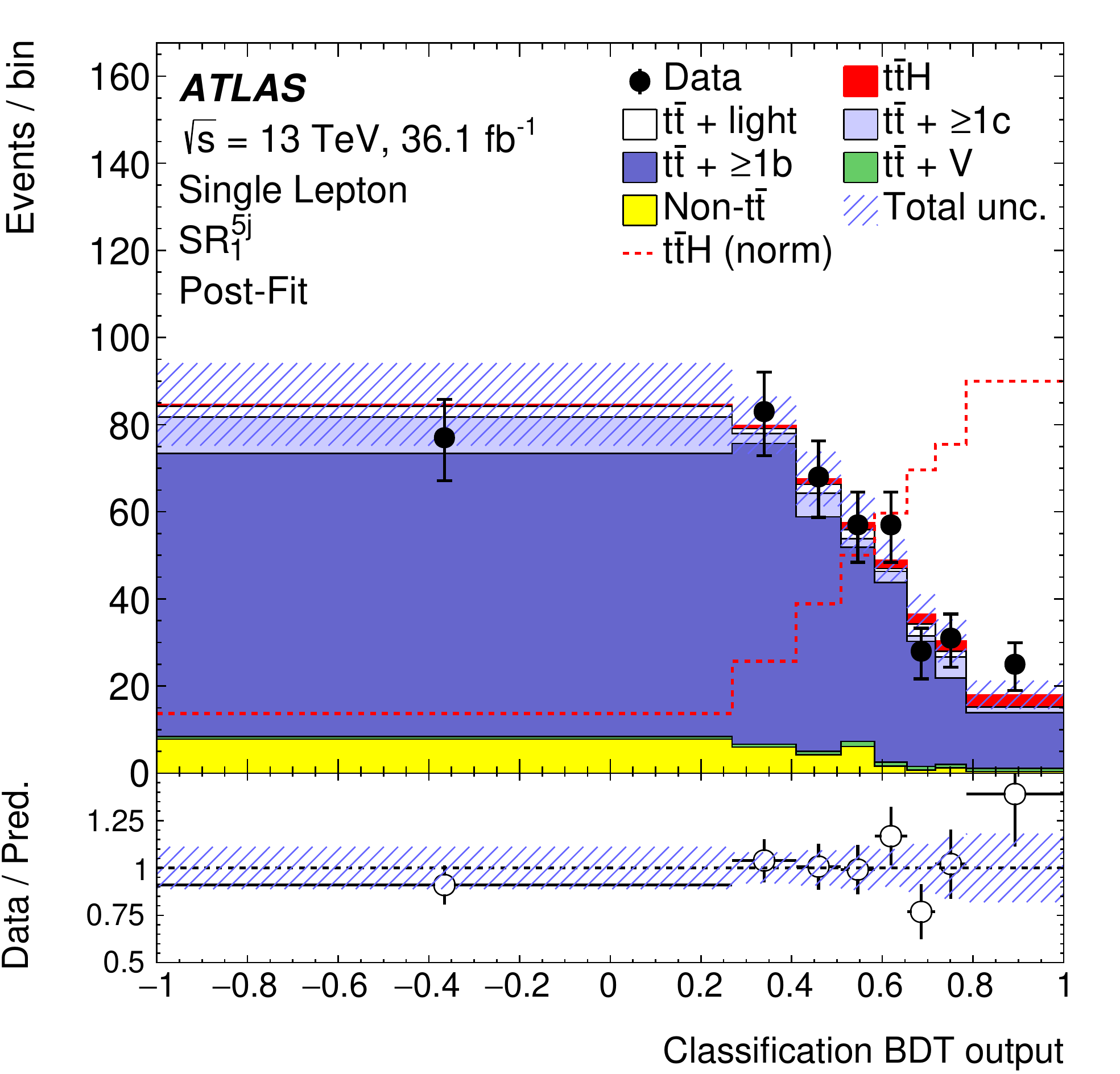}  }\label{fig:SR_sl5j_d}\\[-0.3cm]
\subfigure[]{\includegraphics[width=0.42\textwidth]{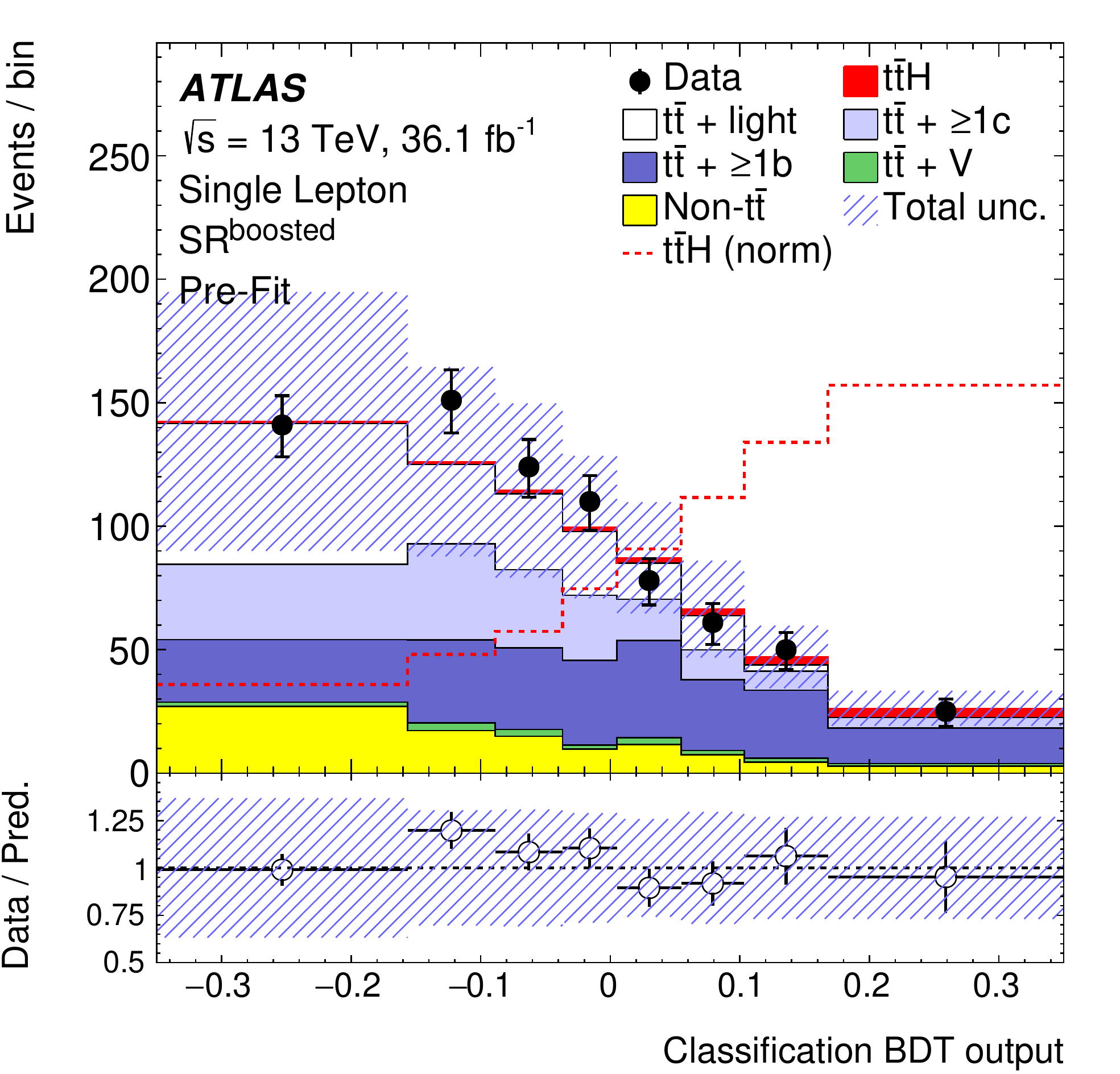}        }\label{fig:SR_sl5j_e}\hspace{1cm}
\subfigure[]{\includegraphics[width=0.42\textwidth]{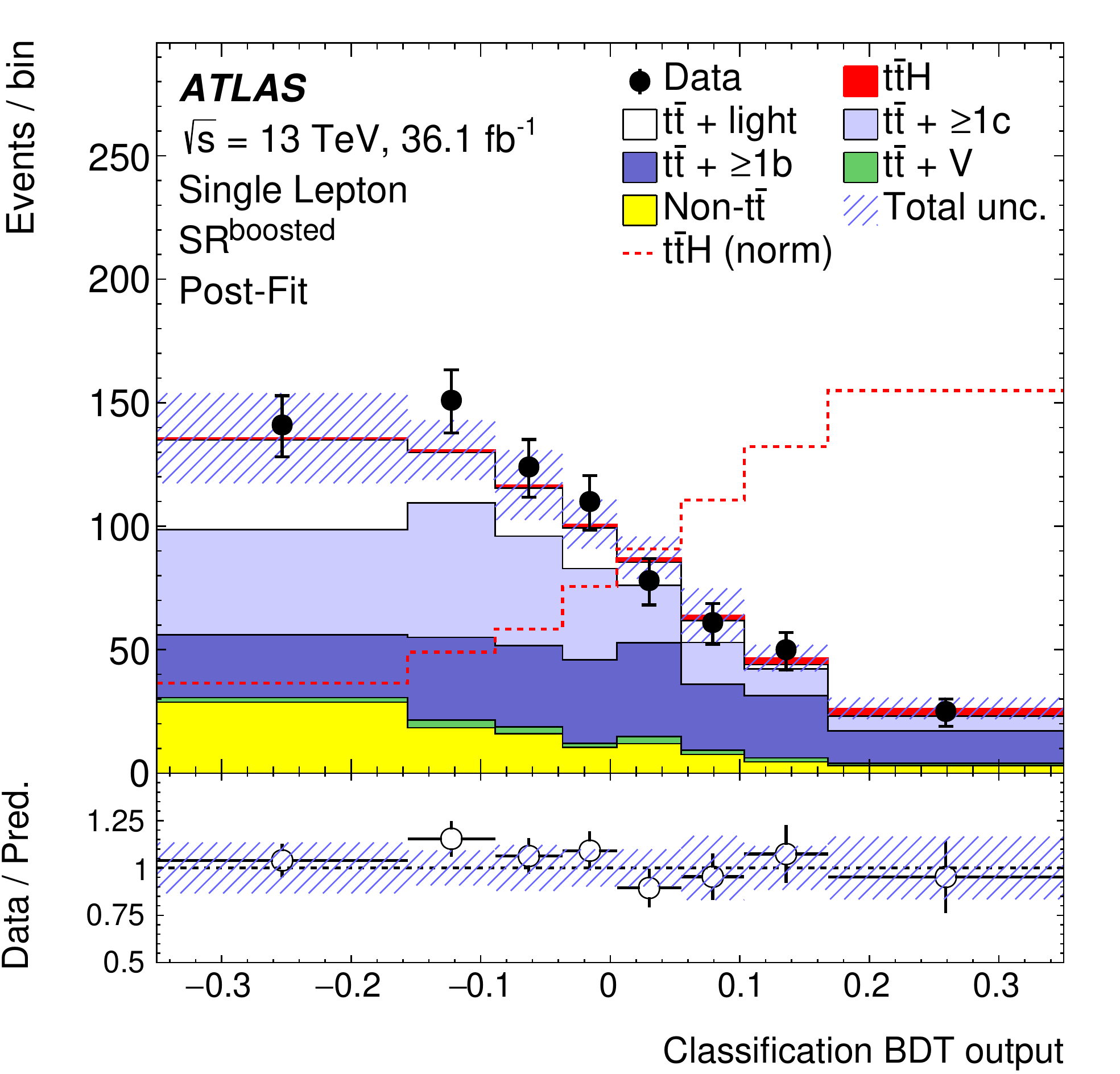}}\label{fig:SR_sl5j_f}
\vspace{-0.5cm}
\caption{Comparison between data and prediction for the BDT discriminant in the single-lepton channel
five-jet and boosted signal regions 
(a, c, e) before, and (b, d, f) after the combined dilepton and single-lepton fit to the data. 
The \ttH\ signal yield (solid red) is normalized to the SM cross-section before the fit and to the 
fitted $\mu$ after the fit. The dashed line shows the \ttH\ signal distribution normalized to the total 
background prediction.  The pre-fit plots do not include an uncertainty for the \ttbin\ or \ttcin\ normalization.}
\label{fig:SR_sl5j} 
\end{center}
\end{figure}

\begin{figure}[ht!]
\begin{center}
\subfigure[]{\includegraphics[width=0.42\textwidth]{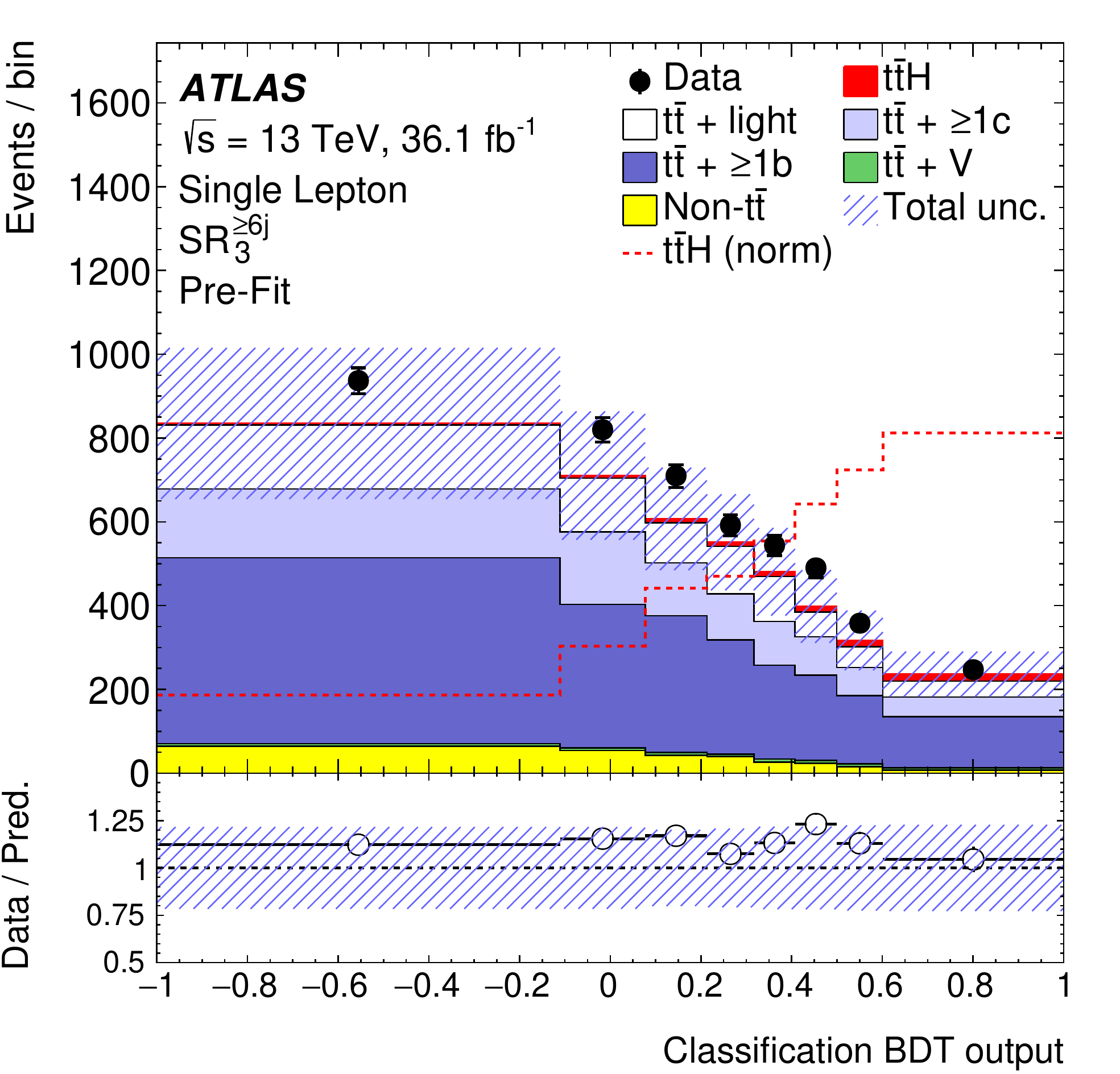}        }\label{fig:SR_sl6j_a}\hspace{1cm}
\subfigure[]{\includegraphics[width=0.42\textwidth]{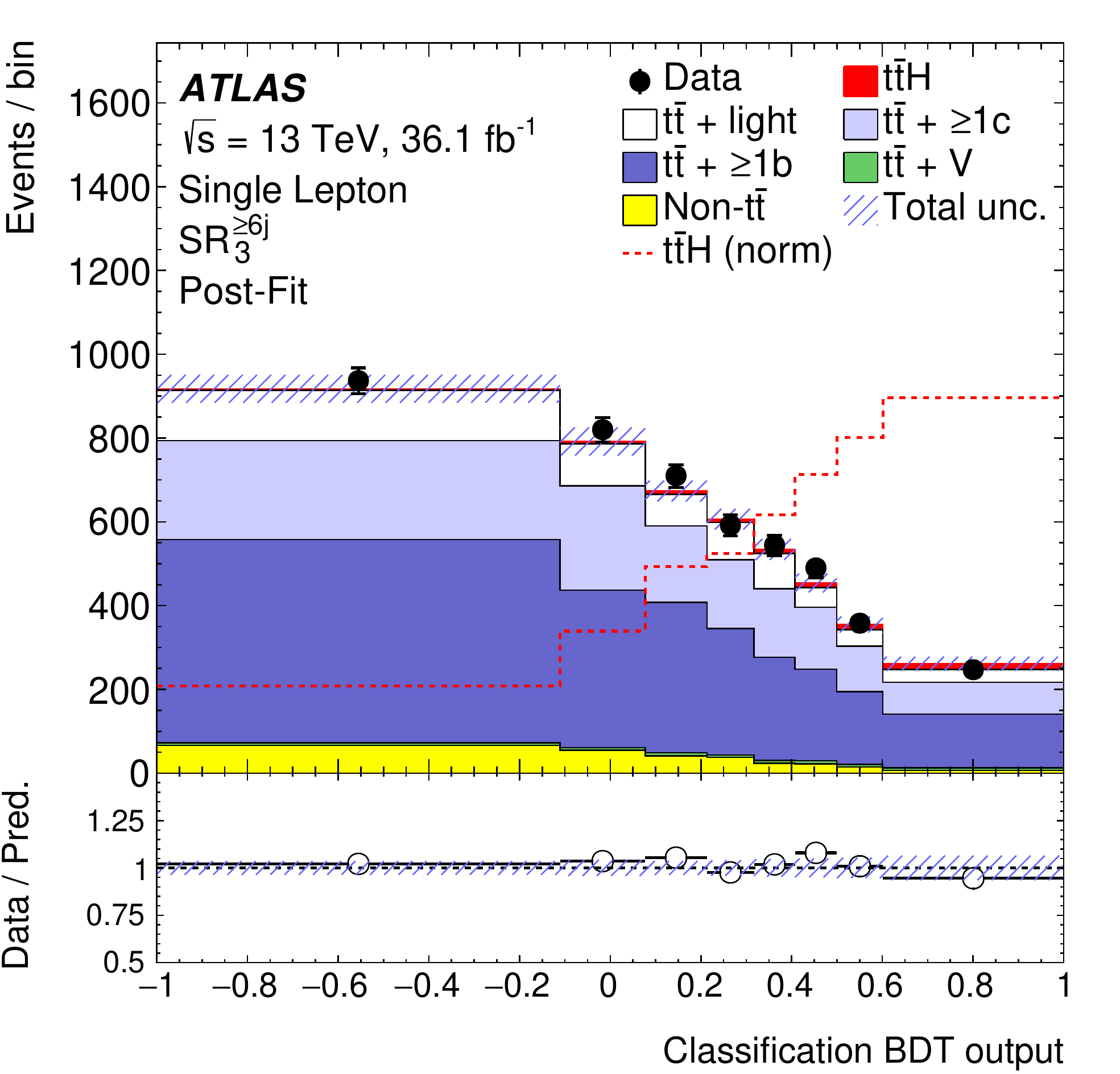}}\label{fig:SR_sl6j_b}\\[-0.3cm]
\subfigure[]{\includegraphics[width=0.42\textwidth]{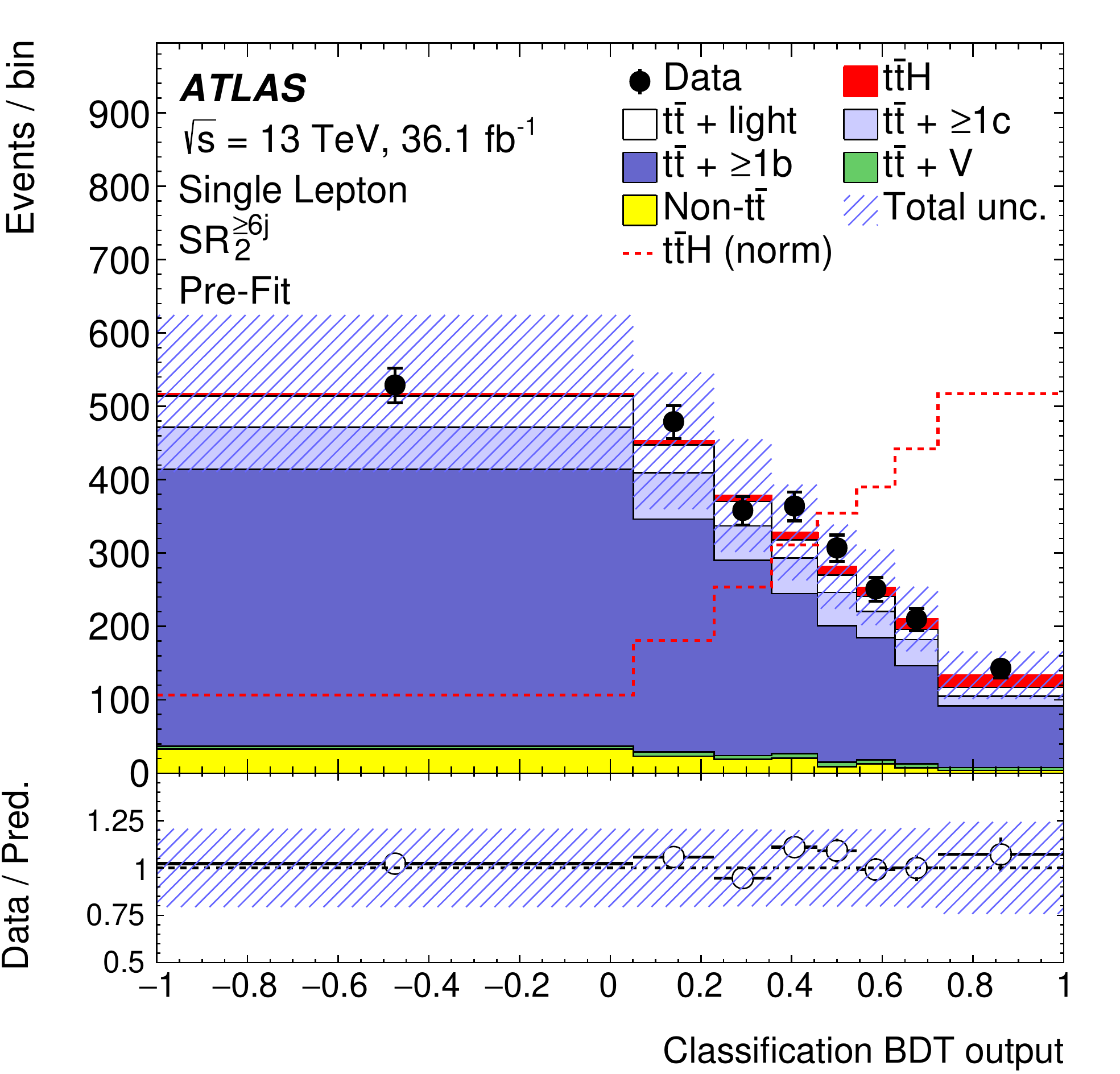}        }\label{fig:SR_sl6j_c}\hspace{1cm}
\subfigure[]{\includegraphics[width=0.42\textwidth]{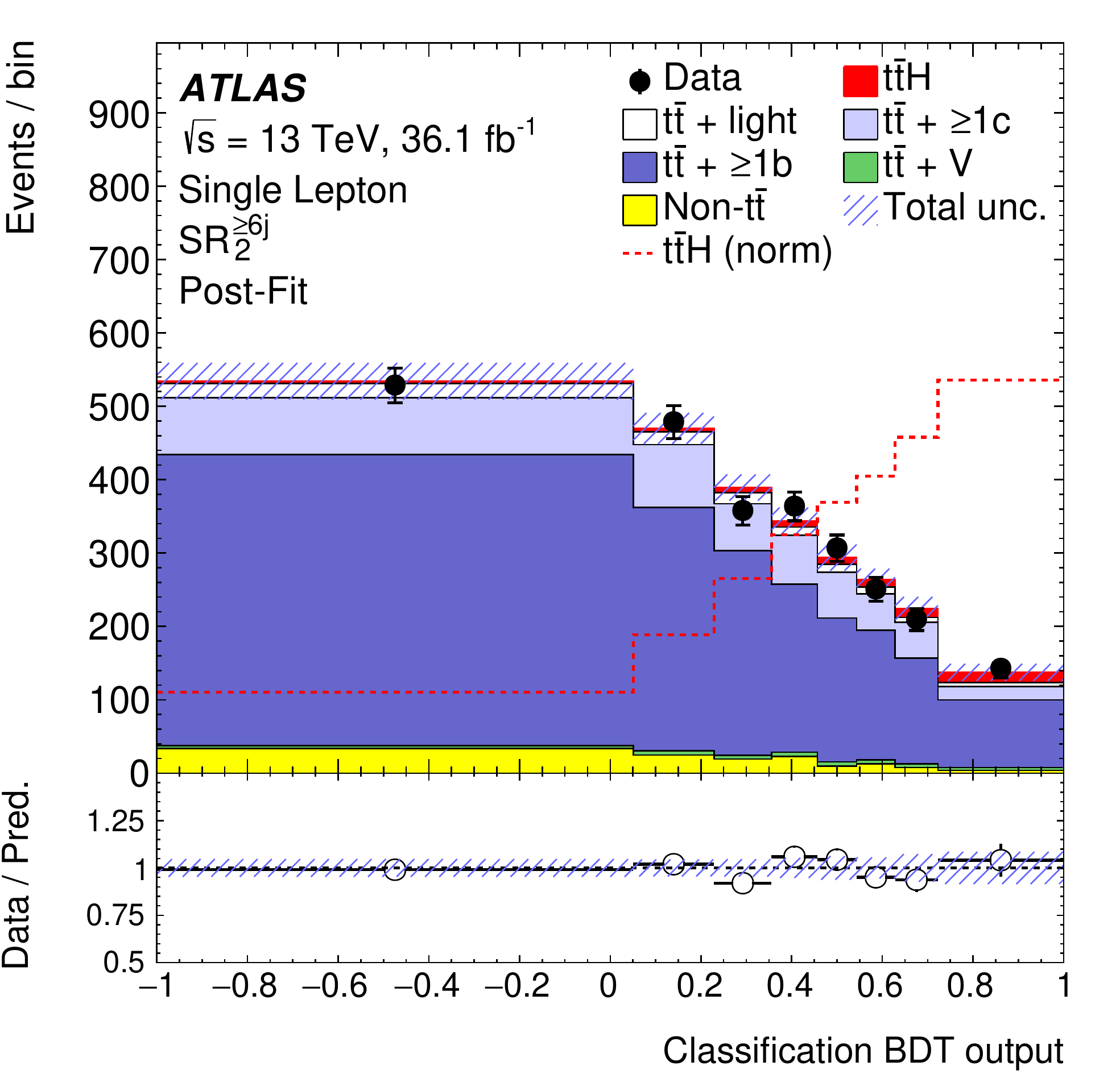}}\label{fig:SR_sl6j_d}\\[-0.3cm]
\subfigure[]{\includegraphics[width=0.42\textwidth]{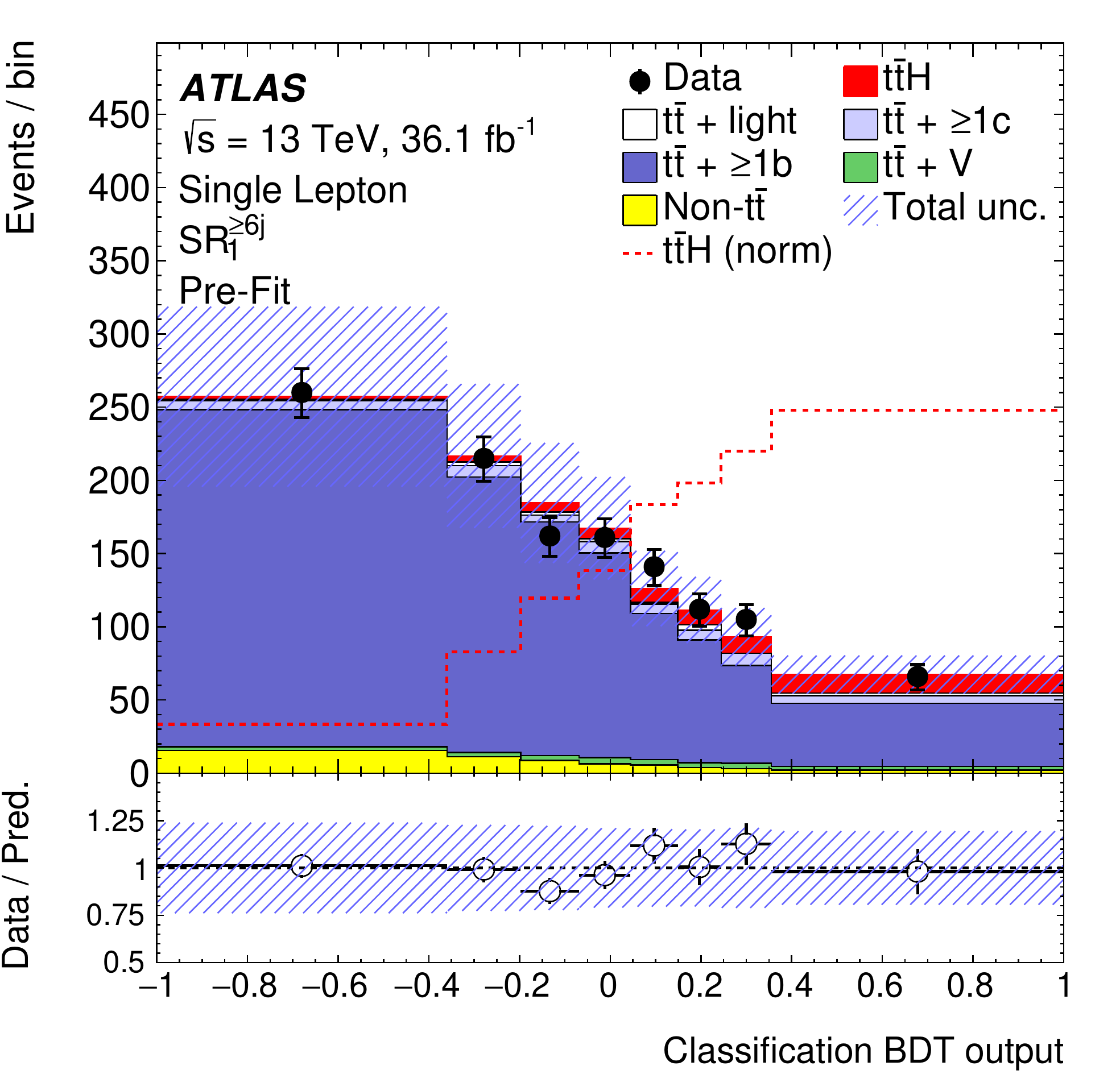}           }\label{fig:SR_sl6j_e}\hspace{1cm}
\subfigure[]{\includegraphics[width=0.42\textwidth]{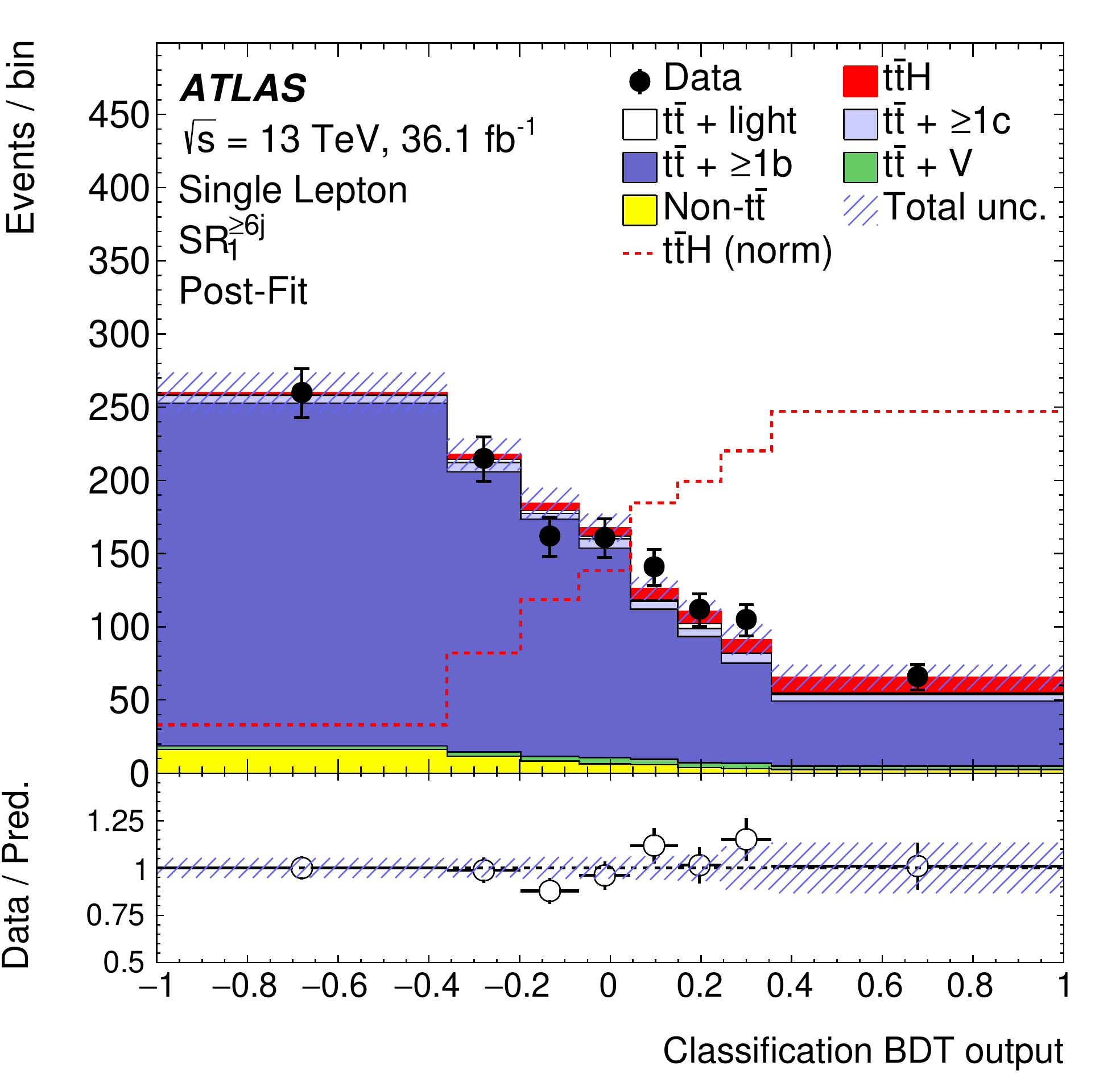}   }\label{fig:SR_sl6j_f}
\vspace{-0.5cm}
\caption{Comparison between data and prediction for the BDT discriminant in the single-lepton channel 
six-jet signal regions 
(a, c, e) before, and (b, d, f) after the combined dilepton and single-lepton fit to the data. 
The \ttH\ signal yield (solid red) is normalized to the SM cross-section before the fit and to the 
fitted $\mu$ after the fit. The dashed line shows the \ttH\ signal distribution normalized to the total 
background prediction.  The pre-fit plots do not include an uncertainty for the \ttbin\ or \ttcin\ normalization.}
\label{fig:SR_sl6j} 
\end{center}
\end{figure}

\begin{figure}[ht!]
\begin{center}
\begin{tabular}{ll}
\subfigure[]{\includegraphics[width=0.42\textwidth]{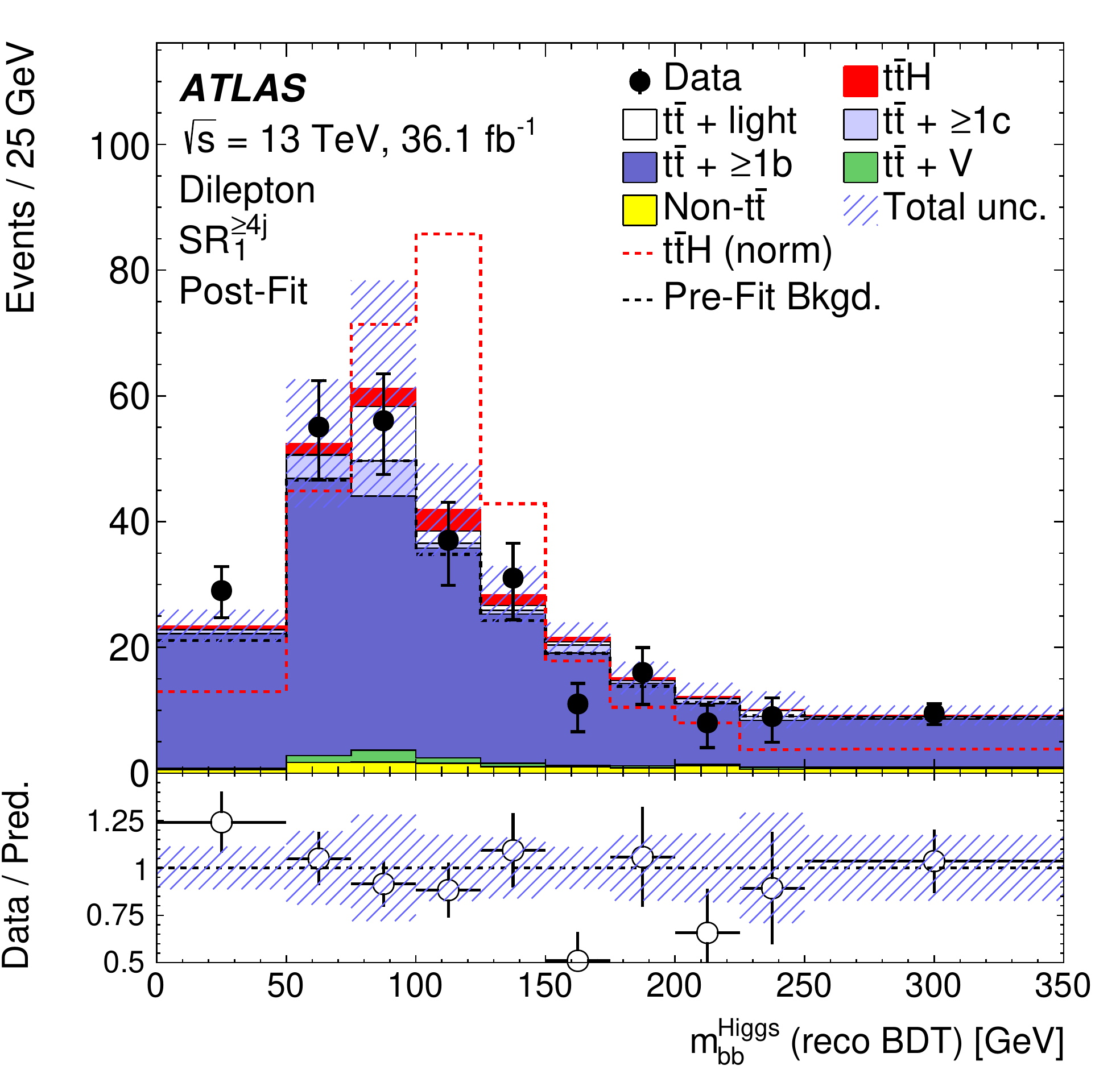}         }\label{fig:mass_a}\hspace{1cm}
&\subfigure[]{\includegraphics[width=0.42\textwidth]{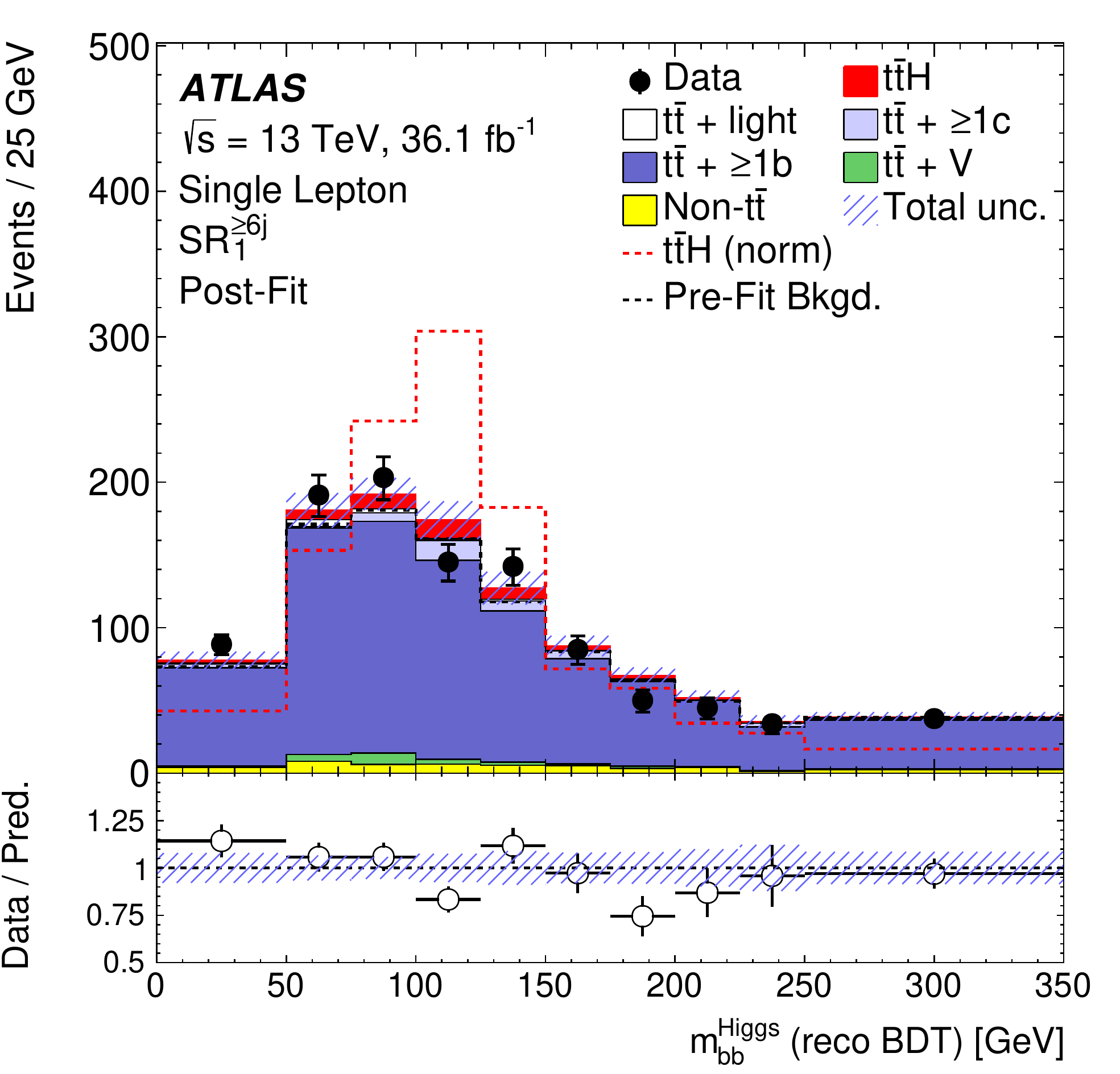}     }\label{fig:mass_b}\\
\subfigure[]{\includegraphics[width=0.42\textwidth]{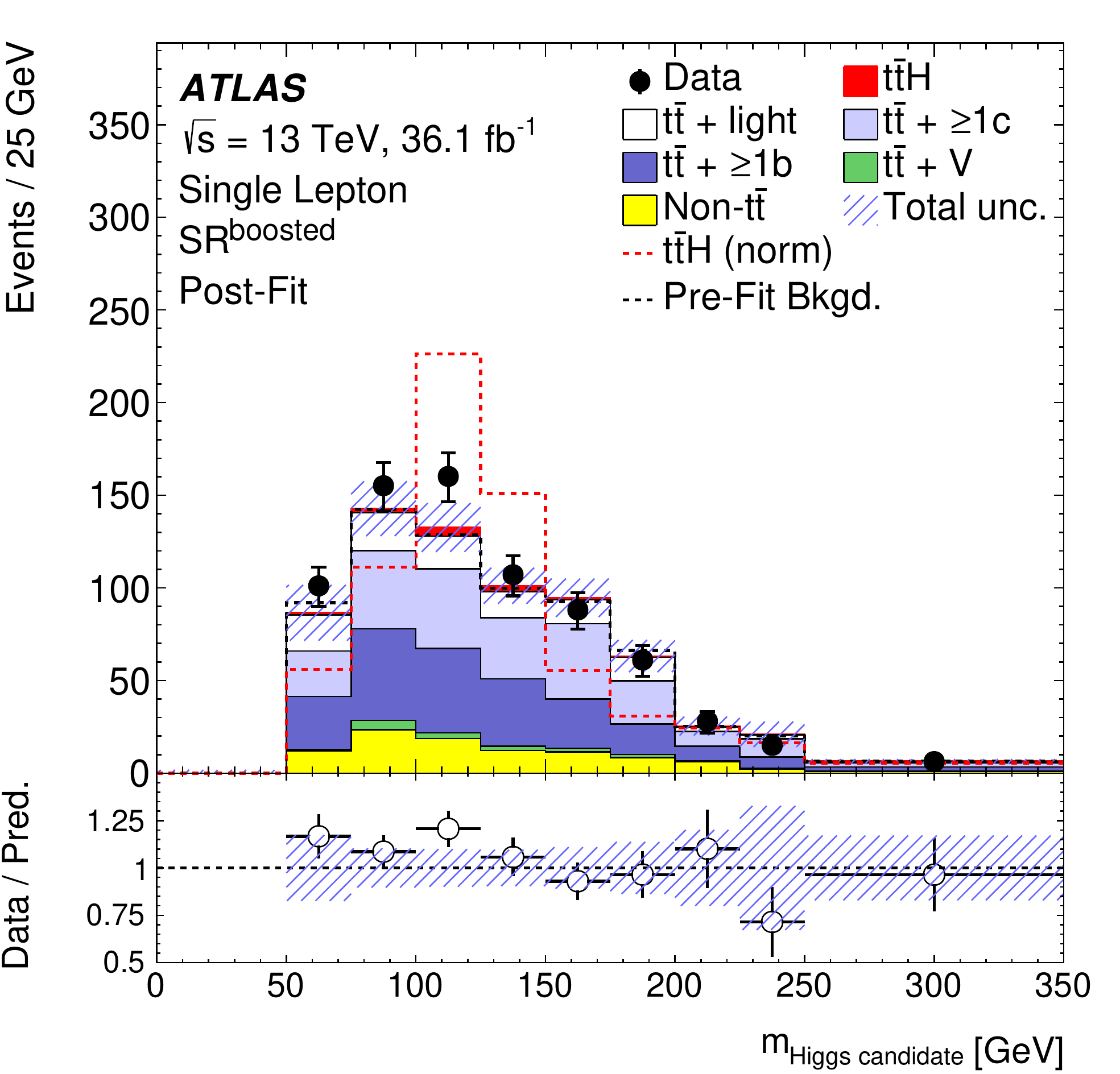}                  }\label{fig:mass_c}\hspace{1cm}
&
\begin{minipage}{0.42\textwidth} 
\vspace{-7cm}
\caption{Comparison between data and prediction for 
the Higgs-boson candidate mass from the reconstruction BDT trained without variables involving the Higgs-boson candidate 
(a) in the dilepton \srfourjone 
and (b) in the single-lepton \srsixjone,
and (c) for the boosted Higgs-boson candidate in \srboosted,
after the combined dilepton and single-lepton fit to the data. 
The \ttH\ signal yield (solid red) is normalized to the 
fitted $\mu$ after the fit. 
The dashed red line shows the \ttH\ signal distribution normalized to the total background yield.
The dashed black line shows the pre-fit total background prediction.
}
\label{fig:mass} 
\end{minipage}
\\
\end{tabular}
\end{center}
\end{figure}

The best-fit $\mu$ value is:
\[
\mu = 0.84 \pm 0.29 \textrm{ (stat.) }^{+0.57}_{-0.54}\textrm{ (syst.)} = 0.84^{+0.64}_{-0.61},
\]
determined by 
the combined fit in all signal and control regions in the two channels. 
The expected uncertainty of the signal strength is identical to the measured one.
An alternative combined fit is also performed in which the dilepton and single-lepton channels are assigned two independent signal strengths.
The corresponding fitted values of $\mu$ are $-0.24^{+1.02}_{-1.05}$ in the dilepton channel and $0.95^{+0.65}_{-0.62}$ in the single-lepton channel. 
The probability of obtaining a discrepancy between these two signal-strength parameters equal to or larger than the one observed is 19\%.
Figure~\ref{fig:muPlot} shows the comparison between 
the combined $\mu$ and the two independent signal-strength parameters from the combined fit, 
with their uncertainties split into the statistical and systematic components. 
The statistical uncertainty is obtained by redoing the fit to data after fixing all the nuisance parameters to their post-fit values,
with the exception of the free normalization factors in the fit: $k(\ttcin)$, $k(\ttbin)$ and $\mu$. 
The total systematic uncertainty is obtained from the subtraction in quadrature of the statistical uncertainty from the total uncertainty. 
The statistical uncertainty contributes significantly less than the systematic component to the overall uncertainty of the measurement. 
When fitting the dilepton and single-lepton data separately, the observed signal strengths are $0.11^{+1.36}_{-1.41}$ and $0.67^{+0.71}_{-0.69}$, respectively. 
These two signal-strength values are both lower than the combined measured $\mu$ due to the 
large 
correlations in the systematic uncertainties of the background prediction between the two channels.

The contributions from the different sources of uncertainty in the combined fit to $\mu$ are reported in Table~\ref{tab:systs}.
The total systematic uncertainty is dominated by the uncertainties in the modeling of the \ttbin\ background, 
the second-largest source being the limited number of events in the simulated samples, 
followed by the uncertainties in the $b$-tagging efficiency, the jet energy scale and resolution, and the signal process modeling. 
The 20 nuisance parameters describing the independent sources of systematic uncertainty with the largest contribution to the total uncertainty of the measured signal strength
are reported in Figure~\ref{fig:ranking}, ranked by 
decreasing contribution. 
For each of these nuisance parameters, the best-fit value and the post-fit uncertainty are shown. 
The uncertainty 
coming from the comparison between the \Sh\ and the nominal prediction for the \ttbin{} process, 
related to the choice of the NLO event generator for this background component, 
has the largest impact on the signal strength, followed by three uncertainties also related to the modeling of the \ttbin{} background. 
Systematic uncertainties related to the \tth{} signal modeling, the modeling of the \ttcin{} and \ttlight{} backgrounds, and to experimental sources such as $b$-tagging, jet energy scale and resolution, also appear in Figure~\ref{fig:ranking}; however, their contributions are significantly smaller than the ones from the \ttbin{} background.
The total uncertainty of the signal strength is reduced by 5\% if the fit is performed excluding the systematic uncertainties not shown in this figure.

\clearpage
\newpage

The theoretical predictions for the \ttbin{} process suffer from large uncertainties as reflected in the size of the difference between alternative simulated samples used to model this background. 
The corresponding systematic uncertainties are therefore large and are a crucial limiting factor for this search.
The choice of nuisance parameters for systematic uncertainties 
related to the \ttbin{} background is studied carefully to ensure sufficient flexibility in the fit to correct for possible mis-modeling of this background and avoid any bias in the measured signal strength.
In total, 13 independent nuisance parameters are assigned to \ttbin{} background modeling uncertainties.
The capability of the fit to correct for mis-modeling effects, beyond the ones present in the distributions used in the fit, is confirmed by comparing the predictions of all input variables of the classification BDT obtained post-fit to data.
As mentioned before, no significant deviations of the predictions from data are found and the agreement is improved post-fit. 
Alternative approaches to model the \ttbin{} background, to define the associated uncertainties and to correlate them are also tested, and the corresponding results are found to be compatible with the nominal result.

To further validate the robustness of the fit, a pseudo-data set was built from simulated events by replacing the nominal \ttbar\ background 
by an alternative sample that is not used in the definition of any uncertainty.
This alternative sample was generated with {\textsc{Powheg}}+\pythia{} and is similar to the sample used for the \tth(\htobb) analysis~\cite{HIGG-2013-27} in Run 1 of the LHC. 
The fit to this pseudo-data sample did not reveal any bias in the signal extraction. 

Figure \ref{fig:ranking} shows that some nuisance parameters are shifted in the fit from their nominal values.
To understand the origin of these shifts, the corresponding nuisance parameters are 
switched to be uncorrelated between 
analysis categories 
and samples and the fit is repeated. 
These shifts are found to correct mainly the predictions of the \ttbar\ background to the observed data in various regions.
Similar shifts are observed when a background-only fit is performed after removing the bins with the most significant signal contributions.
Moreover, the variations induced in the signal strength by these shifts are quantified by fixing the corresponding nuisance parameters to their pre-fit values,
repeating the fit, and comparing the obtained $\mu$-value with the one from the nominal fit.
These variations were found to be smaller than the uncertainty in the signal strength.
Independent signal-strength values extracted from different sets of analysis categories and from the two channels are also found to be compatible.

Figure \ref{fig:ranking} also shows that the uncertainties corresponding to some nuisance parameters are reduced by the fit.
When performing the profile likelihood fit, nuisance parameters associated with uncertainties affecting the discriminant distributions by variations that would result in large deviations from data are significantly constrained.
The capability of the fit to constrain systematic uncertainties is validated on the pseudo-data sample described above, and on the pseudo-data sample produced from the nominal predictions, the Asimov dataset \cite{Cowan:2010js}.

An excess of events over the expected SM background is found with an observed (expected) significance of 1.4 (1.6) standard deviations.
A signal strength larger than 2.0 is excluded at the 95\% confidence level, as shown in Figure~\ref{fig:limitPlot}. 
The expected significance and exclusion limits are calculated using the background estimate after the fit to the data.
Figure~\ref{fig:logSBplot} shows the event yield in data compared to the post-fit prediction for all events entering the analysis selection, 
grouped and ordered by the signal-to-background ratio of the corresponding final-discriminant bins. 
The predictions are shown for both the fit with the background-only hypothesis and with the signal-plus-background hypothesis, 
where the signal is scaled to either the measured $\mu$ or the value of the upper limit on $\mu$.

\begin{figure}[!th]
\begin{center}
\includegraphics[width=0.70\textwidth]{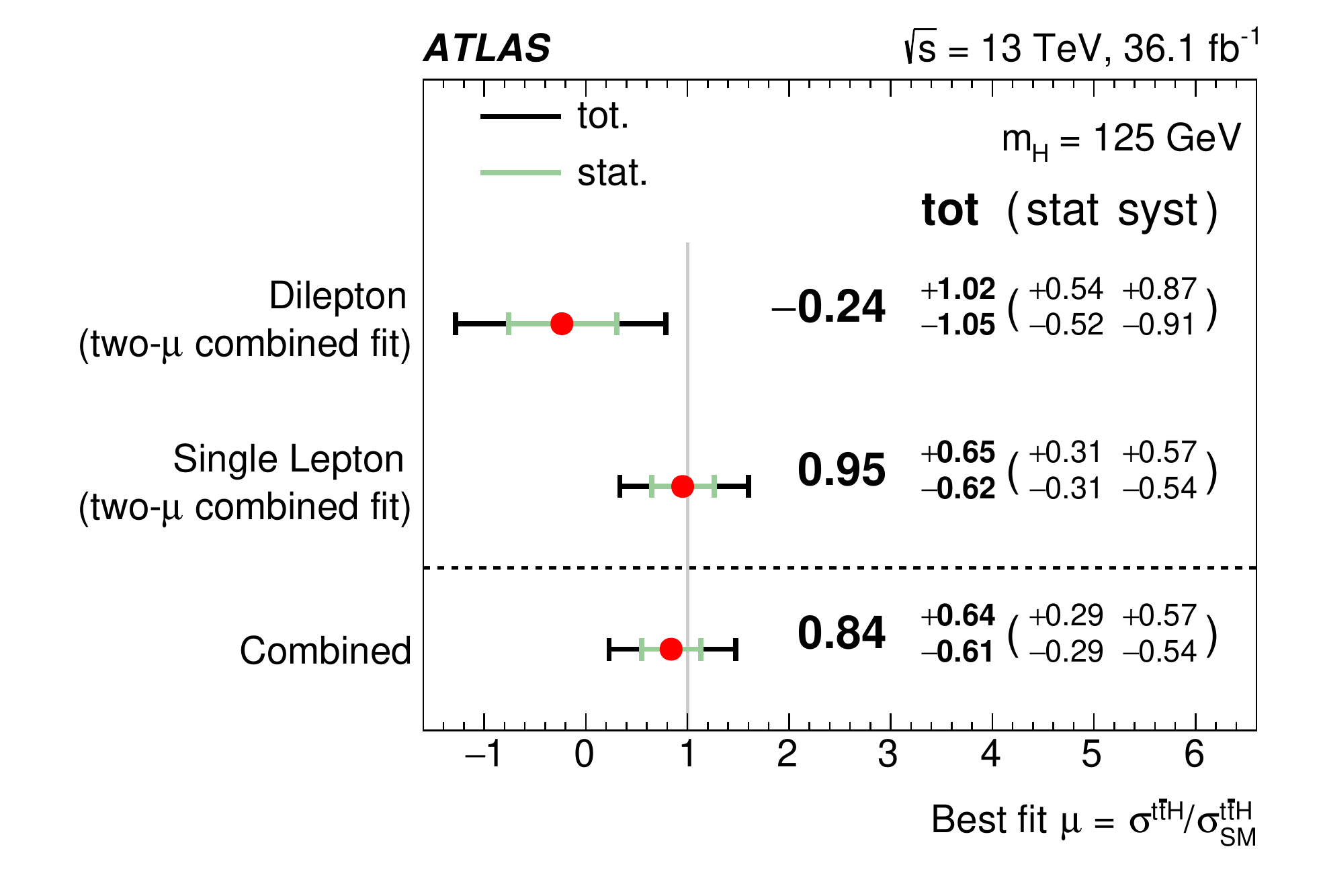} 
\caption{Summary of the signal-strength measurements in the individual channels and for the combination. 
All the numbers are obtained from a simultaneous fit in the two channels, 
but the measurements in the two channels separately are obtained keeping the signal strengths uncorrelated, 
while all the nuisance parameters are kept correlated across channels.}
\label{fig:muPlot} 
\end{center}
\end{figure}

\begin{figure}[!ht]
\begin{center}
\includegraphics[width=0.6\textwidth]{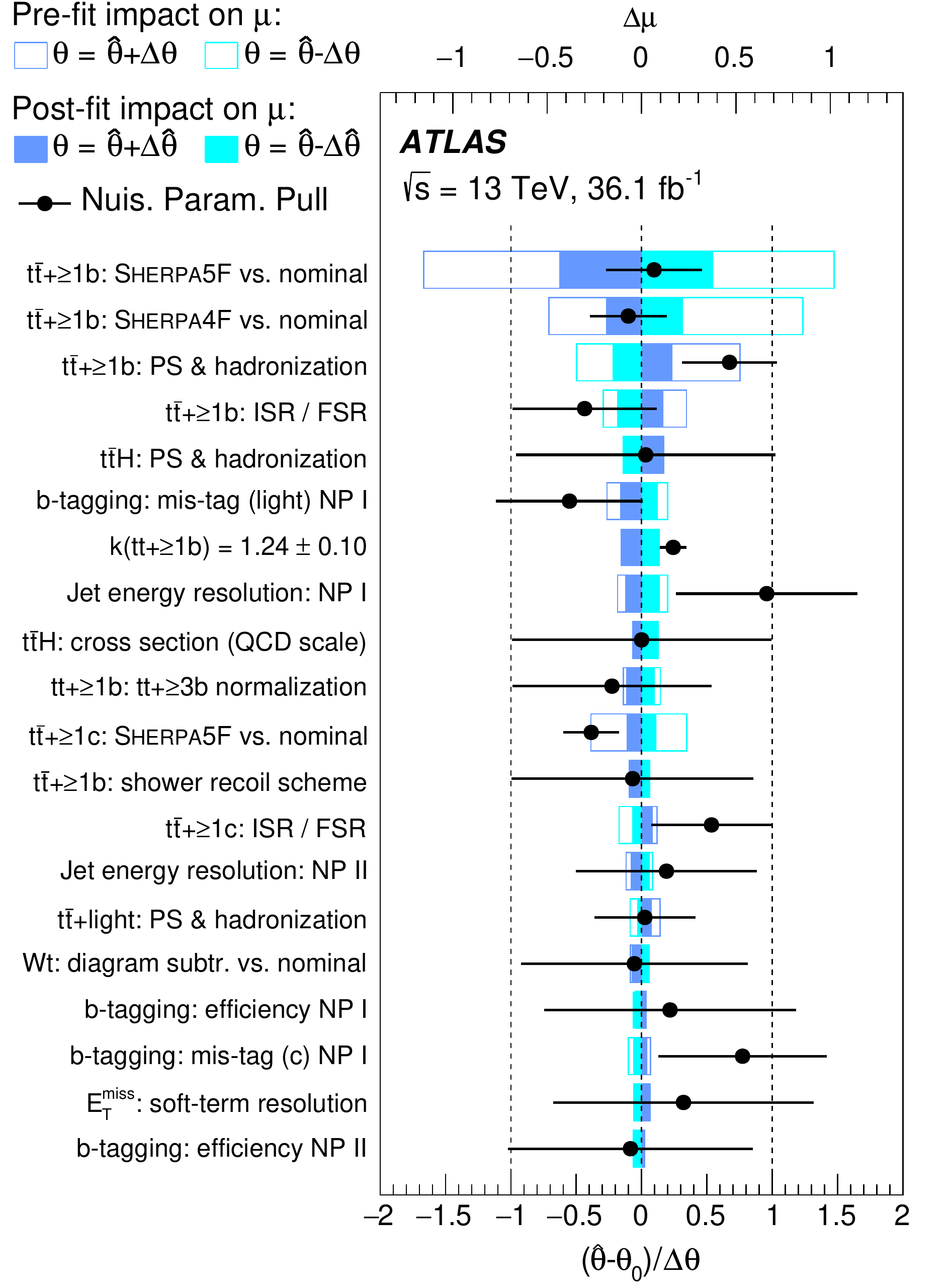}\\
\caption{Ranking of the nuisance parameters included in the fit according to their impact on the measured signal strength $\mu$. 
Only the 
20 most highly ranked parameters are shown.
Nuisance parameters corresponding to MC statistical uncertainties are not included here.
The empty blue rectangles correspond to the pre-fit impact on $\mu$ and the filled blue ones to the post-fit impact on $\mu$, 
both referring to the upper scale. 
The impact of each nuisance parameter, $\Delta\mu$, is computed by comparing the nominal best-fit value of $\mu$ 
with the result of the fit when fixing the considered nuisance parameter to its best-fit value, $\hat{\theta}$, 
shifted by its pre-fit (post-fit) uncertainties $\pm\Delta\theta$ ($\pm\Delta\hat{\theta}$). 
The black points show the pulls of the nuisance parameters relative to their nominal values, $\theta_0$.
These pulls and their relative post-fit errors, $\Delta\hat{\theta}/\Delta\theta$, 
refer to the scale on the bottom axis.
The parameter $k(\ttbin)$ 
refers to the floating normalization of the $\ttbin$ background, 
for which the pre-fit impact on $\mu$ is not defined, 
and for which both $\theta_0$ and $\Delta\theta$ are set to 1. 
For experimental uncertainties that are 
decomposed 
into several independent sources, 
NP I and NP II correspond to the first and second 
nuisance parameters, 
ordered by their impact on $\mu$, 
respectively.
}
\label{fig:ranking} 
\end{center}
\end{figure}

\begin{table}[!ht] 
\caption{\label{tab:systs} 
Breakdown of the contributions to the uncertainties in $\mu$.
The line `background-model statistical uncertainty' refers to the statistical uncertainties 
in the MC events and in the data-driven determination of the non-prompt and fake lepton background component in the single-lepton channel. 
The contribution of the different sources of uncertainty is evaluated after the fit described in Section~\ref{sec:results}. 
The total statistical uncertainty is evaluated, as described in the text, by fixing all the nuisance parameters in the fit 
except for the free-floating normalization factors for the \ttbin\ and \ttcin\ background components. 
The contribution from the uncertainty in the normalization of both \ttbin\ and \ttcin\ is then included in the quoted total statistical uncertainty rather than in the systematic uncertainty component. 
The statistical uncertainty evaluated after also fixing the normalization of \ttbin\ and \ttcin\ is then indicated as
`intrinsic statistical uncertainty'. 
The other quoted numbers are obtained by repeating the fit after having fixed a certain set of nuisance parameters 
corresponding to a group of systematic uncertainty sources, 
and subtracting in quadrature the resulting total uncertainty of $\mu$ from the uncertainty from the full fit. 
The same procedure is followed for quoting the individual effects of the \ttbin\ and the \ttcin\ normalization. 
The total uncertainty is different from the sum in quadrature of the different components due to 
correlations between nuisance parameters built by the fit.
}
\centering 
\begin{tabular}{lcc} 
\toprule
\hline 
Uncertainty source                   &   \multicolumn{2}{c}{$\Delta\mu$} \\ 
\hline 
\hline 
\ttbin\ modeling     & ${+0.46}$ & ${-0.46}$ \\ 
Background-model statistical uncertainty      & ${+0.29}$ & ${-0.31}$ \\ 
$b$-tagging efficiency and mis-tag rates  & ${+0.16}$ & ${-0.16}$ \\ 
Jet energy scale and resolution  & ${+0.14}$ & ${-0.14}$ \\ 
\ttH\ modeling            & ${+0.22}$ & ${-0.05}$ \\ 
\ttcin\ modeling     & ${+0.09}$ & ${-0.11}$ \\ 
JVT, pileup modeling & ${+0.03}$ & ${-0.05}$ \\ 
Other background modeling       & ${+0.08}$ & ${-0.08}$ \\ 
\ttlight\ modeling       & ${+0.06}$ & ${-0.03}$ \\ 
Luminosity                       & ${+0.03}$ & ${-0.02}$ \\ 
Light lepton ($e, \mu$) id., isolation, trigger & ${+0.03}$ & ${-0.04}$ \\ 
\hline 
Total systematic uncertainty     & ${+0.57}$ & ${-0.54}$ \\  
\hline \hline 
\ttbin\ normalization & ${+0.09}$ & ${-0.10}$ \\ 
\ttcin\ normalization & ${+0.02}$ & ${-0.03}$ \\ 
Intrinsic statistical uncertainty & ${+0.21}$ & ${-0.20}$ \\ 
\hline 
Total statistical uncertainty          & ${+0.29}$ & ${ -0.29}$ \\ 
\hline \hline
Total uncertainty                & ${+0.64 }$ & ${-0.61}$ \\  
\hline
\bottomrule
\end{tabular} 
\end{table}

\begin{figure}[!ht]
\begin{center}
\includegraphics[width=0.70\textwidth]{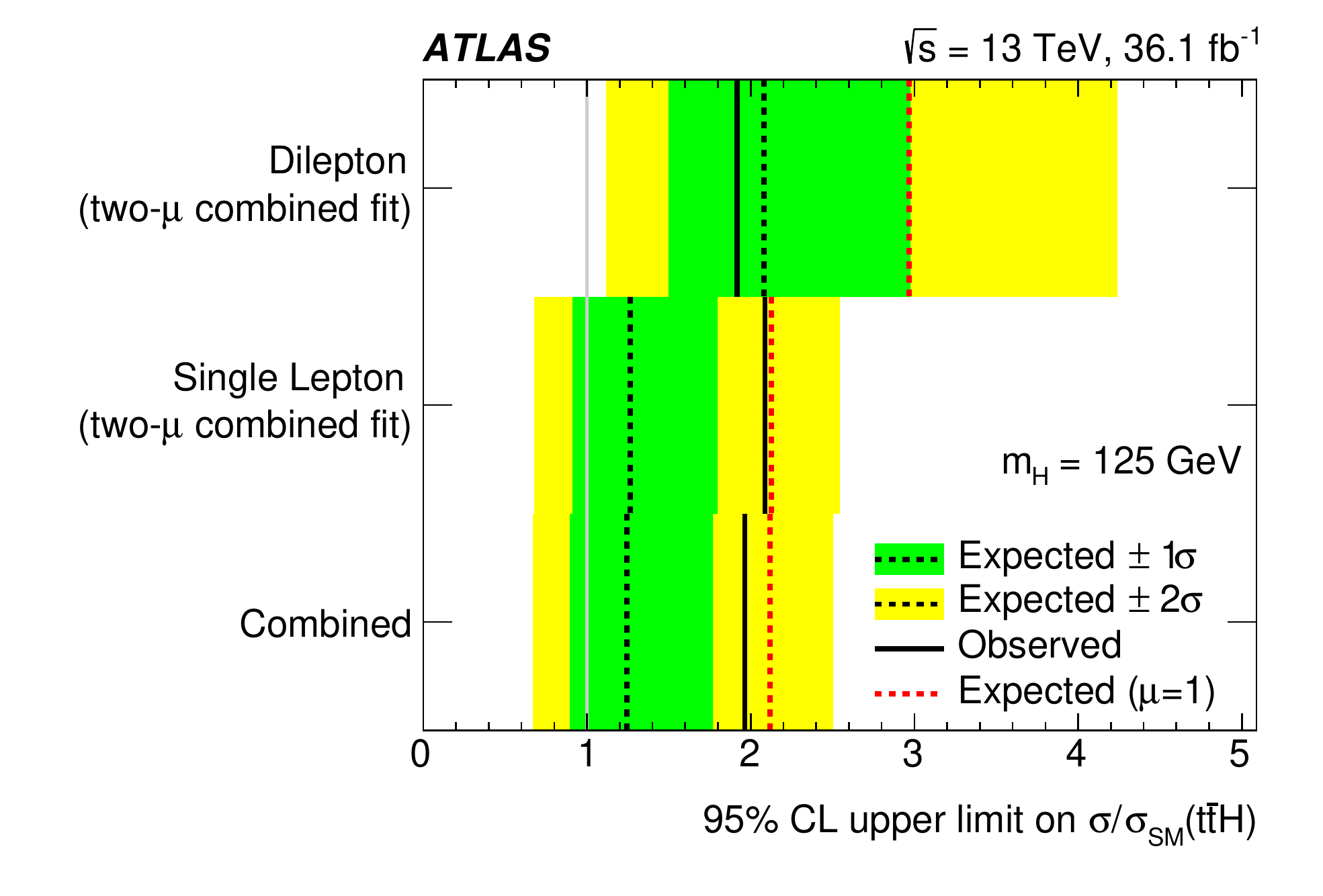}
\caption{Summary of the 95\% confidence level (CL) upper limits on $\sigma(\ttH)$ relative to the SM prediction
in the individual channels and for the combination. 
The observed limits are shown, together with the expected limits both in the background-only hypothesis (dotted black lines) 
and in the SM hypothesis (dotted red lines). 
In the case of the expected limits in the background-only hypothesis, 
one- and two-standard-deviation uncertainty bands are also shown. 
The limits for the two individual channels are derived consistently with Figure~\ref{fig:muPlot}, 
both extracted from the profile likelihood including the data in both channels, 
but with independent signal strengths in the two channels.
}
\label{fig:limitPlot} 
\end{center}
\end{figure}

\begin{figure}[!th]
\begin{center}
\includegraphics[width=0.5\textwidth]{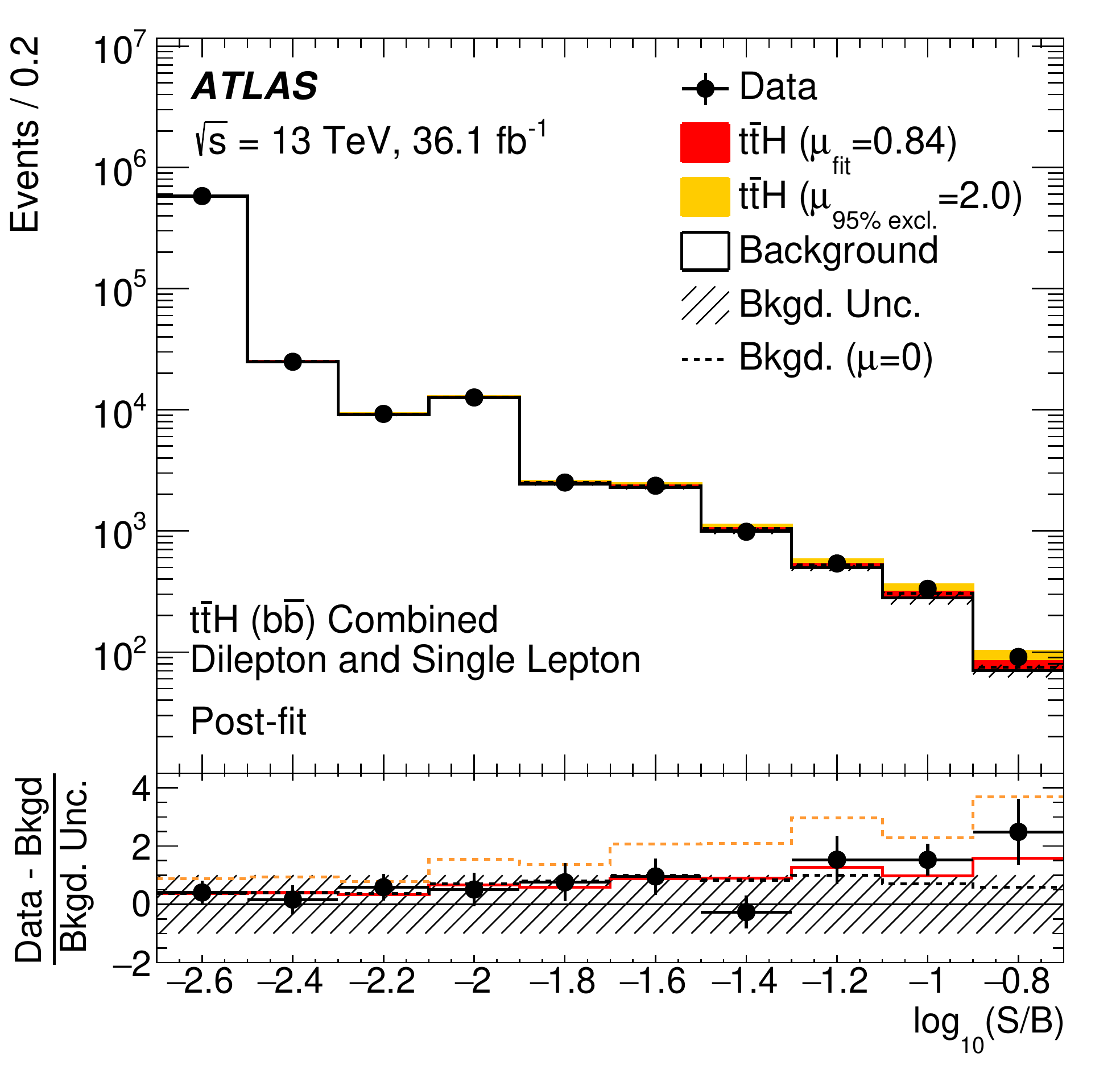}
\caption{Post-fit yields of signal ($S$) and total background ($B$) 
as a function of $\log(S/B)$, compared to data. 
Final-discriminant bins in all dilepton and single-lepton 
analysis categories 
are combined into bins of $\log(S/B)$, 
with the signal normalized to the SM prediction used for the computation of $\log(S/B)$. 
The signal is then shown normalized to the best-fit value and to the value excluded at the 95\% CL, 
in both cases summed to the background prediction from the fit. 
The lower frame reports for each bin the pull (residual divided by its uncertainty) of the data 
relative to the background prediction from the fit. 
These data pulls are compared to the pulls of the signal-plus-background prediction from the fit, 
assuming a signal strength equal to the best-fit value (solid red line) 
and equal to the exclusion limit (dashed orange line). 
The background and its pull are also shown after the fit to data assuming zero signal contribution 
(dashed black line, obscured by solid line in the upper frame). 
The first bin includes the underflow.
}
\label{fig:logSBplot} 
\end{center}
\end{figure}

\FloatBarrier

\section{Conclusion}
\label{sec:conclusion}
A search for the associated production of the Standard Model Higgs boson with a pair of top quarks is presented, based on 36.1~\ifb\ of $pp$ collision data at $\sqrt{s}=13~\tev$, collected with the ATLAS detector at the Large Hadron Collider in 2015 and 2016. The search focuses on decays of the Higgs boson to $b\bar{b}$ and decays of the top quark pair to a final state containing one or two leptons.
Multivariate techniques are used to discriminate between signal and background events, the latter being dominated by $t\bar{t}$ + jets production. 
The observed data are consistent with both the background-only hypothesis and with the Standard Model $\ttH$ prediction. 
A 1.4 $\sigma$ excess above the expected background is observed, while an excess of 1.6 $\sigma$ is expected in the presence of a Standard Model Higgs boson.
The signal strength is measured to be $0.84^{+0.64}_{-0.61}$, 
consistent 
with the expectation from the Standard Model.
A value higher than 2.0 is excluded at the 95\% confidence level, compared to an expected exclusion limit of 1.2 in the absence of signal. 
The measurement uncertainty is presently dominated by systematic uncertainties, and more specifically by the uncertainty in the theoretical knowledge of the \ttbin\ production process. 
An improved understanding of this background will be important for future efforts to observe the \tth(\htobb) process.

\clearpage
\appendix
\part*{Appendix}
\addcontentsline{toc}{section}{Appendix}

\section{Yield tables}
\label{app:yields}
The predicted event yields in each of the analysis categories, 
broken down into the different signal and background contributions 
and compared to the observed yields in data, 
are reported in Tables~\ref{tab:DLyields}, \ref{tab:SLyields5j} and \ref{tab:SLyields6j}. 
Both the pre-fit and post-fit predictions are shown, 
where post-fit refers to the combined fit to the dilepton and single-lepton channels 
with the signal-plus-background hypothesis, reported in Section~\ref{sec:results}. 
The total uncertainties of each of the signal and background components, and of the total prediction are also reported.

\begin{table}[htbp]
  \caption{Event yields in the dilepton channel (top) control regions and (bottom) signal regions. 
  Post-fit yields are after the combined fit in all channels to data.
  The uncertainties are the sum in quadrature of statistical and systematic uncertainties in the yields. 
  In the post-fit case, these uncertainties are computed taking into account correlations among nuisance parameters and among the normalization of different processes.
  The uncertainty in the \ttbin{} and \ttcin{} normalization is not defined pre-fit and therefore only included in the post-fit uncertainties;
  the reported prefit uncertainties on the \ttbin{} and \ttcin{} components arise only from acceptance effects.
  For the \ttH{} signal, the pre-fit yield values correspond to the theoretical prediction and corresponding uncertainties, while the post-fit yield and uncertainties correspond to those in the signal-strength measurement.
  }
  \centering
  \begin{footnotesize}{
\setlength\tabcolsep{2.2pt}
  \begin{tabular}{l | r@{$~\pm~$}l r@{$~\pm~$}l | r@{$~\pm~$}l r@{$~\pm~$}l | r@{$~\pm~$}l  r@{$~\pm~$}l | r@{$~\pm~$}l  r@{$~\pm~$}l }
    \toprule
    \hline
  \multirow{2}{*}{Sample}   & \multicolumn{4}{c|}{\crthreejttlight\rule{0pt}{3ex}}  &  \multicolumn{4}{c|}{\crthreejttb} &  \multicolumn{4}{c|}{\crfourjttlight} &  \multicolumn{4}{c}{\crfourjttc}   \\
                           & \multicolumn{2}{c}{Pre-fit} & \multicolumn{2}{c|}{Post-fit} & \multicolumn{2}{c}{Pre-fit} & \multicolumn{2}{c|}{Post-fit} & \multicolumn{2}{c}{Pre-fit} & \multicolumn{2}{c|}{Post-fit} & \multicolumn{2}{c}{Pre-fit} & \multicolumn{2}{c}{Post-fit}\\
    \hline
    \hline
 $t\bar{t}H$ &                 	\num[group-minimum-digits=5,round-mode=figures,round-precision=3]{32.1653} & \num[group-minimum-digits=5,round-mode=figures,round-precision=2]{3.78742} & 	\num[group-minimum-digits=5,round-mode=figures,round-precision=2]{27.0428} & \num[group-minimum-digits=5,round-mode=figures,round-precision=2]{19.956} & 	\num[group-minimum-digits=5,round-mode=figures,round-precision=2]{8.65542} & \num[group-minimum-digits=5,round-mode=figures,round-precision=2]{1.05945} & 	\num[group-minimum-digits=5,round-mode=figures,round-precision=2]{7.3006} & \num[group-minimum-digits=5,round-mode=figures,round-precision=2]{5.38054} & 	\num[group-minimum-digits=5,round-mode=figures,round-precision=3]{113.992} & \num[group-minimum-digits=5,round-mode=figures,round-precision=2]{11.1027} & 	\num[group-minimum-digits=5,round-mode=figures,round-precision=2]{95.2721} & \num[group-minimum-digits=5,round-mode=figures,round-precision=2]{70.1423} & 	\num[group-minimum-digits=5,round-mode=figures,round-precision=3]{35.2986} & \num[group-minimum-digits=5,round-mode=figures,round-precision=2]{3.55405} & 	\num[group-minimum-digits=5,round-mode=figures,round-precision=2]{29.3491} & \num[group-minimum-digits=5,round-mode=figures,round-precision=2]{21.7521} \\
 $t\bar{t}$ + light     &     	\num[group-minimum-digits=5,round-mode=figures,round-precision=3]{63053.3} & \num[group-minimum-digits=5,round-mode=figures,round-precision=2]{5511.24} & 	\num[group-minimum-digits=5,round-mode=figures,round-precision=3]{59056.6} & \num[group-minimum-digits=5,round-mode=figures,round-precision=2]{1353.84} & 	\num[group-minimum-digits=5,round-mode=figures,round-precision=2]{291.621} & \num[group-minimum-digits=5,round-mode=figures,round-precision=2]{107.569} & 	\num[group-minimum-digits=5,round-mode=figures,round-precision=3]{255.066} & \num[group-minimum-digits=5,round-mode=figures,round-precision=2]{44.1202} & 	\num[group-minimum-digits=5,round-mode=figures,round-precision=3]{42457.8} & \num[group-minimum-digits=5,round-mode=figures,round-precision=2]{9703.66} & 	\num[group-minimum-digits=5,round-mode=figures,round-precision=3]{37086.4} & \num[group-minimum-digits=5,round-mode=figures,round-precision=2]{1309.76} & 	\num[group-minimum-digits=5,round-mode=figures,round-precision=3]{1726.95} & \num[group-minimum-digits=5,round-mode=figures,round-precision=2]{726.606} & 	\num[group-minimum-digits=5,round-mode=figures,round-precision=3]{1411.05} & \num[group-minimum-digits=5,round-mode=figures,round-precision=2]{178.863} \\
 $t\bar{t}$ + $\geq$1$c$ &    	\num[group-minimum-digits=5,round-mode=figures,round-precision=2]{4773.73} & \num[group-minimum-digits=5,round-mode=figures,round-precision=2]{2124.34} & 	\num[group-minimum-digits=5,round-mode=figures,round-precision=2]{7702.63} & \num[group-minimum-digits=5,round-mode=figures,round-precision=2]{1102.83} & 	\num[group-minimum-digits=5,round-mode=figures,round-precision=2]{363.909} & \num[group-minimum-digits=5,round-mode=figures,round-precision=2]{155.822} & 	\num[group-minimum-digits=5,round-mode=figures,round-precision=3]{536.18} & \num[group-minimum-digits=5,round-mode=figures,round-precision=2]{89.4054} & 	\num[group-minimum-digits=5,round-mode=figures,round-precision=2]{6314.84} & \num[group-minimum-digits=5,round-mode=figures,round-precision=2]{2803.94} & 	\num[group-minimum-digits=5,round-mode=figures,round-precision=3]{10348.8} & \num[group-minimum-digits=5,round-mode=figures,round-precision=2]{1379.56} & 	\num[group-minimum-digits=5,round-mode=figures,round-precision=3]{1405.84} & \num[group-minimum-digits=5,round-mode=figures,round-precision=2]{585.636} & 	\num[group-minimum-digits=5,round-mode=figures,round-precision=3]{2162.1} & \num[group-minimum-digits=5,round-mode=figures,round-precision=2]{294.714} \\
 $t\bar{t}$ + $\geq$1$b$ &    	\num[group-minimum-digits=5,round-mode=figures,round-precision=3]{2129.69} & \num[group-minimum-digits=5,round-mode=figures,round-precision=2]{227.502} & 	\num[group-minimum-digits=5,round-mode=figures,round-precision=3]{2623.58} & \num[group-minimum-digits=5,round-mode=figures,round-precision=2]{241.306} & 	\num[group-minimum-digits=5,round-mode=figures,round-precision=2]{714.448} & \num[group-minimum-digits=5,round-mode=figures,round-precision=2]{142.156} & 	\num[group-minimum-digits=5,round-mode=figures,round-precision=3]{847.829} & \num[group-minimum-digits=5,round-mode=figures,round-precision=2]{74.6965} & 	\num[group-minimum-digits=5,round-mode=figures,round-precision=3]{2509.6} & \num[group-minimum-digits=5,round-mode=figures,round-precision=2]{275.736} & 	\num[group-minimum-digits=5,round-mode=figures,round-precision=3]{2853.4} & \num[group-minimum-digits=5,round-mode=figures,round-precision=2]{293.427} & 	\num[group-minimum-digits=5,round-mode=figures,round-precision=3]{1080.84} & \num[group-minimum-digits=5,round-mode=figures,round-precision=2]{116.724} & 	\num[group-minimum-digits=5,round-mode=figures,round-precision=3]{1241.89} & \num[group-minimum-digits=5,round-mode=figures,round-precision=2]{106.821} \\
 $t\bar{t}$ + $V$            &	\num[group-minimum-digits=5,round-mode=figures,round-precision=3]{113.258} & \num[group-minimum-digits=5,round-mode=figures,round-precision=2]{30.5066} & 	\num[group-minimum-digits=5,round-mode=figures,round-precision=3]{112.251} & \num[group-minimum-digits=5,round-mode=figures,round-precision=2]{29.2398} & 	\num[group-minimum-digits=5,round-mode=figures,round-precision=1]{6.7602} & \num[group-minimum-digits=5,round-mode=figures,round-precision=2]{26.9027} & 	\num[group-minimum-digits=5,round-mode=figures,round-precision=1]{7.34613} & \num[group-minimum-digits=5,round-mode=figures,round-precision=2]{29.7848} & 	\num[group-minimum-digits=5,round-mode=figures,round-precision=2]{345.722} & \num[group-minimum-digits=5,round-mode=figures,round-precision=2]{182.942} & 	\num[group-minimum-digits=5,round-mode=figures,round-precision=2]{333.386} & \num[group-minimum-digits=5,round-mode=figures,round-precision=2]{172.124} & 	\num[group-minimum-digits=5,round-mode=figures,round-precision=2]{52.0463} & \num[group-minimum-digits=5,round-mode=figures,round-precision=2]{41.3655} & 	\num[group-minimum-digits=5,round-mode=figures,round-precision=2]{49.9355} & \num[group-minimum-digits=5,round-mode=figures,round-precision=2]{38.6015} \\
 Non-$t\bar{t}$              &	\num[group-minimum-digits=5,round-mode=figures,round-precision=2]{6260.11} & \num[group-minimum-digits=5,round-mode=figures,round-precision=2]{1457.2} & 	\num[group-minimum-digits=5,round-mode=figures,round-precision=2]{6490.27} & \num[group-minimum-digits=5,round-mode=figures,round-precision=2]{1181.17} & 	\num[group-minimum-digits=5,round-mode=figures,round-precision=3]{109.899} & \num[group-minimum-digits=5,round-mode=figures,round-precision=2]{29.0272} & 	\num[group-minimum-digits=5,round-mode=figures,round-precision=3]{111.56} & \num[group-minimum-digits=5,round-mode=figures,round-precision=2]{22.5845} & 	\num[group-minimum-digits=5,round-mode=figures,round-precision=2]{4701.91} & \num[group-minimum-digits=5,round-mode=figures,round-precision=2]{1099.21} & 	\num[group-minimum-digits=5,round-mode=figures,round-precision=3]{4929.13} & \num[group-minimum-digits=5,round-mode=figures,round-precision=2]{906.408} & 	\num[group-minimum-digits=5,round-mode=figures,round-precision=2]{415.674} & \num[group-minimum-digits=5,round-mode=figures,round-precision=2]{116.386} & 	\num[group-minimum-digits=5,round-mode=figures,round-precision=2]{458.928} & \num[group-minimum-digits=5,round-mode=figures,round-precision=2]{100.392} \\
\hline 								
 Total                       &	\num[group-minimum-digits=5,round-mode=figures,round-precision=3]{76362.3} & \num[group-minimum-digits=5,round-mode=figures,round-precision=2]{6544.02} & 	\num[group-minimum-digits=5,round-mode=figures,round-precision=4]{76012.4} & \num[group-minimum-digits=5,round-mode=figures,round-precision=2]{387.835} & 	\num[group-minimum-digits=5,round-mode=figures,round-precision=3]{1495.29} & \num[group-minimum-digits=5,round-mode=figures,round-precision=2]{258.701} & 	\num[group-minimum-digits=5,round-mode=figures,round-precision=4]{1765.28} & \num[group-minimum-digits=5,round-mode=figures,round-precision=2]{59.7715} & 	\num[group-minimum-digits=5,round-mode=figures,round-precision=2]{56443.8} & \num[group-minimum-digits=5,round-mode=figures,round-precision=2]{10751.4} & 	\num[group-minimum-digits=5,round-mode=figures,round-precision=4]{55646.4} & \num[group-minimum-digits=5,round-mode=figures,round-precision=2]{423.059} & 	\num[group-minimum-digits=5,round-mode=figures,round-precision=2]{4716.65} & \num[group-minimum-digits=5,round-mode=figures,round-precision=2]{1063.15} & 	\num[group-minimum-digits=5,round-mode=figures,round-precision=3]{5353.25} & \num[group-minimum-digits=5,round-mode=figures,round-precision=2]{117.821} \\
  \hline
  Data                     & \multicolumn{4}{c|}{\num[group-minimum-digits=5]{76025}} & \multicolumn{4}{c|}{\num[group-minimum-digits=5]{1744}} & \multicolumn{4}{c|}{\num[group-minimum-digits=5]{55627}} & \multicolumn{4}{c}{\num[group-minimum-digits=5]{5389}} \\ 
  \hline
  \bottomrule
  \end{tabular}
  
  \vspace{5pt}
  
\setlength\tabcolsep{6pt}
  \begin{tabular}{l | r@{$~\pm~$}l r@{$~\pm~$}l | r@{$~\pm~$}l r@{$~\pm~$}l | r@{$~\pm~$}l  r@{$~\pm~$}l }
    \toprule
    \hline
  \multirow{2}{*}{Sample}   & \multicolumn{4}{c|}{\srfourjthree\rule{0pt}{3ex}}  &  \multicolumn{4}{c|}{\srfourjtwo} &  \multicolumn{4}{c}{\srfourjone}   \\
                            & \multicolumn{2}{c}{Pre-fit} & \multicolumn{2}{c|}{Post-fit} & \multicolumn{2}{c}{Pre-fit} & \multicolumn{2}{c|}{Post-fit} & \multicolumn{2}{c}{Pre-fit} & \multicolumn{2}{c}{Post-fit}\\
    \hline
    \hline
 $t\bar{t}H$ &                 	\num[group-minimum-digits=5,round-mode=figures,round-precision=3]{21.9074} & \num[group-minimum-digits=5,round-mode=figures,round-precision=2]{2.48392} & 	\num[group-minimum-digits=5,round-mode=figures,round-precision=2]{18.035} & \num[group-minimum-digits=5,round-mode=figures,round-precision=2]{13.4112} & 	\num[group-minimum-digits=5,round-mode=figures,round-precision=3]{29.1457} & \num[group-minimum-digits=5,round-mode=figures,round-precision=2]{4.23487} & 	\num[group-minimum-digits=5,round-mode=figures,round-precision=2]{24.5909} & \num[group-minimum-digits=5,round-mode=figures,round-precision=2]{17.9009} & 	\num[group-minimum-digits=5,round-mode=figures,round-precision=3]{15.5833} & \num[group-minimum-digits=5,round-mode=figures,round-precision=2]{2.50607} & 	\num[group-minimum-digits=5,round-mode=figures,round-precision=3]{12.8693} & \num[group-minimum-digits=5,round-mode=figures,round-precision=2]{9.52298} \\
 $t\bar{t}$ + light     &     	\num[group-minimum-digits=5,round-mode=figures,round-precision=2]{83.3657} & \num[group-minimum-digits=5,round-mode=figures,round-precision=2]{41.404} & 	\num[group-minimum-digits=5,round-mode=figures,round-precision=2]{95.1591} & \num[group-minimum-digits=5,round-mode=figures,round-precision=2]{30.0879} & 	\num[group-minimum-digits=5,round-mode=figures,round-precision=2]{247.528} & \num[group-minimum-digits=5,round-mode=figures,round-precision=2]{114.679} & 	\num[group-minimum-digits=5,round-mode=figures,round-precision=3]{215.117} & \num[group-minimum-digits=5,round-mode=figures,round-precision=2]{42.5693} & 	\num[group-minimum-digits=5,round-mode=figures,round-precision=2]{6.41323} & \num[group-minimum-digits=5,round-mode=figures,round-precision=2]{9.94919} & 	\num[group-minimum-digits=5,round-mode=figures,round-precision=3]{11.1242} & \num[group-minimum-digits=5,round-mode=figures,round-precision=2]{9.25468} \\
 $t\bar{t}$ + $\geq$1$c$ &    	\num[group-minimum-digits=5,round-mode=figures,round-precision=3]{235.14} & \num[group-minimum-digits=5,round-mode=figures,round-precision=2]{60.8425} & 	\num[group-minimum-digits=5,round-mode=figures,round-precision=3]{313.047} & \num[group-minimum-digits=5,round-mode=figures,round-precision=2]{52.9514} & 	\num[group-minimum-digits=5,round-mode=figures,round-precision=2]{344.538} & \num[group-minimum-digits=5,round-mode=figures,round-precision=2]{205.206} & 	\num[group-minimum-digits=5,round-mode=figures,round-precision=3]{426.961} & \num[group-minimum-digits=5,round-mode=figures,round-precision=2]{88.569} & 	\num[group-minimum-digits=5,round-mode=figures,round-precision=3]{12.5625} & \num[group-minimum-digits=5,round-mode=figures,round-precision=2]{9.38285} & 	\num[group-minimum-digits=5,round-mode=figures,round-precision=3]{25.7684} & \num[group-minimum-digits=5,round-mode=figures,round-precision=2]{7.82152} \\
 $t\bar{t}$ + $\geq$1$b$ &    	\num[group-minimum-digits=5,round-mode=figures,round-precision=3]{818.92} & \num[group-minimum-digits=5,round-mode=figures,round-precision=2]{85.2269} & 	\num[group-minimum-digits=5,round-mode=figures,round-precision=3]{917.264} & \num[group-minimum-digits=5,round-mode=figures,round-precision=2]{71.2829} & 	\num[group-minimum-digits=5,round-mode=figures,round-precision=3]{589.645} & \num[group-minimum-digits=5,round-mode=figures,round-precision=2]{96.3323} & 	\num[group-minimum-digits=5,round-mode=figures,round-precision=3]{669.436} & \num[group-minimum-digits=5,round-mode=figures,round-precision=2]{58.6732} & 	\num[group-minimum-digits=5,round-mode=figures,round-precision=3]{247.038} & \num[group-minimum-digits=5,round-mode=figures,round-precision=2]{60.573} & 	\num[group-minimum-digits=5,round-mode=figures,round-precision=3]{262.873} & \num[group-minimum-digits=5,round-mode=figures,round-precision=2]{19.7303} \\
 $t\bar{t}$ + $V$            &	\num[group-minimum-digits=5,round-mode=figures,round-precision=2]{15.1297} & \num[group-minimum-digits=5,round-mode=figures,round-precision=2]{35.2759} & 	\num[group-minimum-digits=5,round-mode=figures,round-precision=2]{14.6051} & \num[group-minimum-digits=5,round-mode=figures,round-precision=2]{34.3763} & 	\num[group-minimum-digits=5,round-mode=figures,round-precision=2]{21.6906} & \num[group-minimum-digits=5,round-mode=figures,round-precision=2]{38.1271} & 	\num[group-minimum-digits=5,round-mode=figures,round-precision=2]{21.9787} & \num[group-minimum-digits=5,round-mode=figures,round-precision=2]{38.5151} & 	\num[group-minimum-digits=5,round-mode=figures,round-precision=1]{6.62169} & \num[group-minimum-digits=5,round-mode=figures,round-precision=2]{55.9657} & 	\num[group-minimum-digits=5,round-mode=figures,round-precision=1]{6.68506} & \num[group-minimum-digits=5,round-mode=figures,round-precision=2]{56.973} \\
 Non-$t\bar{t}$              &	\num[group-minimum-digits=5,round-mode=figures,round-precision=2]{75.1849} & \num[group-minimum-digits=5,round-mode=figures,round-precision=2]{17.2933} & 	\num[group-minimum-digits=5,round-mode=figures,round-precision=2]{77.8672} & \num[group-minimum-digits=5,round-mode=figures,round-precision=2]{15.6498} & 	\num[group-minimum-digits=5,round-mode=figures,round-precision=3]{115.047} & \num[group-minimum-digits=5,round-mode=figures,round-precision=2]{36.1619} & 	\num[group-minimum-digits=5,round-mode=figures,round-precision=3]{121.409} & \num[group-minimum-digits=5,round-mode=figures,round-precision=2]{29.2294} & 	\num[group-minimum-digits=5,round-mode=figures,round-precision=3]{13.5533} & \num[group-minimum-digits=5,round-mode=figures,round-precision=2]{3.7744} & 	\num[group-minimum-digits=5,round-mode=figures,round-precision=3]{14.5636} & \num[group-minimum-digits=5,round-mode=figures,round-precision=2]{3.80557} \\
\hline 						
 Total                       &	\num[group-minimum-digits=5,round-mode=figures,round-precision=3]{1249.65} & \num[group-minimum-digits=5,round-mode=figures,round-precision=2]{137.688} & 	\num[group-minimum-digits=5,round-mode=figures,round-precision=4]{1435.98} & \num[group-minimum-digits=5,round-mode=figures,round-precision=2]{54.6648} & 	\num[group-minimum-digits=5,round-mode=figures,round-precision=3]{1347.59} & \num[group-minimum-digits=5,round-mode=figures,round-precision=2]{321.782} & 	\num[group-minimum-digits=5,round-mode=figures,round-precision=4]{1479.49} & \num[group-minimum-digits=5,round-mode=figures,round-precision=2]{66.1833} & 	\num[group-minimum-digits=5,round-mode=figures,round-precision=3]{301.772} & \num[group-minimum-digits=5,round-mode=figures,round-precision=2]{85.2694} & 	\num[group-minimum-digits=5,round-mode=figures,round-precision=3]{333.883} & \num[group-minimum-digits=5,round-mode=figures,round-precision=2]{59.3016} \\
    \hline
  Data                     & \multicolumn{4}{c|}{\num[group-minimum-digits=5]{1467}} & \multicolumn{4}{c|}{\num[group-minimum-digits=5]{1444}} & \multicolumn{4}{c}{\num[group-minimum-digits=5]{319}} \\ 
    \hline
    \bottomrule
  \end{tabular}
  }\end{footnotesize}
  \label{tab:DLyields}
\end{table}

\begin{table}[htbp]
  \caption{
  Event yields in the single-lepton channel five-jet (top) control regions and (bottom) signal regions, including the boosted signal region.
  Post-fit yields are after the combined fit in all channels to data. 
  The uncertainties are the sum in quadrature of statistical and systematic uncertainties in the yields. 
  In the post-fit case, these uncertainties are computed taking into account correlations among nuisance parameters and among the normalization of different processes. 
  The uncertainty in the \ttbin{} and \ttcin{} normalization is not defined pre-fit and therefore only included in the post-fit uncertainties;
  the reported prefit uncertainties on the \ttbin{} and \ttcin{} components arise only from acceptance effects.
  For the \ttH{} signal, the pre-fit yield values correspond to the theoretical prediction and corresponding uncertainties, while the post-fit yield and uncertainties correspond to those in the signal-strength measurement. 
  }
  \centering
  \begin{footnotesize}{
  \begin{tabular}{l | r@{$~\pm~$}l r@{$~\pm~$}l | r@{$~\pm~$}l r@{$~\pm~$}l | r@{$~\pm~$}l r@{$~\pm~$}l }
    \toprule
    \hline
  \multirow{2}{*}{Sample}    & \multicolumn{4}{c|}{\crfivejttlight\rule{0pt}{3ex}} & \multicolumn{4}{c|}{\crfivejttc} & \multicolumn{4}{c}{\crfivejttoneb}\\
                             & \multicolumn{2}{c}{Pre-fit} & \multicolumn{2}{c|}{Post-fit} & \multicolumn{2}{c}{Pre-fit} & \multicolumn{2}{c|}{Post-fit} & \multicolumn{2}{c}{Pre-fit} & \multicolumn{2}{c}{Post-fit}\\
    \hline
\hline 
 $t\bar{t}H$ &                 	\num[group-minimum-digits=5,round-mode=figures,round-precision=3]{224.279} & \num[group-minimum-digits=5,round-mode=figures,round-precision=2]{22.4716} & 	\num[group-minimum-digits=5,round-mode=figures,round-precision=2]{188.205} & \num[group-minimum-digits=5,round-mode=figures,round-precision=2]{138.255} & 	\num[group-minimum-digits=5,round-mode=figures,round-precision=3]{18.6732} & \num[group-minimum-digits=5,round-mode=figures,round-precision=2]{2.53459} & 	\num[group-minimum-digits=5,round-mode=figures,round-precision=2]{15.4365} & \num[group-minimum-digits=5,round-mode=figures,round-precision=2]{11.5263} & 	\num[group-minimum-digits=5,round-mode=figures,round-precision=3]{68.0149} & \num[group-minimum-digits=5,round-mode=figures,round-precision=2]{7.60018} & 	\num[group-minimum-digits=5,round-mode=figures,round-precision=2]{56.6619} & \num[group-minimum-digits=5,round-mode=figures,round-precision=2]{41.8026} \\
 $t\bar{t}$ + light     &     	\num[group-minimum-digits=5,round-mode=figures,round-precision=3]{197214} & \num[group-minimum-digits=5,round-mode=figures,round-precision=2]{25565.3} & 	\num[group-minimum-digits=5,round-mode=figures,round-precision=4]{179934} & \num[group-minimum-digits=5,round-mode=figures,round-precision=2]{4895.57} & 	\num[group-minimum-digits=5,round-mode=figures,round-precision=3]{2582.22} & \num[group-minimum-digits=5,round-mode=figures,round-precision=2]{716.08} & 	\num[group-minimum-digits=5,round-mode=figures,round-precision=3]{2299.84} & \num[group-minimum-digits=5,round-mode=figures,round-precision=2]{211.161} & 	\num[group-minimum-digits=5,round-mode=figures,round-precision=3]{4246.6} & \num[group-minimum-digits=5,round-mode=figures,round-precision=2]{920.888} & 	\num[group-minimum-digits=5,round-mode=figures,round-precision=3]{3557.43} & \num[group-minimum-digits=5,round-mode=figures,round-precision=2]{244.272} \\
 $t\bar{t}$ + $\geq$1$c$ &    	\num[group-minimum-digits=5,round-mode=figures,round-precision=3]{27496.7} & \num[group-minimum-digits=5,round-mode=figures,round-precision=2]{4314.11} & 	\num[group-minimum-digits=5,round-mode=figures,round-precision=3]{44149.3} & \num[group-minimum-digits=5,round-mode=figures,round-precision=2]{5531.54} & 	\num[group-minimum-digits=5,round-mode=figures,round-precision=3]{1283.26} & \num[group-minimum-digits=5,round-mode=figures,round-precision=2]{502.028} & 	\num[group-minimum-digits=5,round-mode=figures,round-precision=3]{1835.75} & \num[group-minimum-digits=5,round-mode=figures,round-precision=2]{254.087} & 	\num[group-minimum-digits=5,round-mode=figures,round-precision=3]{1770.57} & \num[group-minimum-digits=5,round-mode=figures,round-precision=2]{272.971} & 	\num[group-minimum-digits=5,round-mode=figures,round-precision=3]{2587.51} & \num[group-minimum-digits=5,round-mode=figures,round-precision=2]{393.773} \\
 $t\bar{t}$ + $\geq$1$b$ &    	\num[group-minimum-digits=5,round-mode=figures,round-precision=3]{11311.4} & \num[group-minimum-digits=5,round-mode=figures,round-precision=2]{1115.32} & 	\num[group-minimum-digits=5,round-mode=figures,round-precision=3]{13514.5} & \num[group-minimum-digits=5,round-mode=figures,round-precision=2]{1258.36} & 	\num[group-minimum-digits=5,round-mode=figures,round-precision=2]{791.057} & \num[group-minimum-digits=5,round-mode=figures,round-precision=2]{125.507} & 	\num[group-minimum-digits=5,round-mode=figures,round-precision=3]{944.247} & \num[group-minimum-digits=5,round-mode=figures,round-precision=2]{93.6252} & 	\num[group-minimum-digits=5,round-mode=figures,round-precision=3]{3397.64} & \num[group-minimum-digits=5,round-mode=figures,round-precision=2]{443.08} & 	\num[group-minimum-digits=5,round-mode=figures,round-precision=3]{4034.24} & \num[group-minimum-digits=5,round-mode=figures,round-precision=2]{322.383} \\
 $t\bar{t}$ + $V$            &	\num[group-minimum-digits=5,round-mode=figures,round-precision=3]{588.928} & \num[group-minimum-digits=5,round-mode=figures,round-precision=2]{55.4784} & 	\num[group-minimum-digits=5,round-mode=figures,round-precision=3]{584.045} & \num[group-minimum-digits=5,round-mode=figures,round-precision=2]{53.8891} & 	\num[group-minimum-digits=5,round-mode=figures,round-precision=3]{23.2163} & \num[group-minimum-digits=5,round-mode=figures,round-precision=2]{4.14723} & 	\num[group-minimum-digits=5,round-mode=figures,round-precision=3]{21.3158} & \num[group-minimum-digits=5,round-mode=figures,round-precision=2]{2.85324} & 	\num[group-minimum-digits=5,round-mode=figures,round-precision=3]{48.1068} & \num[group-minimum-digits=5,round-mode=figures,round-precision=2]{5.85252} & 	\num[group-minimum-digits=5,round-mode=figures,round-precision=3]{46.5794} & \num[group-minimum-digits=5,round-mode=figures,round-precision=2]{5.38135} \\
 Non-$t\bar{t}$              &	\num[group-minimum-digits=5,round-mode=figures,round-precision=3]{21298.7} & \num[group-minimum-digits=5,round-mode=figures,round-precision=2]{4149.11} & 	\num[group-minimum-digits=5,round-mode=figures,round-precision=3]{20949.6} & \num[group-minimum-digits=5,round-mode=figures,round-precision=2]{3182.19} & 	\num[group-minimum-digits=5,round-mode=figures,round-precision=2]{518.509} & \num[group-minimum-digits=5,round-mode=figures,round-precision=2]{176.094} & 	\num[group-minimum-digits=5,round-mode=figures,round-precision=2]{442.271} & \num[group-minimum-digits=5,round-mode=figures,round-precision=2]{104.203} & 	\num[group-minimum-digits=5,round-mode=figures,round-precision=2]{957.088} & \num[group-minimum-digits=5,round-mode=figures,round-precision=2]{191.076} & 	\num[group-minimum-digits=5,round-mode=figures,round-precision=2]{855.8} & \num[group-minimum-digits=5,round-mode=figures,round-precision=2]{160.845} \\
\hline 						
 Total                       &	\num[group-minimum-digits=5,round-mode=figures,round-precision=3]{257872} & \num[group-minimum-digits=5,round-mode=figures,round-precision=2]{28874.2} & 	\num[group-minimum-digits=5,round-mode=figures,round-precision=5]{259320} & \num[group-minimum-digits=5,round-mode=figures,round-precision=2]{911.779} & 	\num[group-minimum-digits=5,round-mode=figures,round-precision=2]{5193.39} & \num[group-minimum-digits=5,round-mode=figures,round-precision=2]{1129.5} & 	\num[group-minimum-digits=5,round-mode=figures,round-precision=3]{5558.86} & \num[group-minimum-digits=5,round-mode=figures,round-precision=2]{157.52} & 	\num[group-minimum-digits=5,round-mode=figures,round-precision=3]{10398.1} & \num[group-minimum-digits=5,round-mode=figures,round-precision=2]{1296.48} & 	\num[group-minimum-digits=5,round-mode=figures,round-precision=4]{11138.2} & \num[group-minimum-digits=5,round-mode=figures,round-precision=2]{287.76} \\
\hline 
  Data   & \multicolumn{4}{c|}{\num[group-minimum-digits=5]{259320}} & \multicolumn{4}{c|}{\num[group-minimum-digits=5]{5465}} & \multicolumn{4}{c}{\num[group-minimum-digits=5]{11095}} \\ 
    \hline
    \bottomrule
  \end{tabular}
  
  \vspace{5pt}

  \begin{tabular}{l | r@{$~\pm~$}l r@{$~\pm~$}l | r@{$~\pm~$}l r@{$~\pm~$}l | r@{$~\pm~$}l r@{$~\pm~$}l }
    \toprule
    \hline
  \multirow{2}{*}{Sample}    & \multicolumn{4}{c|}{\srfivejtwo\rule{0pt}{3ex}} & \multicolumn{4}{c|}{\srfivejone} & \multicolumn{4}{c}{\srboosted}\\
                             & \multicolumn{2}{c}{Pre-fit} & \multicolumn{2}{c|}{Post-fit} & \multicolumn{2}{c}{Pre-fit} & \multicolumn{2}{c|}{Post-fit} & \multicolumn{2}{c}{Pre-fit} & \multicolumn{2}{c}{Post-fit}\\
    \hline
\hline 
 $t\bar{t}H$ &                 	\num[group-minimum-digits=5,round-mode=figures,round-precision=3]{40.1009} & \num[group-minimum-digits=5,round-mode=figures,round-precision=2]{5.0734} & 	\num[group-minimum-digits=5,round-mode=figures,round-precision=2]{33.8187} & \num[group-minimum-digits=5,round-mode=figures,round-precision=2]{24.6877} & 	\num[group-minimum-digits=5,round-mode=figures,round-precision=3]{15.9118} & \num[group-minimum-digits=5,round-mode=figures,round-precision=2]{2.09902} & 	\num[group-minimum-digits=5,round-mode=figures,round-precision=3]{13.2514} & \num[group-minimum-digits=5,round-mode=figures,round-precision=2]{9.78121} & 	\num[group-minimum-digits=5,round-mode=figures,round-precision=3]{16.8745} & \num[group-minimum-digits=5,round-mode=figures,round-precision=2]{1.93592} & 	\num[group-minimum-digits=5,round-mode=figures,round-precision=2]{14.3306} & \num[group-minimum-digits=5,round-mode=figures,round-precision=2]{10.4861} \\
 $t\bar{t}$ + light     &     	\num[group-minimum-digits=5,round-mode=figures,round-precision=2]{501.903} & \num[group-minimum-digits=5,round-mode=figures,round-precision=2]{208.272} & 	\num[group-minimum-digits=5,round-mode=figures,round-precision=3]{392.859} & \num[group-minimum-digits=5,round-mode=figures,round-precision=2]{66.6946} & 	\num[group-minimum-digits=5,round-mode=figures,round-precision=2]{15.2111} & \num[group-minimum-digits=5,round-mode=figures,round-precision=2]{32.593} & 	\num[group-minimum-digits=5,round-mode=figures,round-precision=3]{12.4607} & \num[group-minimum-digits=5,round-mode=figures,round-precision=2]{9.30982} & 	\num[group-minimum-digits=5,round-mode=figures,round-precision=2]{176.689} & \num[group-minimum-digits=5,round-mode=figures,round-precision=2]{122.87} & 	\num[group-minimum-digits=5,round-mode=figures,round-precision=3]{112.378} & \num[group-minimum-digits=5,round-mode=figures,round-precision=2]{32.4161} \\
 $t\bar{t}$ + $\geq$1$c$ &    	\num[group-minimum-digits=5,round-mode=figures,round-precision=3]{435.694} & \num[group-minimum-digits=5,round-mode=figures,round-precision=2]{92.0925} & 	\num[group-minimum-digits=5,round-mode=figures,round-precision=2]{610.38} & \num[group-minimum-digits=5,round-mode=figures,round-precision=2]{103.533} & 	\num[group-minimum-digits=5,round-mode=figures,round-precision=2]{29.6094} & \num[group-minimum-digits=5,round-mode=figures,round-precision=2]{16.6473} & 	\num[group-minimum-digits=5,round-mode=figures,round-precision=2]{27.916} & \num[group-minimum-digits=5,round-mode=figures,round-precision=2]{14.3705} & 	\num[group-minimum-digits=5,round-mode=figures,round-precision=3]{168.012} & \num[group-minimum-digits=5,round-mode=figures,round-precision=2]{69.9454} & 	\num[group-minimum-digits=5,round-mode=figures,round-precision=3]{235.325} & \num[group-minimum-digits=5,round-mode=figures,round-precision=2]{39.3901} \\
 $t\bar{t}$ + $\geq$1$b$ &    	\num[group-minimum-digits=5,round-mode=figures,round-precision=3]{1226.84} & \num[group-minimum-digits=5,round-mode=figures,round-precision=2]{200.536} & 	\num[group-minimum-digits=5,round-mode=figures,round-precision=3]{1447.27} & \num[group-minimum-digits=5,round-mode=figures,round-precision=2]{109.03} & 	\num[group-minimum-digits=5,round-mode=figures,round-precision=3]{272.641} & \num[group-minimum-digits=5,round-mode=figures,round-precision=2]{52.6214} & 	\num[group-minimum-digits=5,round-mode=figures,round-precision=3]{335.129} & \num[group-minimum-digits=5,round-mode=figures,round-precision=2]{25.459} & 	\num[group-minimum-digits=5,round-mode=figures,round-precision=3]{236.413} & \num[group-minimum-digits=5,round-mode=figures,round-precision=2]{88.9326} & 	\num[group-minimum-digits=5,round-mode=figures,round-precision=3]{228.749} & \num[group-minimum-digits=5,round-mode=figures,round-precision=2]{33.014} \\
 $t\bar{t}$ + $V$            &	\num[group-minimum-digits=5,round-mode=figures,round-precision=3]{19.87} & \num[group-minimum-digits=5,round-mode=figures,round-precision=2]{2.87385} & 	\num[group-minimum-digits=5,round-mode=figures,round-precision=3]{19.7209} & \num[group-minimum-digits=5,round-mode=figures,round-precision=2]{2.44187} & 	\num[group-minimum-digits=5,round-mode=figures,round-precision=2]{6.41633} & \num[group-minimum-digits=5,round-mode=figures,round-precision=2]{1.31504} & 	\num[group-minimum-digits=5,round-mode=figures,round-precision=2]{6.37021} & \num[group-minimum-digits=5,round-mode=figures,round-precision=2]{1.17492} & 	\num[group-minimum-digits=5,round-mode=figures,round-precision=3]{16.1459} & \num[group-minimum-digits=5,round-mode=figures,round-precision=2]{2.93673} & 	\num[group-minimum-digits=5,round-mode=figures,round-precision=3]{16.5591} & \num[group-minimum-digits=5,round-mode=figures,round-precision=2]{2.43688} \\
 Non-$t\bar{t}$              &	\num[group-minimum-digits=5,round-mode=figures,round-precision=3]{268.982} & \num[group-minimum-digits=5,round-mode=figures,round-precision=2]{63.8536} & 	\num[group-minimum-digits=5,round-mode=figures,round-precision=3]{220.135} & \num[group-minimum-digits=5,round-mode=figures,round-precision=2]{51.8756} & 	\num[group-minimum-digits=5,round-mode=figures,round-precision=2]{54.0414} & \num[group-minimum-digits=5,round-mode=figures,round-precision=2]{11.3665} & 	\num[group-minimum-digits=5,round-mode=figures,round-precision=3]{28.0832} & \num[group-minimum-digits=5,round-mode=figures,round-precision=2]{8.37219} & 	\num[group-minimum-digits=5,round-mode=figures,round-precision=3]{103.965} & \num[group-minimum-digits=5,round-mode=figures,round-precision=2]{30.0542} & 	\num[group-minimum-digits=5,round-mode=figures,round-precision=3]{100.743} & \num[group-minimum-digits=5,round-mode=figures,round-precision=2]{25.6287} \\
\hline 						
 Total                       &	\num[group-minimum-digits=5,round-mode=figures,round-precision=3]{2439.81} & \num[group-minimum-digits=5,round-mode=figures,round-precision=2]{386.142} & 	\num[group-minimum-digits=5,round-mode=figures,round-precision=4]{2724.18} & \num[group-minimum-digits=5,round-mode=figures,round-precision=2]{70.0559} & 	\num[group-minimum-digits=5,round-mode=figures,round-precision=3]{370.696} & \num[group-minimum-digits=5,round-mode=figures,round-precision=2]{68.0694} & 	\num[group-minimum-digits=5,round-mode=figures,round-precision=3]{423.21} & \num[group-minimum-digits=5,round-mode=figures,round-precision=2]{23.2951} & 	\num[group-minimum-digits=5,round-mode=figures,round-precision=2]{709.623} & \num[group-minimum-digits=5,round-mode=figures,round-precision=2]{198.989} & 	\num[group-minimum-digits=5,round-mode=figures,round-precision=3]{708.085} & \num[group-minimum-digits=5,round-mode=figures,round-precision=2]{39.8522} \\
\hline 
  Data   & \multicolumn{4}{c|}{\num[group-minimum-digits=5]{2798}} & \multicolumn{4}{c|}{\num[group-minimum-digits=5]{426}} & \multicolumn{4}{c}{\num[group-minimum-digits=5]{740}} \\ 
    \hline
    \bottomrule
  \end{tabular}
  }\end{footnotesize}
  \label{tab:SLyields5j}
\end{table}

\begin{table}[htbp]
  \caption{
  Event yields in the single-lepton channel six-jet (top) control regions and (bottom) signal regions.
  Post-fit yields are after the combined fit in all channels to data.
  The uncertainties are the sum in quadrature of statistical and systematic uncertainties in the yields. 
  In the post-fit case, these uncertainties are computed taking into account correlations among nuisance parameters and among the normalization of different processes.
  The uncertainty in the \ttbin{} and \ttcin{} normalization is not defined pre-fit and therefore only included in the post-fit uncertainties;
  the reported prefit uncertainties on the \ttbin{} and \ttcin{} components arise only from acceptance effects.
  For the \ttH{} signal, the pre-fit yield values correspond to the theoretical prediction and corresponding uncertainties, while the post-fit yield and uncertainties correspond to those in the signal-strength measurement.
  }
  \centering
  \begin{footnotesize}{
   \begin{tabular}{l | r@{$~\pm~$}l r@{$~\pm~$}l | r@{$~\pm~$}l r@{$~\pm~$}l | r@{$~\pm~$}l r@{$~\pm~$}l }
    \toprule
    \hline
  \multirow{2}{*}{Sample}    & \multicolumn{4}{c|}{\crsixjttlight\rule{0pt}{3ex}} & \multicolumn{4}{c|}{\crsixjttc} & \multicolumn{4}{c}{\crsixjttoneb}\\
                             & \multicolumn{2}{c}{Pre-fit} & \multicolumn{2}{c|}{Post-fit} & \multicolumn{2}{c}{Pre-fit} & \multicolumn{2}{c|}{Post-fit} & \multicolumn{2}{c}{Pre-fit} & \multicolumn{2}{c}{Post-fit}\\
    \hline
\hline 
 $t\bar{t}H$ &                 	\num[group-minimum-digits=5,round-mode=figures,round-precision=3]{450.323} & \num[group-minimum-digits=5,round-mode=figures,round-precision=2]{48.3643} & 	\num[group-minimum-digits=5,round-mode=figures,round-precision=2]{373.182} & \num[group-minimum-digits=5,round-mode=figures,round-precision=2]{278.372} & 	\num[group-minimum-digits=5,round-mode=figures,round-precision=3]{102.155} & \num[group-minimum-digits=5,round-mode=figures,round-precision=2]{13.1493} & 	\num[group-minimum-digits=5,round-mode=figures,round-precision=2]{86.796} & \num[group-minimum-digits=5,round-mode=figures,round-precision=2]{63.6031} & 	\num[group-minimum-digits=5,round-mode=figures,round-precision=3]{100.322} & \num[group-minimum-digits=5,round-mode=figures,round-precision=2]{12.4322} & 	\num[group-minimum-digits=5,round-mode=figures,round-precision=2]{82.9173} & \num[group-minimum-digits=5,round-mode=figures,round-precision=2]{61.4214} \\
 $t\bar{t}$ + light     &     	\num[group-minimum-digits=5,round-mode=figures,round-precision=3]{125318} & \num[group-minimum-digits=5,round-mode=figures,round-precision=2]{34476.2} & 	\num[group-minimum-digits=5,round-mode=figures,round-precision=4]{108239} & \num[group-minimum-digits=5,round-mode=figures,round-precision=2]{4332.35} & 	\num[group-minimum-digits=5,round-mode=figures,round-precision=2]{4330.03} & \num[group-minimum-digits=5,round-mode=figures,round-precision=2]{2005.43} & 	\num[group-minimum-digits=5,round-mode=figures,round-precision=3]{3354.65} & \num[group-minimum-digits=5,round-mode=figures,round-precision=2]{426.331} & 	\num[group-minimum-digits=5,round-mode=figures,round-precision=3]{2219.42} & \num[group-minimum-digits=5,round-mode=figures,round-precision=2]{520.887} & 	\num[group-minimum-digits=5,round-mode=figures,round-precision=3]{1820.89} & \num[group-minimum-digits=5,round-mode=figures,round-precision=2]{165.904} \\
 $t\bar{t}$ + $\geq$1$c$ &    	\num[group-minimum-digits=5,round-mode=figures,round-precision=3]{28395.3} & \num[group-minimum-digits=5,round-mode=figures,round-precision=2]{7221.43} & 	\num[group-minimum-digits=5,round-mode=figures,round-precision=3]{45694.5} & \num[group-minimum-digits=5,round-mode=figures,round-precision=2]{5138.66} & 	\num[group-minimum-digits=5,round-mode=figures,round-precision=2]{3560.58} & \num[group-minimum-digits=5,round-mode=figures,round-precision=2]{1348.95} & 	\num[group-minimum-digits=5,round-mode=figures,round-precision=3]{5299.82} & \num[group-minimum-digits=5,round-mode=figures,round-precision=2]{678.724} & 	\num[group-minimum-digits=5,round-mode=figures,round-precision=3]{1455.44} & \num[group-minimum-digits=5,round-mode=figures,round-precision=2]{333.429} & 	\num[group-minimum-digits=5,round-mode=figures,round-precision=3]{2077.78} & \num[group-minimum-digits=5,round-mode=figures,round-precision=2]{304.513} \\
 $t\bar{t}$ + $\geq$1$b$ &    	\num[group-minimum-digits=5,round-mode=figures,round-precision=3]{13111.7} & \num[group-minimum-digits=5,round-mode=figures,round-precision=2]{1804.35} & 	\num[group-minimum-digits=5,round-mode=figures,round-precision=3]{14584.3} & \num[group-minimum-digits=5,round-mode=figures,round-precision=2]{1370.47} & 	\num[group-minimum-digits=5,round-mode=figures,round-precision=3]{2660.31} & \num[group-minimum-digits=5,round-mode=figures,round-precision=2]{537.822} & 	\num[group-minimum-digits=5,round-mode=figures,round-precision=3]{2949.77} & \num[group-minimum-digits=5,round-mode=figures,round-precision=2]{282.539} & 	\num[group-minimum-digits=5,round-mode=figures,round-precision=3]{3672.75} & \num[group-minimum-digits=5,round-mode=figures,round-precision=2]{502.575} & 	\num[group-minimum-digits=5,round-mode=figures,round-precision=3]{4083.85} & \num[group-minimum-digits=5,round-mode=figures,round-precision=2]{317.383} \\
 $t\bar{t}$ + $V$            &	\num[group-minimum-digits=5,round-mode=figures,round-precision=3]{1011.58} & \num[group-minimum-digits=5,round-mode=figures,round-precision=2]{117.089} & 	\num[group-minimum-digits=5,round-mode=figures,round-precision=3]{995.749} & \num[group-minimum-digits=5,round-mode=figures,round-precision=2]{91.1708} & 	\num[group-minimum-digits=5,round-mode=figures,round-precision=3]{117.94} & \num[group-minimum-digits=5,round-mode=figures,round-precision=2]{20.6181} & 	\num[group-minimum-digits=5,round-mode=figures,round-precision=3]{118.299} & \num[group-minimum-digits=5,round-mode=figures,round-precision=2]{13.7512} & 	\num[group-minimum-digits=5,round-mode=figures,round-precision=3]{70.4815} & \num[group-minimum-digits=5,round-mode=figures,round-precision=2]{8.54457} & 	\num[group-minimum-digits=5,round-mode=figures,round-precision=3]{67.9376} & \num[group-minimum-digits=5,round-mode=figures,round-precision=2]{7.21382} \\
 Non-$t\bar{t}$              &	\num[group-minimum-digits=5,round-mode=figures,round-precision=3]{12585} & \num[group-minimum-digits=5,round-mode=figures,round-precision=2]{2976.83} & 	\num[group-minimum-digits=5,round-mode=figures,round-precision=3]{11800.7} & \num[group-minimum-digits=5,round-mode=figures,round-precision=2]{2042.87} & 	\num[group-minimum-digits=5,round-mode=figures,round-precision=3]{1056.09} & \num[group-minimum-digits=5,round-mode=figures,round-precision=2]{343.486} & 	\num[group-minimum-digits=5,round-mode=figures,round-precision=3]{1002.8} & \num[group-minimum-digits=5,round-mode=figures,round-precision=2]{214.407} & 	\num[group-minimum-digits=5,round-mode=figures,round-precision=2]{706.479} & \num[group-minimum-digits=5,round-mode=figures,round-precision=2]{156.99} & 	\num[group-minimum-digits=5,round-mode=figures,round-precision=2]{597.933} & \num[group-minimum-digits=5,round-mode=figures,round-precision=2]{111.751} \\
\hline 						
 Total                       &	\num[group-minimum-digits=5,round-mode=figures,round-precision=3]{180654} & \num[group-minimum-digits=5,round-mode=figures,round-precision=2]{39332.8} & 	\num[group-minimum-digits=5,round-mode=figures,round-precision=5]{181687} & \num[group-minimum-digits=5,round-mode=figures,round-precision=2]{856.74} & 	\num[group-minimum-digits=5,round-mode=figures,round-precision=3]{11789.2} & \num[group-minimum-digits=5,round-mode=figures,round-precision=2]{3199.41} & 	\num[group-minimum-digits=5,round-mode=figures,round-precision=4]{12812.1} & \num[group-minimum-digits=5,round-mode=figures,round-precision=2]{260.693} & 	\num[group-minimum-digits=5,round-mode=figures,round-precision=2]{8181.77} & \num[group-minimum-digits=5,round-mode=figures,round-precision=2]{1141.36} & 	\num[group-minimum-digits=5,round-mode=figures,round-precision=3]{8731.3} & \num[group-minimum-digits=5,round-mode=figures,round-precision=2]{233.144} \\
\hline 
  Data   & \multicolumn{4}{c|}{\num[group-minimum-digits=5]{181706}} & \multicolumn{4}{c|}{\num[group-minimum-digits=5]{12778}} & \multicolumn{4}{c}{\num[group-minimum-digits=5]{8576}} \\ 
    \hline
    \bottomrule
  \end{tabular}
  
  \vspace{5pt}
  
  \begin{tabular}{l | r@{$~\pm~$}l r@{$~\pm~$}l | r@{$~\pm~$}l r@{$~\pm~$}l | r@{$~\pm~$}l r@{$~\pm~$}l }
    \toprule
    \hline
  \multirow{2}{*}{Sample}    & \multicolumn{4}{c|}{\srsixjthree\rule{0pt}{3ex}} & \multicolumn{4}{c|}{\srsixjtwo} & \multicolumn{4}{c}{\srsixjone}\\
                             & \multicolumn{2}{c}{Pre-fit} & \multicolumn{2}{c|}{Post-fit} & \multicolumn{2}{c}{Pre-fit} & \multicolumn{2}{c|}{Post-fit} & \multicolumn{2}{c}{Pre-fit} & \multicolumn{2}{c}{Post-fit}\\
    \hline
 $t\bar{t}H$ &                 	\num[group-minimum-digits=5,round-mode=figures,round-precision=2]{84.6591} & \num[group-minimum-digits=5,round-mode=figures,round-precision=2]{10.3387} & 	\num[group-minimum-digits=5,round-mode=figures,round-precision=2]{71.0574} & \num[group-minimum-digits=5,round-mode=figures,round-precision=2]{52.1681} & 	\num[group-minimum-digits=5,round-mode=figures,round-precision=2]{81.2654} & \num[group-minimum-digits=5,round-mode=figures,round-precision=2]{10.2132} & 	\num[group-minimum-digits=5,round-mode=figures,round-precision=2]{67.8669} & \num[group-minimum-digits=5,round-mode=figures,round-precision=2]{49.9384} & 	\num[group-minimum-digits=5,round-mode=figures,round-precision=2]{62.4568} & \num[group-minimum-digits=5,round-mode=figures,round-precision=2]{10.8859} & 	\num[group-minimum-digits=5,round-mode=figures,round-precision=2]{51.4507} & \num[group-minimum-digits=5,round-mode=figures,round-precision=2]{38.049} \\
 $t\bar{t}$ + light     &     	\num[group-minimum-digits=5,round-mode=figures,round-precision=2]{745.435} & \num[group-minimum-digits=5,round-mode=figures,round-precision=2]{369.899} & 	\num[group-minimum-digits=5,round-mode=figures,round-precision=3]{586.346} & \num[group-minimum-digits=5,round-mode=figures,round-precision=2]{97.7145} & 	\num[group-minimum-digits=5,round-mode=figures,round-precision=2]{208.743} & \num[group-minimum-digits=5,round-mode=figures,round-precision=2]{208.739} & 	\num[group-minimum-digits=5,round-mode=figures,round-precision=2]{96.1476} & \num[group-minimum-digits=5,round-mode=figures,round-precision=2]{32.9777} & 	\num[group-minimum-digits=5,round-mode=figures,round-precision=2]{14.0702} & \num[group-minimum-digits=5,round-mode=figures,round-precision=2]{10.1681} & 	\num[group-minimum-digits=5,round-mode=figures,round-precision=3]{12.0807} & \num[group-minimum-digits=5,round-mode=figures,round-precision=2]{5.7507} \\
 $t\bar{t}$ + $\geq$1$c$ &    	\num[group-minimum-digits=5,round-mode=figures,round-precision=2]{883.684} & \num[group-minimum-digits=5,round-mode=figures,round-precision=2]{349.495} & 	\num[group-minimum-digits=5,round-mode=figures,round-precision=3]{1327.41} & \num[group-minimum-digits=5,round-mode=figures,round-precision=2]{189.636} & 	\num[group-minimum-digits=5,round-mode=figures,round-precision=2]{346.47} & \num[group-minimum-digits=5,round-mode=figures,round-precision=2]{104.113} & 	\num[group-minimum-digits=5,round-mode=figures,round-precision=3]{472.735} & \num[group-minimum-digits=5,round-mode=figures,round-precision=2]{99.4264} & 	\num[group-minimum-digits=5,round-mode=figures,round-precision=2]{53.2957} & \num[group-minimum-digits=5,round-mode=figures,round-precision=2]{32.8553} & 	\num[group-minimum-digits=5,round-mode=figures,round-precision=2]{44.3455} & \num[group-minimum-digits=5,round-mode=figures,round-precision=2]{19.9891} \\
 $t\bar{t}$ + $\geq$1$b$ &    	\num[group-minimum-digits=5,round-mode=figures,round-precision=3]{2097.03} & \num[group-minimum-digits=5,round-mode=figures,round-precision=2]{415.627} & 	\num[group-minimum-digits=5,round-mode=figures,round-precision=3]{2290.28} & \num[group-minimum-digits=5,round-mode=figures,round-precision=2]{168.689} & 	\num[group-minimum-digits=5,round-mode=figures,round-precision=3]{1749.46} & \num[group-minimum-digits=5,round-mode=figures,round-precision=2]{365.663} & 	\num[group-minimum-digits=5,round-mode=figures,round-precision=3]{1845.7} & \num[group-minimum-digits=5,round-mode=figures,round-precision=2]{130.049} & 	\num[group-minimum-digits=5,round-mode=figures,round-precision=3]{1011.46} & \num[group-minimum-digits=5,round-mode=figures,round-precision=2]{237.751} & 	\num[group-minimum-digits=5,round-mode=figures,round-precision=4]{1031.92} & \num[group-minimum-digits=5,round-mode=figures,round-precision=2]{59.4868} \\
 $t\bar{t}$ + $V$            &	\num[group-minimum-digits=5,round-mode=figures,round-precision=3]{51.1805} & \num[group-minimum-digits=5,round-mode=figures,round-precision=2]{7.43556} & 	\num[group-minimum-digits=5,round-mode=figures,round-precision=3]{50.7763} & \num[group-minimum-digits=5,round-mode=figures,round-precision=2]{5.91533} & 	\num[group-minimum-digits=5,round-mode=figures,round-precision=3]{40.7825} & \num[group-minimum-digits=5,round-mode=figures,round-precision=2]{5.71793} & 	\num[group-minimum-digits=5,round-mode=figures,round-precision=3]{40.3376} & \num[group-minimum-digits=5,round-mode=figures,round-precision=2]{4.8496} & 	\num[group-minimum-digits=5,round-mode=figures,round-precision=3]{25.8246} & \num[group-minimum-digits=5,round-mode=figures,round-precision=2]{3.69936} & 	\num[group-minimum-digits=5,round-mode=figures,round-precision=3]{25.2609} & \num[group-minimum-digits=5,round-mode=figures,round-precision=2]{3.24034} \\
 Non-$t\bar{t}$              &	\num[group-minimum-digits=5,round-mode=figures,round-precision=3]{302.626} & \num[group-minimum-digits=5,round-mode=figures,round-precision=2]{82.0408} & 	\num[group-minimum-digits=5,round-mode=figures,round-precision=3]{267.434} & \num[group-minimum-digits=5,round-mode=figures,round-precision=2]{63.4893} & 	\num[group-minimum-digits=5,round-mode=figures,round-precision=3]{155.339} & \num[group-minimum-digits=5,round-mode=figures,round-precision=2]{52.3197} & 	\num[group-minimum-digits=5,round-mode=figures,round-precision=3]{134.117} & \num[group-minimum-digits=5,round-mode=figures,round-precision=2]{45.9116} & 	\num[group-minimum-digits=5,round-mode=figures,round-precision=2]{75.3606} & \num[group-minimum-digits=5,round-mode=figures,round-precision=2]{19.8684} & 	\num[group-minimum-digits=5,round-mode=figures,round-precision=2]{57.9805} & \num[group-minimum-digits=5,round-mode=figures,round-precision=2]{17.4628} \\
\hline 						
 Total                       &	\num[group-minimum-digits=5,round-mode=figures,round-precision=3]{4135.37} & \num[group-minimum-digits=5,round-mode=figures,round-precision=2]{849.466} & 	\num[group-minimum-digits=5,round-mode=figures,round-precision=3]{4593.3} & \num[group-minimum-digits=5,round-mode=figures,round-precision=2]{105.702} & 	\num[group-minimum-digits=5,round-mode=figures,round-precision=3]{2554.96} & \num[group-minimum-digits=5,round-mode=figures,round-precision=2]{506.198} & 	\num[group-minimum-digits=5,round-mode=figures,round-precision=4]{2656.9} & \num[group-minimum-digits=5,round-mode=figures,round-precision=2]{81.9838} & 	\num[group-minimum-digits=5,round-mode=figures,round-precision=3]{1223.78} & \num[group-minimum-digits=5,round-mode=figures,round-precision=2]{252.236} & 	\num[group-minimum-digits=5,round-mode=figures,round-precision=4]{1223.04} & \num[group-minimum-digits=5,round-mode=figures,round-precision=2]{42.3982} \\
\hline 
  Data   & \multicolumn{4}{c|}{\num[group-minimum-digits=5]{4698}} & \multicolumn{4}{c|}{\num[group-minimum-digits=5]{2641}} & \multicolumn{4}{c}{\num[group-minimum-digits=5]{1222}} \\ 
    \hline
    \bottomrule
  \end{tabular}
  }\end{footnotesize}
  \label{tab:SLyields6j}
\end{table}

\clearpage
\section{Input variables to the classification BDTs}
\label{sec:app_classBDTinputs}
In this appendix, 
the full list of variables used as inputs to the classification BDT, described in Section~\ref{sec:mvas}, in each of the signal regions is reported. 
Variables are listed separately in Table~\ref{tab:dlClassBDTvars} for the dilepton channel, 
in Table~\ref{tab:inputVarsClassSL} for the resolved single-lepton channel
and in Table~\ref{tab:inputVarsClassBoosted} for the boosted category.
Variables are grouped according to the type of information that is exploited. 
The variables from the reconstruction BDT exploit the chosen jet--parton assignments described in Section~\ref{sub:recobdt}.
The $b$-tagging discriminant assigned to each jet is defined in Section~\ref{sec:selection}.
The most powerful variables in the classification BDT are the reconstruction BDT output, the LHD (Section~\ref{sub:lhd}) and the MEM$_{D1}$ (Section~\ref{sub:mem}).
The large-$R$ jets used to build the Higgs-boson and top-quark candidates in the boosted category are defined in Section~\ref{sec:selection}.

Some kinematic and topological variables are built considering only $b$-tagged-jets in the event. The $b$-tagging requirements for these jets are optimized separately for each variable in each region to improve the classification BDT performance.
In the resolved single-lepton channel, $b$-tagged-jets are defined as the four jets with the largest value of the $b$-tagging discriminant. If two jets have the same $b$-tagging discriminant value, they are ordered by decreasing jet \pt{} value.
In the dilepton channel, the $b$-tagging requirements depend on the signal region: in \srfourjone\ the \textit{tight} working point is used, in \srfourjthree\ the \textit{very tight} working point is used and in \srfourjtwo\ the \textit{loose} working point is used with the exception of \nhiggsthirtybb{}, which uses the \textit{medium} working point, and \aplab{}, which uses the \textit{tight} working point.
The \textit{loose} working point is used in the boosted signal region.

\begin{table}[!htb]
\caption{Variables used in the classification BDTs in the dilepton signal regions. 
For variables from the reconstruction BDT, those with a $^{*}$ are from the BDT using Higgs-boson information, 
those with no $^{*}$ are from the BDT without Higgs-boson information 
while for those with a $^{\textrm{**}}$ both versions are used.
These two versions of the reconstruction BDT are described in Section~\ref{sub:recobdt}.}
\begin{center}
\begin{tabular}{l|l|ccc}
\hline
\hline
Variable & Definition & \srfourjone & \srfourjtwo & \srfourjthree \rule{0pt}{3ex} \\[0.5ex]
\hline
\hline
\multicolumn{5}{l}{General kinematic variables} \\
\hline
\mbbminM \rule{0pt}{3ex}  & \begin{minipage}[l]{0.55\textwidth}Minimum invariant mass of a $b$-tagged jet pair                   \end{minipage}& \checkmark & \checkmark & - \\[1.2ex]
\mbbmaxM                  & \begin{minipage}[l]{0.55\textwidth}Maximum invariant mass of a $b$-tagged jet pair                   \end{minipage}& -          & -          & \checkmark \\[1.2ex]
\mbbmindr                 & \begin{minipage}[l]{0.55\textwidth}Invariant mass of the $b$-tagged jet pair with minimum $\Delta R$ \end{minipage}& \checkmark & -          & \checkmark \\[1.2ex]
\mjjmaxpt                 & \begin{minipage}[l]{0.55\textwidth}Invariant mass of the jet pair with maximum \pt                 \end{minipage}& \checkmark & -          & -          \\[1.2ex]
\mbbmaxpt                 & \begin{minipage}[l]{0.55\textwidth}Invariant mass of the $b$-tagged jet pair with maximum \pt      \end{minipage}& \checkmark & -          & \checkmark \\[1.2ex]
\detabbav                 & \begin{minipage}[l]{0.55\textwidth}Average $\Delta\eta$ for all $b$-tagged jet pairs                 \end{minipage}& \checkmark & \checkmark & \checkmark \\[1.2ex]
\maxdetalj                & \begin{minipage}[l]{0.55\textwidth}Maximum $\Delta\eta$ between a jet and a lepton                   \end{minipage}& -          & \checkmark & \checkmark \\[1.2ex]
\drbbmaxpt                & \begin{minipage}[l]{0.55\textwidth}$\Delta R$ between the $b$-tagged jet pair with maximum \pt     \end{minipage}& -          & \checkmark & \checkmark \\[1.2ex]
\nhiggsthirtybb           & \begin{minipage}[l]{0.55\textwidth}Number of $b$-tagged jet pairs with invariant mass within 30~\GeV\ of the Higgs-boson mass\end{minipage}  & \checkmark & \checkmark & -     \\[2ex]
$n^{\pt > 40}_{\mathrm{jets}}$        & \begin{minipage}[l]{0.55\textwidth}Number of jets with $\pt >$ 40~\GeV\                                \end{minipage}& -          & \checkmark & \checkmark \\[1.2ex]
\aplab                    & \begin{minipage}[l]{0.55\textwidth}$1.5 \lambda_2$, where $\lambda_2$ is the second eigenvalue of the momentum tensor~\cite{tensor} built with all $b$-tagged jets\end{minipage} & -     & \checkmark & -     \\[2ex]
\htall                    & \begin{minipage}[l]{0.55\textwidth}Scalar sum of \pt\ of all jets and leptons                       \end{minipage}& -          & -          & \checkmark \\[1.2ex]
\hline
\hline
\multicolumn{5}{l}{Variables from reconstruction BDT} \\
\hline
BDT output \rule{0pt}{3ex}               & \begin{minipage}[l]{0.55\textwidth}Output of the reconstruction BDT                                                    \end{minipage}& \checkmark$^\textrm{**}$ & \checkmark$^\textrm{**}$ & \checkmark \\[1.2ex]
$m_{bb}^{\mathrm{Higgs}}$            & \begin{minipage}[l]{0.55\textwidth}Higgs candidate mass                                    \end{minipage}& \checkmark & -          & \checkmark \\[1.2ex]
$\Delta R_{H,\ttbar}$     & \begin{minipage}[l]{0.55\textwidth}$\Delta R$ between Higgs candidate and \ttbar\ candidate system   \end{minipage}& \checkmark$^\textrm{*}$ & - & -  \\[2ex]
$\Delta R_{H,\ell}^{\mathrm{min}}$ & \begin{minipage}[l]{0.55\textwidth}Minimum $\Delta R$ between Higgs candidate and lepton     \end{minipage}& \checkmark & \checkmark & \checkmark \\[1.2ex]
$\Delta R_{H,b}^{\mathrm{min}}$    & \begin{minipage}[l]{0.55\textwidth}Minimum $\Delta R$ between Higgs candidate and $b$-jet from top \end{minipage}& \checkmark & \checkmark & -      \\[2ex]
$\Delta R_{H,b}^{\mathrm{max}}$    & \begin{minipage}[l]{0.55\textwidth}Maximum $\Delta R$ between Higgs candidate and $b$-jet from top \end{minipage}& -          & \checkmark & -      \\[2ex]
$\Delta R^{\mathrm{Higgs}}_{bb}$     & \begin{minipage}[l]{0.55\textwidth}$\Delta R$ between the two jets matched to the Higgs candidate      \end{minipage}& -          & \checkmark & -          \\[1.2ex]
\hline
\hline
\multicolumn{5}{l} {Variables from $b$-tagging}\\
\hline

$w_{b\textrm{-tag}}^{\mathrm{Higgs}}$ \rule{0pt}{3ex}                          & \begin{minipage}[l]{0.55\textwidth}Sum of $b$-tagging discriminants of jets from best Higgs candidate from the reconstruction BDT\end{minipage} & - & \checkmark & -\\[2ex]
\hline
\hline
\end{tabular}
\label{tab:dlClassBDTvars}
\end{center}
\end{table}

\begin{table}[!htb]
\centering     
\caption{Input variables to the classification BDTs in the single-lepton signal regions. 
For variables from the reconstruction BDT, those with a $^{*}$ are from the BDT using Higgs-boson information, 
those with no $^{*}$ are from the BDT without Higgs-boson information. 
These two versions of the reconstruction BDT are described in Section~\ref{sub:recobdt}.
The MEM$_{D1}$ variable is only used in \srsixjone, while variables based on the $b$-tagging discriminant are not used in this region.}
\vspace{-0.25cm}
\begin{center}
\begin{tabular}{l|l|cc}
\hline       
\hline
Variable & Definition &  SR$^{\geq6\textrm{j}}_{1,2,3}$ & SR$^{5\textrm{j}}_{1,2}$ \rule{0pt}{3ex} \\[0.5ex]
\hline
\hline
\multicolumn{4}{l}{General kinematic variables} \\ 
\hline
\drbbav       \rule{0pt}{3ex} & \begin{minipage}[l]{0.65\textwidth}Average $\Delta R$ for all $b$-tagged jet pairs                                                                      \end{minipage}& \checkmark & \checkmark \\[1ex]
\drbbmaxpt     & \begin{minipage}[l]{0.65\textwidth}$\Delta R$ between the two $b$-tagged jets with the largest vector sum $\pt$                                         \end{minipage}& \checkmark & --         \\[1ex]
\maxdeta       & \begin{minipage}[l]{0.65\textwidth}Maximum $\Delta\eta$ between any two jets                                                                            \end{minipage}& \checkmark & \checkmark \\[1ex]
\mbbmindr      & \begin{minipage}[l]{0.65\textwidth}Mass of the combination of two $b$-tagged jets with the smallest $\Delta R$                                      \end{minipage}& \checkmark & --         \\[2ex]
\mjjmindr      & \begin{minipage}[l]{0.65\textwidth}Mass of the combination of any two jets with the smallest $\Delta R$                                                 \end{minipage}& --         & \checkmark \\[1ex]
\nhiggsthirtybb  & \begin{minipage}[l]{0.65\textwidth}Number of $b$-tagged jet pairs with invariant mass within 30~\GeV\ of the Higgs-boson mass                                   \end{minipage}& \checkmark & \checkmark \\[2ex]
\hthad         & \begin{minipage}[l]{0.65\textwidth}Scalar sum of jet $\pt$                                                                                              \end{minipage}& --         & \checkmark \\[1ex]
\drlepbbmindr  & \begin{minipage}[l]{0.65\textwidth}$\Delta R$ between the lepton and the combination of the two $b$-tagged jets with the smallest $\Delta R$            \end{minipage}& --         & \checkmark \\[2.5ex]
\apla          & \begin{minipage}[l]{0.65\textwidth}$1.5 \lambda_2$, where $\lambda_2$ is the second eigenvalue of the momentum tensor~\cite{tensor} built with all jets \end{minipage}& \checkmark & \checkmark \\[2ex]
$H_1$           & \begin{minipage}[l]{0.65\textwidth}Second Fox--Wolfram moment computed using all jets and the lepton                                                    \end{minipage}& \checkmark & \checkmark \\[0.75ex]
\hline
\hline
\multicolumn{4}{l}{Variables from reconstruction BDT} \\
\hline
BDT output \rule{0pt}{3ex}                        &  \begin{minipage}[l]{0.65\textwidth}  Output of the reconstruction BDT     \end{minipage} & \checkmark$^*$ & \checkmark$^*$ \\[1ex]
$m_{bb}^{\mathrm{Higgs}}$          & \begin{minipage}[l]{0.65\textwidth}Higgs candidate mass                                             \end{minipage}& \checkmark     & \checkmark     \\[1ex] 
$m_{{H},b_{\mathrm{lep~top}}}$     & \begin{minipage}[l]{0.65\textwidth}Mass of Higgs candidate and $b$-jet from leptonic top candidate           \end{minipage}& \checkmark     & --             \\[1ex]
$\Delta R^{\mathrm{Higgs}}_{bb}$   & \begin{minipage}[l]{0.65\textwidth}$\Delta R$ between $b$-jets from the Higgs candidate            \end{minipage}& \checkmark     & \checkmark     \\[1ex] 
$\Delta R_{H,\ttbar}$              & \begin{minipage}[l]{0.65\textwidth}$\Delta R$ between Higgs candidate and \ttbar{} candidate system           \end{minipage}& \checkmark$^*$ & \checkmark$^*$ \\[1ex]
$\Delta R_{H,\mathrm{lep~top}}$    & \begin{minipage}[l]{0.65\textwidth}$\Delta R$ between Higgs candidate and leptonic top candidate          \end{minipage}& \checkmark     & --             \\[1ex]
$\Delta R_{H,b_{\mathrm{had~top}}}$& \begin{minipage}[l]{0.65\textwidth}$\Delta R$ between Higgs candidate and $b$-jet from hadronic top candidate\end{minipage}& --             & \checkmark$^*$ \\[0.75ex]
\hline
\hline
\multicolumn{4}{l} {Variables from likelihood and matrix element method calculations} \\
\hline
LHD  \rule{0pt}{3ex}                              & \begin{minipage}[l]{0.4\textwidth}Likelihood discriminant     \end{minipage}& \checkmark & \checkmark \\[1ex]
MEM$_{D1}$                        & \begin{minipage}[l]{0.5\textwidth}Matrix element discriminant (in \srsixjone{} only) \end{minipage}& \checkmark & --         \\[0.75ex]
\hline
\hline
\multicolumn{4}{l} {Variables from $b$-tagging (not in \srsixjone)}\\[0.1ex]
\hline
$w_{b\textrm{-tag}}^{\mathrm{Higgs}}$ \rule{0pt}{3ex}        & \begin{minipage}[l]{0.65\textwidth}Sum of $b$-tagging discriminants of jets from best Higgs candidate from the reconstruction BDT\end{minipage}& \checkmark & \checkmark \\[2ex]
$B_{\mathrm{jet}}^{3}$         & \begin{minipage}[l]{0.65\textwidth}3$^\mathrm{rd}$ largest jet $b$-tagging discriminant                                 \end{minipage}& \checkmark & \checkmark \\[1ex]
$B_{\mathrm{jet}}^{4}$         & \begin{minipage}[l]{0.65\textwidth}4$^\mathrm{th}$ largest jet $b$-tagging discriminant                                 \end{minipage}& \checkmark & \checkmark \\[1ex]
$B_{\mathrm{jet}}^{5}$         & \begin{minipage}[l]{0.65\textwidth}5$^\mathrm{th}$ largest jet $b$-tagging discriminant                                 \end{minipage}& \checkmark & \checkmark \\[0.75ex]
\hline
\hline
\end{tabular}
\end{center}
\label{tab:inputVarsClassSL}
\end{table}

\begin{table}[!htb]
\centering    
\caption{Input variables to the classification BDT in the boosted single-lepton signal region. Additional $b$-jets are $b$-jets not contained in the Higgs-boson and top-quark candidates.}
\begin{center}
\begin{tabular}{l|l}
\hline       
\hline
Variable & Definition \\
\hline
\hline
\multicolumn{2}{l}{Variables from jet reclustering} \\
\hline
$\Delta R_{H,t}$ \rule{0pt}{3ex}        & $\Delta R$ between the Higgs-boson and top-quark candidates               \\[1ex]
$\Delta R_{t,b^{\text{add}}}$   & $\Delta R$ between the top-quark candidate and additional $b$-jet   \\[1ex]
$\Delta R_{H,b^{\text{add}}}$   & $\Delta R$ between the Higgs-boson candidate and additional $b$-jet \\[1ex]
$\Delta R_{H,\ell}$     & $\Delta R$ between the Higgs-boson candidate and lepton             \\[1ex]
$m_{\mathrm{Higgs~candidate}}$                    & Higgs-boson candidate mass                                          \\[1ex]
$\sqrt{d_{12}}$          & Top-quark candidate first splitting scale \cite{STDM-2011-19}                          \\[1ex]
\hline
\hline
\multicolumn{2}{l} {Variables from $b$-tagging}\\
\hline
$w_{b\textrm{-tag}}^{ }$ \rule{0pt}{3ex}                & Sum of $b$-tagging discriminants of all $b$-jets                          \\[1ex]
$w_{b\textrm{-tag}}^{\text{add}} / w_{b\textrm{-tag}}^{ }$   & Ratio of sum of $b$-tagging discriminants of additional $b$-jets to all $b$-jets \\[1.5ex]
\hline
\hline
\end{tabular}
\end{center}
\label{tab:inputVarsClassBoosted}
\end{table}
 
\clearpage

\part*{Acknowledgments}

We thank CERN for the very successful operation of the LHC, as well as the
support staff from our institutions without whom ATLAS could not be
operated efficiently.

We acknowledge the support of ANPCyT, Argentina; YerPhI, Armenia; ARC, Australia; BMWFW and FWF, Austria; ANAS, Azerbaijan; SSTC, Belarus; CNPq and FAPESP, Brazil; NSERC, NRC and CFI, Canada; CERN; CONICYT, Chile; CAS, MOST and NSFC, China; COLCIENCIAS, Colombia; MSMT CR, MPO CR and VSC CR, Czech Republic; DNRF and DNSRC, Denmark; IN2P3-CNRS, CEA-DRF/IRFU, France; SRNSFG, Georgia; BMBF, HGF, and MPG, Germany; GSRT, Greece; RGC, Hong Kong SAR, China; ISF, I-CORE and Benoziyo Center, Israel; INFN, Italy; MEXT and JSPS, Japan; CNRST, Morocco; NWO, Netherlands; RCN, Norway; MNiSW and NCN, Poland; FCT, Portugal; MNE/IFA, Romania; MES of Russia and NRC KI, Russian Federation; JINR; MESTD, Serbia; MSSR, Slovakia; ARRS and MIZ\v{S}, Slovenia; DST/NRF, South Africa; MINECO, Spain; SRC and Wallenberg Foundation, Sweden; SERI, SNSF and Cantons of Bern and Geneva, Switzerland; MOST, Taiwan; TAEK, Turkey; STFC, United Kingdom; DOE and NSF, United States of America. In addition, individual groups and members have received support from BCKDF, the Canada Council, CANARIE, CRC, Compute Canada, FQRNT, and the Ontario Innovation Trust, Canada; EPLANET, ERC, ERDF, FP7, Horizon 2020 and Marie Sk{\l}odowska-Curie Actions, European Union; Investissements d'Avenir Labex and Idex, ANR, R{\'e}gion Auvergne and Fondation Partager le Savoir, France; DFG and AvH Foundation, Germany; Herakleitos, Thales and Aristeia programmes co-financed by EU-ESF and the Greek NSRF; BSF, GIF and Minerva, Israel; BRF, Norway; CERCA Programme Generalitat de Catalunya, Generalitat Valenciana, Spain; the Royal Society and Leverhulme Trust, United Kingdom.

The crucial computing support from all WLCG partners is acknowledged gratefully, in particular from CERN, the ATLAS Tier-1 facilities at TRIUMF (Canada), NDGF (Denmark, Norway, Sweden), CC-IN2P3 (France), KIT/GridKA (Germany), INFN-CNAF (Italy), NL-T1 (Netherlands), PIC (Spain), ASGC (Taiwan), RAL (UK) and BNL (USA), the Tier-2 facilities worldwide and large non-WLCG resource providers. Major contributors of computing resources are listed in Ref.~\cite{ATL-GEN-PUB-2016-002}.

\clearpage
\printbibliography

\clearpage
\begin{flushleft}
{\Large The ATLAS Collaboration}

\bigskip

M.~Aaboud$^\textrm{\scriptsize 137d}$,
G.~Aad$^\textrm{\scriptsize 88}$,
B.~Abbott$^\textrm{\scriptsize 115}$,
O.~Abdinov$^\textrm{\scriptsize 12}$$^{,*}$,
B.~Abeloos$^\textrm{\scriptsize 119}$,
S.H.~Abidi$^\textrm{\scriptsize 161}$,
O.S.~AbouZeid$^\textrm{\scriptsize 139}$,
N.L.~Abraham$^\textrm{\scriptsize 151}$,
H.~Abramowicz$^\textrm{\scriptsize 155}$,
H.~Abreu$^\textrm{\scriptsize 154}$
% ...
\bigskip
\\
% ...
$^{12}$ Institute of Physics, Azerbaijan Academy of Sciences, Baku, Azerbaijan\\
% ...
$^{88}$ CPPM, Aix-Marseille Universit{\'e} and CNRS/IN2P3, Marseille, France\\
% ...
$^{115}$ Homer L. Dodge Department of Physics and Astronomy, University of Oklahoma, Norman OK, United States of America\\
% ...
$^{119}$ LAL, Univ. Paris-Sud, CNRS/IN2P3, Universit{\'e} Paris-Saclay, Orsay, France\\
% ...
$^{137}$ $^{(a)}$ Facult{\'e} des Sciences Ain Chock, R{\'e}seau Universitaire de Physique des Hautes Energies - Universit{\'e} Hassan II, Casablanca; $^{(b)}$ Centre National de l'Energie des Sciences Techniques Nucleaires, Rabat; $^{(c)}$ Facult{\'e} des Sciences Semlalia, Universit{\'e} Cadi Ayyad, LPHEA-Marrakech; $^{(d)}$ Facult{\'e} des Sciences, Universit{\'e} Mohamed Premier and LPTPM, Oujda; $^{(e)}$ Facult{\'e} des sciences, Universit{\'e} Mohammed V, Rabat, Morocco\\
% ...
$^{139}$ Santa Cruz Institute for Particle Physics, University of California Santa Cruz, Santa Cruz CA, United States of America\\
% ...
$^{151}$ Department of Physics and Astronomy, University of Sussex, Brighton, United Kingdom\\
% ...
$^{154}$ Department of Physics, Technion: Israel Institute of Technology, Haifa, Israel\\
$^{155}$ Raymond and Beverly Sackler School of Physics and Astronomy, Tel Aviv University, Tel Aviv, Israel\\
% ...
$^{161}$ Department of Physics, University of Toronto, Toronto ON, Canada\\
% ...
$^{a}$ Also at Department of Physics, King's College London, London, United Kingdom\\
$^{b}$ Also at Institute of Physics, Azerbaijan Academy of Sciences, Baku, Azerbaijan\\
$^{c}$ Also at Novosibirsk State University, Novosibirsk, Russia\\
$^{d}$ Also at TRIUMF, Vancouver BC, Canada\\
$^{e}$ Also at Department of Physics {\&} Astronomy, University of Louisville, Louisville, KY, United States of America\\
% ...
$^{*}$ Deceased
\end{flushleft}


\end{document}